\documentclass[a4paper]{aa} 

\usepackage{epstopdf}
\usepackage{graphicx}
\usepackage{grffile}
\usepackage[varg]{txfonts}
\usepackage{natbib}
\usepackage{lscape}
\usepackage[font={small,rm}]{caption}
\usepackage{multicol}
\usepackage{array}
\usepackage{bm}
\usepackage{subfig}
\usepackage{setspace}
\usepackage{float}
\usepackage{bm}
\usepackage{color}

\begin{document}

  \title{Understanding the water emission in the mid- and far-IR\\ from protoplanetary disks around T~Tauri stars}
   \subtitle{}
   \author{Antonellini$^{1}$, S.\and Kamp, I.$^{1}$\and P., Riviere-Marichalar$^{2}$\and Meijerink, R.$^{3}$\and  Woitke, P.$^{4}$\and Thi, W.-F.$^{5,6}$\and Spaans, M.$^{1}$\and Aresu, G.$^{1,7}$\and Lee, G.$^{4}$}
\authorrunning{Antonellini et al.}

           \institute{Kapteyn Astronomical Institute, University of Groningen, Postbus 800, 9700 AV Groningen, The Netherlands\\
                      \email{antonellini@astro.rug.nl}\and
                      Centro de Astrobiolog\'ia (INTA-CSIC) - Depto. Astrof\'isica, POB 78, ESAC Campus, 28691 Villanueva de la Ca\~nada, Spain\and
                      Leiden Observatory, Leiden University, PO Box, 2300 RA Leiden, The Netherlands\and      
                      SUPA, School of Physics and Astronomy, University of St. Andrews, St. Andrews KY16 9SS, UK\and
                      Univ. Grenoble Alpes, IPAG, F-38000 Grenoble, France CNRS, IPAG, F-38000 Grenoble, France\and
                      Max-Planck-Institut f$\rm\ddot{u}$r extraterrestrische Physisk, Giessenbachstrasse 1, 85748 Garching, Germany\and
                      INAF, Osservatorio Astronomico di Cagliari, via della Scienza 5, 09047 Selargius, Italy
}

           \date{}

        % \abstract{}{}{}{}{} 
        % 5 {} token are mandatory
         
          \abstract
          % context heading (optional)
          %{} leave it empty if necessary  
{}
% aims heading (mandatory)
{We investigate which properties of protoplanetary disks around T~Tauri stars affect the physics and chemistry in the regions where mid- and far-IR water lines originate and their respective line fluxes. We 
search for diagnostics for future observations.}   
          % methods heading (mandatory)
{With the code ProDiMo, we build a series of models exploring a large parameter space, computing rotational and ro-vibrational transitions of water in nonlocal thermodynamic equilibrium (non-LTE). 
We select a sample of transitions in the mid-IR regime and the fundamental ortho and para water transitions in the far-IR. We investigate the chemistry and the local physical conditions in the line emitting regions.
%We analyzed dependencies of the sample of real observed water transitions in the Spitzer/IRS and Herschel/HIFI regime with Pearson statistics that cross-correlate the fluxes with physical properties of the emitting region.
We calculate Spitzer spectra for each model and compare far-IR and mid-IR lines. In addition, we use mid-IR colors to tie the water line predictions to the dust continuum.} 
        % results heading (mandatory)
{Parameters affecting the water line fluxes in disks by more than a factor of three are : the disk gas mass, the dust-to-gas mass ratio, the dust maximum grain size, ISM (InterStellar Medium) UV radiation field, the mixing 
parameter of Dubrulle settling, the disk flaring parameter, and the dust size distribution. The first four parameters affect the mid-IR lines much more  than the far-IR lines.}
        % conclusions heading (optional), leave it empty if necessary 
{A key driver behind water spectroscopy is the dust opacity, which sets the location of the water line emitting region. We identify three types of parameters, including those (1) affecting global disk opacity and opacity function 
(maximum dust size and dust size distribution), (2) affecting global disk opacity (dust-to-gas mass ratio, Dubrulle settling, disk gas mass), and (3) not affecting disk opacity (flaring parameter, ISM UV radiation field, fraction of PAHs).
Parameters, such  as dust-to-gas ratio, ISM radiation field, and dust size distribution, affect  the mid-IR lines more, while the far-IR transitions are more affected by the flaring index. The gas mass greatly affects  lines in both
regimes.
Higher spectral resolution and line sensitivities, like from the James Webb Space Telescope, are needed to detect a statistically relevant sample of individual water lines to distinguish further between these types of 
parameters.}

{}

{}
           \maketitle
        %
        %________________________________________________________________

\section{Introduction}\vspace{5mm}

Water is a fundamental component of protoplanetary disks. Its role is already pivotal in the disk formation phase as an important coolant in the cloud collapse \citep[see, e.g.,\ recent works,][]{zinnecker,vandertak,vandishoeck2}. 
Water ice mantles can facilitate the sticking between dust grains \citep[][]{machida1}. To date water has even been detected in several exoplanets \citep[e.g.,\ ][]{brogi,fraine,crouzet}, and is common throughout the 
whole solar system in comets, meteorites, moons, and rocky planets \citep[][]{encrenaz}. The observed abundance of water on Earth, Mars, and maybe Venus (combined with D/H ratio measurements) suggests a delivery process by 
comets \citep{vandishoeck2}.

Water in protoplanetary disks around T~Tauri stars has been detected over a wide spectral range: near-IR, ground-based spectra \citep[][]{salyk}, mid-IR Spitzer, VISIR spectra \citep[][]{carr,pontoppidan1,pontoppidan2}, 
and far-IR Herschel spectra \citep[][]{hogerheijde,riviere-marichalar}. For Spitzer, the detected transitions span from 5 to 36 $\mu$m, and they originate from warm water gas layers ($T$ $\simeq$~100-2000~K) close to the 
planet-forming region. At the low resolution of Spitzer, these lines are blended in complexes with very close upper level energies ($E_\mathrm{up}$).
It is possible to use different $E_\mathrm{up}$ of the transitions to probe the radial distribution of water in disks \citep[][]{pontoppidan2,zhang}.
Spitzer observations found a correlation between water and HCN in T~Tauri stars. The presence of these molecular features can be explained with low continuum opacity \citep[possibly due to settling;][]{najita1}.
\citet{riviere-marichalar} reported Herschel/PACS observations of o-H$_{2}$O 63.32~$\mu$m (8$_{18}$-7$_{07}$, $E_\mathrm{up}$~=~1071 K). %and [OI] 63 $\mu$m. %The two line fluxes 
They found them to be consistent with emission from within 3 AU, as predicted by ProDiMo modeling of a typical T~Tauri disk. %\citep{riviere-marichalar}.
The Herschel/HIFI water lines have only been detected  for TW~Hya \citep{hogerheijde} and DG~Tau \citep{podio}. \citet{bergin2} report a tentative detection toward DM Tau. The reason for this low detection 
rate can be due to water first freezing on dust, then settling and possibly ice transportation to the midplane beyond the snow line \citep[][]{meijerink,hogerheijde,hogerheijde-TWHya}, or because of photodissociation in 
the outer disk \citep{dominik}. Two modeling studies for TW Hya and DG~Tau performed with ProDiMo show different gas temperatures in these regions (\citealp[$\simeq$ 30 K for TW Hya;][]{kamp}; \citealp[$\simeq$~50-600~K for DG~Tau;][]{podio}). 

Next to these observational studies, disk modeling has been carried out using various degrees of complexity and self-consistency in the numerical codes. \citet{meijerink2} modeled far-IR water transitions with an X-ray 
thermochemical code using a typical T~Tauri disk model. They show that the far-IR lines probe the physical conditions in the outer disk (10-100 AU) and the level of turbulence. 
These lines are typically in non-LTE because of their large critical densities \citep[$n_\mathrm{cr}\!>\!10^{8-12}\!\mathrm{cm}^{-3}$][]{meijerink}.
Using a simplified water abundance prescription, \citet{meijerink} find that dust depletion (due to settling, dust growth, lower dust-to-gas ratio) is required to produce the observed line-to-continuum ratio in the far-IR line 
emitting regions.
\citet{glassgold}, using an X-ray irradiated chemical slab model at 1 AU, reproduced molecular column densities of water consistent with the observations of IR lines toward T~Tauri disks when H$_2$ is efficient formation on dust 
grains is considered.
A DM Tau study performed with a 1+1D disk model and very simple chemical network \citep{dominik} found an extended cold water reservoir (up to the outer disk, 800~AU). This reservoir is unaffected by the external FUV 
radiation field because of a balance between photodesorption and photodissociation. One of the main formation channels for water is through ion molecule chemistry. The particular disk model for DM Tau predicted the sub-mm water
lines to be in absorption and undetectable by the Herschel Space Telescope because of beam dilution.

\citet{woitke2} used a radiation thermochemical disk model for a typical disk around a Herbig star to identify the presence of three water reservoirs: an inner one in LTE (log$\epsilon(\mathrm{H_2O})\!\sim\!10^{-4}$; 
$\epsilon(\mathrm{H_2O})\!=n_\mathrm{H_2O}/n_\mathrm{tot}$), extended up to 10 AU dominated by equilibrium chemistry and endothermic reactions; a cold outer belt ($R \sim$~20-150~AU; $z/r~\lesssim$~0.05) with desorption 
and adsorption in tight equilibrium, where the water vapour is mainly photodissociated by the stellar FUV photons; and an intermediate irradiated region ($R~\sim$~1-30 AU; $z/r~\lesssim$~0.1-0.3), where water is mainly 
produced by neutral-neutral channels and destroyed by photodissociation. Far-IR rotational lines with higher $E_\mathrm{up}$ are produced from surface layers ($z/r$~$\sim$~0.1-0.3). Woitke et al. also found that a computation of the 
level populations using vertical escape probability underestimates the population compared to a detailed Monte Carlo computation, affecting the final line fluxes by 2-45\%.

\citet{meijerink} found that the role of X-ray heating (10-40\% efficiency) for neutral-neutral water formation is more important than FUV heating. If X-ray heating dominates locally, then far-IR lines are stronger. 
Simultaneously, also ion-molecule reactions become more important \citep{meijerink1}.

Recent modeling by \citet{heinzeller} performed with a thermochemical code found that mixing and settling are important in limiting the water formation if the high H$_2$ formation rate on dust grains from \citet{cazaux} is used.
Including these effects, they find a better agreement with the observed column densities. Non-LTE is required to reproduce the observed mid-IR line fluxes.

\citet{du} use radiation thermochemical disk models to study the impact of the stellar radiation field and disk properties on the water column densities/masses in a typical T~Tauri disk with a focus on the inner warm reservoir
($\lesssim5$~AU). At very low dust opacities (dust-to-gas mass ratio $<~10^{-4}$), photodissociation can supress the gas-phase water reservoir inside 1.3~AU. Large dust grains with a smaller scale height than the small grain 
population (mimicking settling) produces a similar effect. 

What is still lacking after the history of modeling described above is a detailed systematic study of how the disk and stellar parameters affect the key water emission lines in the mid- and far-IR. We want to 
understand what conclusions on specific pure disk properties can be drawn, given the existing Spitzer and Herschel observations, and also in the light of the new capabilities with JWST/MIRI. This kind of study requires a 
combination of radiation thermochemical disk models with non-LTE treatment of water level populations, thus taking  the dust for the underlying physical disk structure into account, as well as the chemical structure for 
the subsequent computation of emission lines (fraction of water column above the dust continuum). This is the aim of the work presented here.
In a next paper (Antonellini et al., in prep.), we will address the effect of a different central star on the disk mid-IR water spectroscopy, trying to find the reason for the different behavior in observations toward 
low-mass and high-mass pre-main-sequence stars.

In Section~\ref{2}, we describe our code, the line selection, the standard disk, and the series of models, also motivating  the explored parameter space. In Section~\ref{3}, we present the results of the standard model, its 
chemistry, and water spectroscopy. In Section~\ref{4}, we present the results of our parameter study, and finally, in Section~\ref{5} and \ref{6}, we present the discussion and conclusions.

        %__________________________________________________________________

\section{Modeling procedure}\vspace{5mm}
\label{2}

Our code ProDiMo (Protoplanetary Disk Model) is a radiation thermochemical code, which self-consistently computes the physical and chemical structure of disks. The code includes 2D radiative transfer, and gas chemistry, 
including freeze-out and desorption processes; it can perform hydrostatic or parametric description of an axisymmetric disk structure \citep{woitke}. X-ray chemistry has been implemented by \citet{aresu}.
The code computes the energy balance of the gas, including photoelectric heating, C photoionization, H$_2$ photodissociation and formation heating, collisional de-excitation of H$_2$, cosmic ray heating, viscous 
heating, cooling via Ly-$\alpha$, [OI] 6300~\AA, and more than 10000 atomic/molecular emission lines. Thermalization between dust and gas and free-free processes complete the energy balance.  

The radiation field comprises the central star and the ISM (InterStellar Medium) UV field. The multiwavelength radiative transfer is performed using accelerated $\Lambda$ iteration, and ray tracing (generally 100 rays). 
The statistical equilibrium of atoms/molecules, however, is computed considering a two-directional escape probability, assuming that the line source function varies slowly in local environments, where the optical depths grows 
quickly, and photon escape is mainly in the vertical direction. For each species treated in the spectroscopy, the code ProDiMo computes the statistical equilibrium based on a large set of molecular and collisional data. 
Optionally, a detailed line radiative transfer can be performed for a selected subsample of transitions.

The code has been updated with several features: a detailed computation of the photoionization and photodissociation cross-sections from \citet{vandishoeck3}, an extension of the collisional partners for the water statistical 
equilibrium \citep{kamp4}, a soft-edge truncation of the outer disk (Sect. \ref{flare}), PAH (Polycyclic Aromatic Hydrocarbons) ionization balance and heating/cooling, UV fluorescence pumping, a parametric description of 
settling, pumping by OH photodissocation, H$_2$ pumping by formation on dust grains, and chemical heating \citep{woitke1}. 
A prescription for the settling following the formalism of \citet{dubrulle} has been implemented and is described in Section \ref{turbulence}.

The chemical modeling has been performed using the UMIST 2006 database \citep{woodall}. Even though the UMIST 2012 database \citep{mcelroy} is now available, it was not implemented when our work started. If three-body 
reactions are included, however, the results for water are comparable with the older network (Kamp et al. in prep.). A test performed with the UMIST 2012 database and the standard model shows that the disk gaseous water 
distribution is unchanged (except a tiny abundance variation, which is very difficult to quantify in the outer disk), and the line fluxes of the far-IR water lines are affected by less than 25\% and those of the mid-IR lines by 
about 1\%. Our chemical network contains a complete series of gas-phase reactions for 94 atomic, ionic, and molecular species, involving water and related precursors. 
Surface reactions for water are not included in the network, as they are not relevant for the emitting regions of the  transitions that we consider \citep{kamp}. The chemistry is calculated assuming steady state. The 
chemical network and heating/cooling balance are solved iteratively. Gas heating via exothermic reactions is approximated via the chemical heating efficiency \citep{woitke1} using the reaction enthalpies at the reference 
temperature of 0 K from NIST\footnote{\tiny{http://www.nist.gov/}} and \citet{millar}. 
Dependence of the reaction thermodynamics on the real temperature is neglected.

We calculate the detailed water statistical equilibrium using a subset of the first 500 levels, including high-J rotational and ro-vibrational levels.
The ro-vibrational levels are computed from the work of \citet{tennyson} and \citet{barber}; the complete set of levels (824 in total) include 411 o-H$_2$O and 413 p-H$_2$O ro-vibrational levels, in which rotational levels up to
$J$ = 25 (and projections till $K_\mathrm{a,c}$~=~18) and vibrational levels up to $v_1$~=~2, $v_2$~=~2, $v_3$~=~1 are considered. 
Our water energy levels, lines, and collisional data \citep{green,faure,rothman} are the same as adopted in \citet{kamp}.

%\begin{table*}
%\caption{Water collisional data}
%\centering
%\begin{tabular}{|c|c|c|c|c|c|}
%\hline\hline
%Species & Type of transition & Collisional partner & References\\ \hline
%o/p H$_{2}$O & rotational$^{\mathrm{(a)}}$ & o/p H$_{2}$; H (He scaled by 1.39); e$^{-}$ & 1; 2; 1\\
%o/p H$_{2}$O & ro-vibrational$^{\mathrm(b)}$ & o/p H$_{2}$; e$^{-}$; H (H$_2$ scaled by $\sqrt{2}$) & 1; 1; 1\\ \hline%2\\
%\end{tabular}\label{collis}
%\tablefoot{(a)  (1) \citet{faure} (2) \citet{green}}
%\end{table*}
        %______________________________________________________________

The analysis of each model is based on the averaged physical and chemical quantities computed in the line emitting region that produces 50\% of the radially and vertically integrated flux (for details see Appendix \ref{app2}).

\subsection{Standard model and description of the parameter study}\vspace{5mm}

The standard model is a parametrized ``monolithic'' disk (properties defined in Table~\ref{standard} are the same throughout the whole disk structure) surrounding a typical T~Tauri K-type star with the parameters 
reported in Table~\ref{standard}. It represents a typical disk around a T~Tauri star without viscous heating ($\alpha_\mathrm{vis}~=~0.0$), with X-ray irradiation from the central star, a UV excess, a 20\% chemical heating 
efficiency due to exothermic reactions, a settling prescription for the dust according to \citet{dubrulle} (Sect. \ref{turbulence}), and a tapering off outer edge (Sect. \ref{flare}).

We explore individual parameters around this standard model, considering a series of models in which we vary a single parameter at a time. A list of the considered parameters in the series and adopted values are reported in 
Table~\ref{global}. For each parameter, the range spans from a factor of a few up to orders of magnitude around the initial standard value. We choose to model single parameters at a time  to disentangle 
and understand individual effects on both water spectroscopy and continuum fluxes/spectral index indicators in the same wavelength regime. Based on our results from model grids, such as DENT \citep{woitke1,kamp3} and the 
beta2 grid \citep{woitke3}, the interplay of parameters is often well represented by combining the effects of the individual parameters.

%We tested the interplay between gas mass and dust size power law distribution index, 
%using results from a non published grid of ProDiMo models to validate our approach. Modeling grids such as DENT \citep{woitke1,kamp3} and the beta\footnote{http://www.diana-project.com/data-results-downloads/model-grids/} grid 
%show that the interplay of multiple parameters is often well represented by combined effect of the individual ones.

The series deliberately extends beyond real observed disks to facilitate the detection of correlations and to disentangle the effects of the various individual parameters on the water lines. 
In the following subsections, we provide a short description of the different parameters investigated in this modeling.

\begin{table}
\centering
\caption{Overview of the fixed model parameters for the standard T~Tauri model} 
\begin{tabular}{p{4.5cm}p{1.9cm}p{1.5cm}}%{llc}
\hline\hline
\multicolumn{3}{c}{Central star and radiation field parameters}\\
%\multicolumn{3}{c}{for the standard T~Tauri model}\\
\hline
Parameter & Symbol & Value \\ \hline
Photospheric temperature & $T_\mathrm{eff}$ [K] & 4400\\
Stellar mass & $M_\mathrm{*}$ [M$_\mathrm{\odot}$] & 0.8\\
Stellar luminosity & $L_\mathrm{*}$ [L$_\mathrm{\odot}$] & 0.7\\ 
FUV excess & $L_\mathrm{UV}$/$L_\mathrm{*}$ & 0.01\\ 
UV powerlaw exponent & $p_\mathrm{UV}$ & 0.2\\ 
X-ray luminosity & $L_\mathrm{X}$ [erg/s] & 10$^{30}$\\
X-ray minimum energy & $E_\mathrm{min,X}$ [keV] & 0.1\\
X-ray Temperature & $T_\mathrm{X}$ [K] & 10$^7$\\ \hline\hline
\multicolumn{3}{c}{Disk parameters of the standard model fixed in the
series}\\
\hline
Parameter & Symbol & Value\\ \hline
Radial $\times$ vertical grid points & $N_\mathrm{xx}\times\! N_\mathrm{zz}$ & 70 $\times$ 70\\
%Disk mass & $M_\mathrm{gas}$ [M$_\mathrm{\odot}$] & 0.01\\
Outer radius & $R_\mathrm{out}$ [AU] & 300\\ 
Inner radius & $R_\mathrm{in}$ [AU] & 0.1\\
Surface density power law index & $\epsilon$ & 1.0\\ 
Minimum dust size & $a_\mathrm{min}$ [$\mu$m] & 0.05\\ 
Reference radius & $R_\mathrm{0}$ [AU] & 0.1\\ 
Scale height at reference radius & $H_\mathrm{0}$ [AU] &
3.5$\times10^{-3}$\\ 
Tapering-off radius & $R_\mathrm{taper}$ [AU] & 200\\ 
Chemical heating efficiency & - & 0.2\\ 
Settling description & - & Dubrulle \\
Cosmic ray ionization rate & $\zeta_\mathrm{CRs}$ [s$^{-1}$] &
1.7$\times10^{-17}$\\
Distance & $d$ [pc] & 140\\
Turbulence viscosity coefficient & $\alpha_\mathrm{vis}$ & 0\\
Disk inclination & [$\deg$] & 30\\
\hline
\end{tabular}
\label{standard}
\end{table}

\begin{table*}
\centering
\caption{Disk parameters varied in the series}
\begin{tabular}{l l l}
\hline\hline
Parameter & Symbol & Series values\\ \hline
Flaring power law index & $\beta$ & 0.8, 0.85, 0.9, 0.95, 1.0, 1.05,
1.1, $\bm{1.13}$, 1.15, 1.2, 1.25\\ %\hline
Dust-to-gas mass ratio & $d/g$ & 0.001, $\bm{0.01}$, 0.1, 1, 10, 100\\
Gas mass & $M_\mathrm{gas}$ [M$_\mathrm{\odot}$] & 10$^{-5}$,
10$^{-4}$, 0.001, $\bm{0.01}$, 0.05, 0.1 \\ 
Mixing parameter & $\alpha_\mathrm{set}$ & 10$^{-5}$,
10$^{-4}$, 10$^{-3}$, 10$^{-2}$, $\bm{0.05}$, 0.1\\ 
ISM radiation field & $\chi_\mathrm{ISM}$ [$G_{0}^{*}$] & 0.5, $\bm{1}$, 10,
100, 1000, 10$^{4}$, 10$^{5}$, 10$^{6}$ \\
Fraction of PAH$^{**}$ & $f_\mathrm{PAH}$ & 0.0001, 0.001, $\bm{0.01}$, 0.1, 1.0\\
Maximum dust size & $a_\mathrm{max}$ [$\mu$m] & 250, 400, 500, 700,
$\bm{1000}$, 2000, 5000, 10$^{4}$, 10$^{5}$\\
Dust size power law index & $a_\mathrm{pow}$ & 2.0, 2.5, 3.0,
$\bm{3.5}$, 4.0, 4.5\\ \hline
\end{tabular}
\tablefoot{Bold numbers refer to standard model values; (*) Draine units $G_\mathrm{0}~=~1.6\times10^{-3}$~erg~cm$^{-2}$ s$^{-1}$ \citep{draine3}; (**) Relative to ISM abundance 
\citep[$\sim$ 3$\times$10$^{-7}$ per H nucleus;][]{tielens1}} 
\label{global}
\end{table*}

Several other parameters have been studied as well but we do not discuss them because they do not significantly affect the mid-IR lines, such as the cosmic ray ionization rate (less than 1\% variation), the chemical 
heating efficiency (factor less than 3), the disk outer radius (4\% variation), the radioactive decay rate of unstable nuclei and related heating (less than 1\%), the tapering-off radius (about 10\%), and the surface density 
power law index (about 10\%). Some of these parameters particularly affect  the optically thick regions of the disk, and will be discussed in a future paper on ice reservoirs.

\subsubsection{Dust-to-gas mass ratio}\vspace{5mm}

In the intestellar medium, dust represents approximately 1\% of the total mass \citep{flower}, while this value can be different in environments in which dust grows, settles and the gas photoevaporates (as in protoplanetary disks). For example, an extreme ratio can occur in the case of a debris disk, in which the disk is essentially composed of dust, probably of secondary origin \citep[recent review by][]{wyatt}.
A common hypothesis is that the dust-to-gas mass ratio is related to the age of the object, and expected to increase with age. 
However, this picture might be oversimplified, since there are detected objects in millimeter observations with low dust content, but still high accretion rates, for example, toward star formation environments like Cep
OB2 \citep{sicilia-aguilar1}. This suggests that the dust content of the disk alone cannot be considered a diagnostic for the disk age.

In our series of models, we explore a range of dust-to-gas mass ratio spanning from 10$^{-3}$ up to 100.0 to reproduce the situation of dust poor and dust-rich protoplanetary disks. The highest dust-to-gas mass ratio 
represents the most evolved systems.

\subsubsection{Turbulent viscosity and settling}\vspace{5mm}
\label{turbulence}

A description of the dust settling has been implemented in ProDiMo following the approach of \citet{dubrulle}. 
The description of the adopted settling is based on the balance between gravity and vertical mechanical gas mixing effects in disks on dust grains. Since the effect is grain size dependent, it produces a different corresponding 
scale height for each dust grain size. \citet{dubrulle} describe settling as a function of the grain size, density, relative dust and gas velocity, and the mixing parameter ($\alpha_\mathrm{set}$).
The prescription of the gas vertical mixing is conceptually the same as the turbulent motion responsible for the angular momentum redistribution through the disk and accretion, i.e.,\nolinebreak[4]

\begin{equation}\label{vis}
 \nu = \alpha_\mathrm{vis} c_\mathrm{s} H
\end{equation}

\noindent
Here, $\nu$ is the kinematical viscosity of the gas, $c_\mathrm{s}$ is the sound speed, and $H$ the gas scale height \citep{shakura}. The two $\alpha$'s are physically the same quantities, but here we model passive disks 
(without viscous heating and accretion, and so $\alpha_\mathrm{vis}~=~0$) and use $\alpha_\mathrm{set}$ to describe the amount of settling in the disk. 
The relation between settling, viscous accretion, and UV excess is still not fully understood and  additional factors,
such as stellar activity and dead zones, possibly contribute as well.

We consider a range of $\alpha_\mathrm{set}$ that spans from 0.1 to 10$^{-5}$. The  dust settling changes the corresponding dust scale height with respect to the gas scale height. As a consequence, settling produces a local 
variation of the dust-to-gas mass ratio. In particular, a low value of $\alpha_\mathrm{set}$ means that the dust is less homogeneously mixed with the gas and hence strongly settled.

\subsubsection{Structure of the disk}\vspace{5mm}
\label{flare}

The geometrical shape of the disk is determined by the flaring index $\beta$ and the tapering-off radius $R_\mathrm{taper}$.
The scale height $H$ of the disk is parametrized as a power law function of the radius $r$

\begin{equation}\label{scale}
\centering
H(r) =  H_0 \left(\frac{r}{R_0}\right)^\beta
\end{equation}

\noindent
The surface density prescription follows that often used in observational studies \citep[][]{degregorio,guilloteau}, i.e.,

\begin{equation}\label{dens}
\centering
\Sigma(r) = \int_{0}^{z_\mathrm{max}}\rho(r,z)dz \propto r^{-\epsilon}e^{-\left(\frac{r}{R_\mathrm{taper}}\right)^{2-\gamma}}
\end{equation}

%\begin{equation}\label{dens}
%\centering
%\Sigma(r) = (2-\gamma)\frac{M_\mathrm{disk}}{2\pi R_\mathrm{c}^2}\left(\frac{r}{R_\mathrm{taper}}\right)^{-\gamma}\mathrm{exp}\left[-\left(\frac{r}{R_\mathrm{taper}}\right)^{2-\gamma}\right]
%\end{equation}\\
\noindent
It includes an exponential radial cut-off (beyond the tapering-off radius $R_\mathrm{taper}$), which affects the flaring structure in the outer disk. The parameter $\epsilon$ is the power law index of the surface density; $\gamma$ is an 
exponent that regulates the radial cut-off, $\gamma = \mathrm{min}(2,\epsilon)$; and $M_\mathrm{gas}$ is the disk gas mass.
The vertical mass distribution of a disk follow hydrostatics, and the density decreases exponentially with $z$, accordingly to the radial scale height ($H$), 

{\begin{equation}\label{den}
\centering
\rho(r,z) =  \frac{\Sigma(r)}{\sqrt{2\pi}H}\mathrm{exp}\left(-\frac{z^2}{2H^2}\right)
\end{equation}

Observations of protoplanetary disks are often found to be consistent with the previous theoretical description, and have flared structures, with a scale height increasing with the radius \citep{kenyon}. HST scattered 
light images of disks in Taurus \citep{burrows,stapelfeldt,padgett} and in Orion \citep{smith}  show  the flared silhouette directly. Flaring is important for the thermal, ionization, and chemical structure of the disk. It is 
a relevant aspect of disk evolution, since the amount of irradiation of the outer regions affects  photoevaporation \citep[see recent review by][]{williams}.
The tapered disk profile can for example explain the differences in size of CO rotational emission and continuum images \citep[e.g.,][]{pietu,isella1,degregorio}.
Viscous accretion models indeed predict this kind of exponential outer cut-off \citep{lynden,hartmann}. In our series of models
we explore  very flat ($\beta~<~1.0$) to very flared disks (up to $\beta~=~1.25$)  to study a wide dynamic range.

\subsubsection{Dust grain sizes and size distributions}\vspace{5mm}

Dust grows in protoplanetary disks beyond sizes found in the ISM. This can be inferred from the SED slope in the mm range ($F~\propto~\nu^{\alpha_\mathrm{mm}}$). In particular, the slope in the range 0.5-1 mm is much shallower in 
protoplanetary disks \citep[$\rm\alpha_\mathrm{mm}\approx$~2-3][]{beckwith1,mannings1,andrews2,andrews3} than in the diffuse ISM \citep[$\rm\alpha_\mathrm{mm}~\sim$~4; ][]{boulanger}. There is evidence that the growth continues 
during the disk lifetime, since a change in the slope has been detected for class I (median $\alpha_\mathrm{mm}$~=~2.5) and class II objects \citep[median $\alpha_\mathrm{mm}$~=~1.8;][]{andrews2}.
These mm observations, however, only provide  a lower limit to the dust size, since the free-free emission from an ionized stellar wind can contaminate the emission for $\lambda\gtrsim$~1~cm \citep{natta1}. Assuming that 
the 3.5 cm emission detected from TW Hya \citep{wilner} is entirely from thermal emission of dust, a bimodal dust model \citep{draine} suggests the presence of grains up to $\sim$~10~cm, extended out at least to tens of
AU  to match the observations. However, growth of dust grains should proceed faster in the inner disk because of their dependence on orbital timescale and density. Hence the size distribution can also be a function of radius 
\citep[e.g.,\ ][]{dullemond,isella2,blum}.

During the growth of dust particles, competitive processes of depletion start to happen, like settling, differential mixing, and fragmentation, which are all size dependent. Fragmentation can explain the presence of 
small grains in protoplanetary disks ($a~<~100~\mu$m) after 10$^{4}$~yrs \citep{dullemond,zsom}. Small grains experience a complex interplay between collisions and coagulation \citep{dullemond1}. 
Large dust bodies (in particular, around a meter in size) are also affected by radial drift, which can enhance local selective dust depletion \citep[e.g.,][]{brauer,ormel}.
In the interstellar medium and in protoplanetary disks, the dust size distribution is often described as a power law in the form $n(a)~\sim~a^{-a_\mathrm{pow}}$. The previous described effects can lead to a change in the power 
law distribution over the lifetime of protoplanetary disks with clear deviations from the MRN distribution \citep{mathis,testi,birnstiel}.
From the observation of the continuum in the mm, the power law index $\alpha_\mathrm{mm}$ (see previous section) can be used to estimate the opacity function ($\kappa~\propto~\nu^{\beta}$), where 
$\beta$ is the power law index for the opacity function. From this, the dust size power law distribution can be inferred. $\beta$ is smaller than 1 for $a_\mathrm{max}~\!~\gtrsim~\!~0.5$ mm in a range of power law indices 
$a_\mathrm{pow}$ from 2.5 to 3.5 \citep{natta1,draine1}. 
Indexes that are smaller  than 3.5 are to be expected as a consequence of the dust growth, while a value above 3.5 is indicative of a regime of fragmentation \citep{weidenschilling}. 
Many observations are consistent with indexes between 2.5 and 3.3, e.g.,\ objects like TW Hya, CQ Tau, and HD 142666 \citep{natta1}. The dust size power law index is degenerate with the maximum dust size, however, 
values $\ge$ 4.0 do not fit the observations \citep{natta1}. 
To probe slightly beyond the observed range, we extended the upper limit to 4.5. The smaller the 
power law index, the higher the fraction of the total dust mass in big grains, thus lowering the UV opacity of the dust.
%With dust growth, we also expect a flattening in the power law grain size distribution. In order to disentangle these effects, we decide to separate the variation of the power law index from that of the maximum grain size.
%The model series involving this parameter considers different maximum sizes, from {\bf{250~$\mu$m}} up to 10~cm. For the power law index, we choose a range of values around the MRN value from 2.0 to 4.5. 

\subsubsection{The ISM UV radiation field}\vspace{5mm}

Star formation mostly happens in large clusters with hundreds to thousands of stars \citep{mckee,lada} with a relevant population of O and B-type stars. These are intense sources of UV radiation, which generate local UV radiation
fields that are several orders of magnitude stronger than the local UV field in the solar neighborhood. This strong radiation bath can erode the external surface regions of the disk.
Examples of these objects are cocoons embedded in an ionized sheath, like those observed in Orion by the Hubble Space Telescope \citep{bally}. In these objects, the far-UV (FUV) photons produce a thick photon dominated 
region (PDR), whose exposed gas layers reach $T~\ge$ 1000 K. 

The ISM UV radiation field has been extensively studied by \citet{draine3}. He defined the solar neighborhood integrated flux $G_\mathrm{0}~=~1.6\times10^{-3}$~erg~s$^{-1}$~cm$^{-2}$.
To probe the effects of the external radiation field, we create models with $\chi_\mathrm{ISM}$ spanning from less than the local solar neighborhood flux (0.5~$G_\mathrm{0}$) up to 10$^{6}~G_\mathrm{0}$. 
The high limit reproduces extreme cases like the Orion Bar \citep{marconi} or active galaxies \citep{vanderwerf}, which should also host protoplanetary disks as part of the star formation process. 

\subsubsection{PAHs in disks}\vspace{5mm}

From a study of the interstellar medium opacity function \citep{draine2}, dust is assumed to be a mixture of silicates and carbon compounds; however, polycyclic aromatic hydrocarbons (PAHs) contribute as well to the total opacity
function. 
PAHs are of major importance for the gas physics of the disks, since they represent an important source of heating in the irradiated layers \citep{kamp2}.
Usually, for protoplanetary disks, the fraction of PAHs ($f_\mathrm{PAHs}$) is assumed to be 0.01 with respect to the ISM abundance \citep[$\sim$~3$\times$10$^{-7}$ per H nucleus;][]{tielens1}. However, there are no clear 
observational constraints because for T~Tauri stars the PAH features are absent with relatively few exceptions \citep{gurtler}. A more detailed study concerning embedded YSOs suggests that the carriers of the PAHs are  
depleted by at least  an order of magnitude with respect to the ISM \citep{geers}. 

The PAH abundance can affect lines produced in the upper layers of the disk or in optically thin regions.
The model series assumes a range of $f_\mathrm{PAH}$ from 10$^{-4}$ to 1.0. In the current implementation, $f_\mathrm{PAH}$ only affects the heating of the gas and is not treated as an additional opacity source. 

\subsubsection{Gas masses}\vspace{5mm}

Gas generally dominates the total mass of protoplanetary disks, at least for the youngest objects. However, its detection is more challenging than the dust, since the emission is in discrete lines and bands. Their detection 
often requires high spectral resolution. Emission lines can only be detected if the gas excitation temperature ($T_\mathrm{ex}$) is larger than the dust temperature ($T_\mathrm{dust}$). 

The mass of the disk can be inferred from line observations in the (sub)mm, using tracers as CO and CO isotopologues (\element[][13]{C}{O}, C\element[][18]{O}) \citep[e.g.,][]{zuckerman,dent,miotello}. \citet{kamp3} propose a 
method for the gas mass estimation based on the [OI] 63 $\mu$m and CO $J$~=~2-1 transitions for low-mass disks ($M_\mathrm{gas}~<~10^{-3}$~M$_\mathrm{\odot}$). Alternatively, sub-mm continuum observations can be used to derive a 
disk mass from the dust mass, making assumptions on the dust-to-gas mass ratio \citep{beckwith1,andrews2,andrews3}. Continuum observations toward Taurus-Auriga and $\rho$ Oph are consistent with disk masses in the range 
10$^{-4}$~-~0.1~M$_\mathrm{\odot}$. 
To analyze the situation of the most extreme disks observed toward several SFRs, we investigate a range of $M_\mathrm{gas}$ from 10$^{-5}$ to 0.1~M$_\mathrm{\odot}$.

\section{Standard model results}\vspace{5mm}
\label{3}

Our standard disk model (Fig.~\ref{reservoirs}) shows three main water gas reservoirs (A,B,C) and an ice reservoir (I). A similar structure has been  found previously for a generic Herbig disk \citep{woitke2}. The classification 
of the different reservoirs adopted here is, however, different. Because of the overall lower disk temperature compared to Herbig disks, the ice reservoir is more extended in the case of the T~Tauri disks. In general, models of 
circumstellar disks  find similar structures for the water vapour distribution, but the size of the reservoirs  change according to the detailed physical conditions \citep{vandishoeck2}.
The radial extent of each reservoir is  defined here using the density and the emitting regions of the water lines in the standard model. The ice reservoir is shaped by the disk opacity. At $A_\mathrm{V}~\ge$~10, the dust 
temperatures and photodesorption rates are so low that H$_2$O freezes onto dust grains. The exact position of the snow line is a function of the gas temperatures and pressure \citep{lecar}. 

The inner zone of the disk (A) is a warm reservoir with high density ($\ge$~10$^9$~cm$^{-3}$), which produces the Spitzer lines. Disk (A) zone is that with the largest water column density in the disk, and is extended from the 
inner wall up to the snow line. The lowest abundance gas reservoir is the outer reservoir (C). In this region, the density is below 100~cm$^{-3}$, and the optical depth is low. Hence, transitions from this cold outer region of 
the disk probe deeper into the disk; examples are the Herschel/HIFI lines as 538.29~$\mu$m and 269.27~$\mu$m. Reservoir C is extended from the outer end of the snow line (45 AU) up to the outer disk radius. 
The intermediate region above the ice is the reservoir B ($z/r \sim$ 0.1-0.25); it is radially extended between the reservoirs A and C. In this region, physical conditions (densities of H$_{2}$, H, e$^-$, temperatures of dust 
and gas) show large changes with radius. Many of the Herschel PACS and Spitzer Space Telescope high excitation lines (200~$<$~$E_\mathrm{up}$~$<$~1000~ K, not analyzed in this work) arise in this disk zone. 
Below this region, water is predominantly in the form of ice (I). The icy reservoir in Fig.~\ref{reservoirs} is here defined by a combination of the $T_\mathrm{dust}~\simeq~100$~K line and the location of the continuum 
opacity\footnote{$A_\mathrm{V}$ is the optical extinction in magnitudes; $r$ and $z$ are the radial and vertical coordinates: 
$A_\mathrm{V}(r,z)~=~\mathrm{min}(A_\mathrm{V}(r);A_\mathrm{V}(z);A_\mathrm{V}(R_\mathrm{out})-A_\mathrm{V}(r))$.} 
$A_\mathrm{V}$ = 10. In the ice reservoir, the water abundances of the gas phase are depleted by 5 orders of magnitude with respect to the upper reservoir B. The ice reservoir can be indirectly probed through ice features arising from 
dust covered by mantles of frozen water. The 3~$\mu$m NIR ice feature has been detected by \citet{hiroshi}, while \citet{mclure} reported the tentative detection of the 63~$\mu$m far-IR feature.
Gas and dust are thermally coupled in the regions of the line emission of reservoir A and C (Fig.~\ref{Standard model}). In reservoir B, dust and gas temperatures are partially decoupled.

\begin{figure} 
\resizebox{\hsize}{!}{\includegraphics{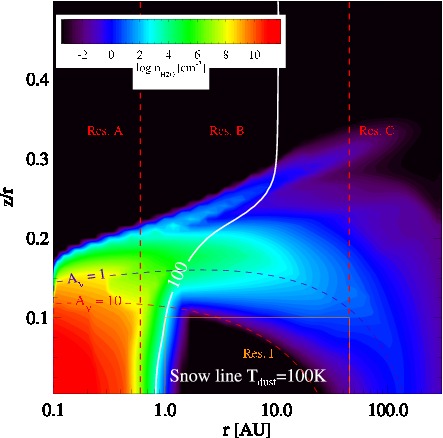}}
\caption{Water reservoirs in the standard disk; the white line represents the contour of $T_\mathrm{dust}~=~100$ K, the black and red dashed curved lines represent  the contours of optical extinction for 
$A_\mathrm{V}~=~1$ and $A_\mathrm{V}~=~10 $, respectively. The water ice reservoir is outlined by the white contour and the dashed line corresponds to $A_\mathrm{V}~=~10$. Vertical red dashed lines and the orange rectangle delimit the 
water reservoirs.} 
\label{reservoirs}
\end{figure}

In general, dust settling produces a lower local dust content in the surface layers of all models (Fig.~\ref{Standard model}) and steepens the local dust power law distribution with respect to a completely 
unsettled case. The effect of settling is particularly relevant in the outer disk, where the mixing is not strong enough to stir up the dust grains. This makes the average grain size $\left<a\right>$ close to the midplane larger 
with respect to an unsettled model.
 In the line emitting region of the 12.407~$\mu$m water line (Fig.~\ref{reserv_plot}) for the standard model the average dust size is 0.45 $\mu$m, while in the region of the 538.29~$\mu$m this size drops to 0.36~$\mu$m. The 
standard model has a high mixing parameter, and is thus close to an unsettled disk model.
In the water line emitting regions of all reservoirs ($z/r$~$\simeq$~0.15-0.2), above the $A_\mathrm{V}~>$ 1 layer, chemical heating and H$_{2}$ formation heating are the main heating processes (Fig.~\ref{Standard model}). The 
cooling is dominated by H$_{2}$O rotational and ro-vibrational emission and CO rotational and ro-vibrational emission. Deeper in the disk, gas-grain collisions couple $T_\mathrm{gas}$ and $T_\mathrm{dust}$.  
Additional plots on the standard model properties can be found in Appendix~\ref{app3} (Fig.~\ref{Standard model}).
 
\subsection{Water chemistry}\vspace{5mm}
\label{wchem}

A complete list of the main reactions involving water in the disk model is reported in Table~\ref{reactions}. The most important reactions in the emitting region of every transition are summarized in Table~\ref{rates}. They are
sorted by their relevance with respect to the timescale and the number of grid points in which they dominate.
The  kinetical reaction constants of the gas phase are expressed in a parametrized manner with three coefficients ($\alpha$, $\beta$, $\gamma$).
The calculation of the various types of reaction rates, using these coefficients, is described in \citet{woodall} and \citet{woitke}.

\begin{table*}
\caption{Main reaction channels in order or relevance for water formation and destruction in the standard disk model}
\centering
\resizebox{\textwidth}{!}{
\begin{tabular}{|c|c|c|c|c|c|c|}
\hline\hline
Reaction number & Reaction & $\rm\alpha$ [cm$^3$ s$^{-1}$] or [s$^{-1}$] & $\rm\beta$ [-] & $\rm\gamma$ [K] & Type & References\\ \hline
1 & \small{H$_2$ + OH $\rightarrow$ H$_2$O + H} & 2.05$\times10^{-12}$ & 1.52 & 1740.00 & \small{Neutral-Neutral} & 1\\
2 & \small{NH + OH $\rightarrow$ H$_2$O + N} & 3.11$\times10^{-12}$ & 1.20 & 0.00 & \small{Neutral-Neutral} & 2,1 \\
3 & \small{NH$_2$ + OH $\rightarrow$ H$_2$O + NH} & 7.78$\times10^{-13}$ & 1.50 & -230.00 & \small{Neutral-Neutral} & 1 \\
4 & \small{NH$_2$ + NO $\rightarrow$ N$_2$ + H$_2$O} & 4.27$\times10^{-11}$ & -2.50 & 331.00 & \small{Neutral-Neutral} & 2,1\\
5 & \small{OH + OH $\rightarrow$ H$_2$O + O} & 1.65$\times10^{-12}$ & 1.14 & 50.00 & \small{Neutral-Neutral} & 3\\
6 & \small{OH + H$_2$CO $\rightarrow$ HCO + H$_2$O} & 2.22$\times10^{-12}$ & 1.42 & -416.00 & \small{Neutral-Neutral} & 1\\
7 & \small{NH$_2$ + H$_3$O$^+$ $\rightarrow$ H$_2$O + NH$_3^+$} & 9.70$\times10^{-10}$ & 0.00 & 0.00 & \small{Ion-Neutral} & 4\\
8 & \small{H$_3$O$^+$ + Si $\rightarrow$ SiH$^+$ + H$_2$O} & 1.80$\times10^{-9}$ & 0.00 & 0.00 & \small{Ion-Neutral} & 4\\
9 & \small{H$_3$O$^+$ + SiO $\rightarrow$ SiOH$^+$ + H$_2$O} & 2.00$\times10^{-9}$ & 0.00 & 0.00 & \small{Ion-Neutral} & 5\\
10 & \small{H$_3$O$^+$ + e$^-$ $\rightarrow$ H$_2$O + H} & 1.08$\times10^{-7}$ & -0.50 & 0.00 & \small{Dissociative Recombination} & 6\\
11 & \small{H$^-$ + OH $\rightarrow$ H$_2$O + e$^-$} & 1.00$\times10^{-10}$ & -5.22 & 90.00 & \small{Associative Detachment} & 4\\
12 & \small{H + OH $\rightarrow$ H$_2$O + h$\nu$} & 5.26$\times10^{-18}$ & -5.22 & 90.00 & \small{Radiative Association} & 7\\
13 & \small{H$_3$O$^+$ + h$\nu$ $\rightarrow$ H$_2$O + H$^+$} & 2.00$\times10^{-11}$ & 0.00 & 2.00 & \small{Photodissociation} & 8\\
14 & \small{H$_2^{*}$ + OH $\rightarrow$ H$_2$O + H} & 3.60$\times10^{-11}$ & 0.00 & 0.00 & \small{Neutral-Neutral}$^{\mathrm{(a)}}$ & 1\\
15 & \small{H$_2$O$_{\rm(ice)}$ + dust $\rightarrow$ H$_2$O + dust} & - & - & 4800.00 & \small{Desorption}$^{\mathrm{(b)}}$ & 9,19 \\
16 & \small{H$_2$O$_{\rm(ice)}$ + C.R.P. $\rightarrow$ H$_2$O + dust} & - & - & 4800.00 & \small{Cosmic-rays-desorption}$^{\mathrm{(b)}}$ & 19 \\
17 & \small{H$_2$O$_{(ice)}$ + h$\nu$ $\rightarrow$ H$_2$O} & - & - & 4800.00 & \small{Photodesorption}$^{\mathrm{(b)}}$ & 9,19\\ 
18 & \small{PAH$^-$ + H$_3$O$^+$ $\rightarrow$ PAH + H$_2$O + H} & 1.70$\times10^{-8}$ & 0.50 & 0.00 & \small{PAH Dissociative Recombination} & 18\\ \hline
1 & \small{H$_2$O + H $\rightarrow$ OH + H$_2$} & 1.59$\times10^{-11}$ & 1.20 & 9610.00 & \small{Neutral-Neutral} & 1 \\
2 & \small{H$_3^+$ + H$_2$O $\rightarrow$ H$_3$O$^+$ + H$_2$} & 5.90$\times10^{-9}$ & 0.00 & 0.00 & \small{Ion-Neutral} & 10\\
3 & \small{C$^+$ + H$_2$O $\rightarrow$ HCO$^+$ + H} & 9.00$\times10^{-10}$ & 0.00 & 0.00 & \small{Ion-Neutral} & 4\\
4 & \small{CH$_5^+$ + H$_2$O $\rightarrow$ H$_3$O$^+$ + CH$_4$} & 3.70$\times10^{-9}$ & 0.00 & 0.00 & \small{Ion-Neutral} & 10\\
5 & \small{H$_2$O + Si$^+$ $\rightarrow$ SiOH$^+$ + H} & 2.30$\times10^{-10}$ & 0.00 & 0.00 & \small{Ion-Neutral} & 11\\
6 & \small{H$_2$O + HCO$^+$ $\rightarrow$ CO + H$_3$O$^+$} & 2.50$\times10^{-9}$ & 0.00 & 0.00 & \small{Ion-Neutral} & 12\\
7 & \small{H$_2$O + HN$_2^+$ $\rightarrow$ N$_2$ + H$_3$O$^+$} & 2.60$\times10^{-9}$ & 0.00 & 0.00 & \small{Ion-Neutral} & 13\\
8 & \small{H$_2$O + SiH$^+$ $\rightarrow$ Si$^+$ + H$_3$O$^+$} & 8.00$\times10^{-10}$ & 0.00 & 0.00 & \small{Ion-Neutral} & 5\\
9 & \small{H$^+$ + H$_2$O $\rightarrow$ H$_2$O$^+$ + H} & 6.90$\times10^{-9}$ & 0.00 & 0.00 & \small{Charge Exchange} & 14\\
10 & \small{He$^+$ + H$_2$O $\rightarrow$ He + H$_2$O$^+$} & 6.05$\times10^{-11}$ & 0.00 & 0.00 & \small{Charge Exchange} & 15\\
11 & \small{H$_2$O + h$\nu$ $\rightarrow$ OH + H} & 5.90$\times10^{-10}$ & 0.00 & 1.70 & \small{Photodissociation} & 16\\
12 & \small{H$_2$O + h$\nu$ $\rightarrow$ H$_2$O$^+$ + e$^-$} & 3.30$\times10^{-11}$ & 0.00 & 3.90 & \small{Photoionization} & 16\\%
13 & \small{H$_2$O + dust $\rightarrow$ H$_2$O$_{\rm(ice)}$} & - & - & - & \small{Adsorption}$^{\mathrm{(b)}}$ & 19\\
14 & \small{Ne$^+$ + H$_2$O $\rightarrow$ Ne + H$_2$O$^+$} & 8.00$\times10^{-10}$ & 0.00 & 0.00 & \small{Charge Exchange} & 17\\ \hline
\end{tabular}
}
\tablebib{(1) \citet{mallard}; (2) \citet{mcelroy}; (3) \citet{woodall}; (4) \citet{prasad}; (5) \citet{herbst}; (6) \citet{tielens}; (7) \citet{field}; (8) \citet{jensen}; (9) \citet{hollenbach1}; (10) \citet{kim}; (11) \citet{fahey}; (12) \citet{adams1}; 
(13) \citet{rakshit1}; (14) \citet{smith1}; (15) \citet{mauclaire}; (16) \citet{vandishoeck3}; (17) \citet{anicich1}; (18) \citet{flower1}; (19) \citet{aikawa}}
\tablefoot{(a) H$_2$ in a vibrationally excited state, populated by fluorescence from the electronically excited states;
(b) For desorption processes, $\gamma$ is the adsorption energy.}
\label{reactions}
\end{table*} 

\begin{table*}
\caption{Main reaction channels in the emitting regions of the considered transitions in the standard model}
\centering
\begin{tabular}{|>{\small}c|>{\small}c|>{\small}c|}
\hline\hline
Species \& wavelength [$\rm\mu m$] & Formation reaction number & Destruction reaction number\\ \hline
o-H$_{2}$O 12.396 & 1,10 & 1,11\\
p-H$_{2}$O 12.407 & 1,10 & 1,11\\
o-H$_{2}$O 12.445 & 1,10 & 1,11\\
o-H$_{2}$O 12.453 & 1,10 & 1,11\\
p-H$_{2}$O 12.832 & 1,10 & 1,11\\
p-H$_{2}$O 12.894 & 1,10 & 1,11\\
o-H$_{2}$O 33.510 & 1,10,4 & 1,11\\
p-H$_{2}$O 269.27 & 12,10,4 & 12,10,13,11,5\\
o-H$_{2}$O 538.29 & 12,10,4 & 12,10,13,11,5\\ \hline
\end{tabular}
\tablefoot{Reaction numbers in order of relevance.}
\label{rates}
\end{table*}

The emitting regions of the Spitzer transitions (except the 33.510~$\mu$m line) are co-spatial (Fig.~\ref{reserv_plot}). Hence the main processes and chemical rates are very similar. For these transitions, the main formation 
channels are the neutral-neutral reactions of H$_2$ and OH, and due to the presence of X-rays, the dissociative recombination of H$_3$O$^+$. For the 33.510~$\mu$m line, there is an additional neutral-neutral nitrogen chemistry 
channel via NH$_2$ and NO. The reaction rates (for a two-body reaction, $k_\mathrm{i,j}\cdot n_\mathrm{i}\cdot n_\mathrm{j}$~[s$^{-1}$]) in these dense and hot regions of the disk are above 10$^3$~s$^{-1}$.
Water in these regions is chemically mainly depleted through neutral-neutral reactions involving atomic hydrogen and through photodissociation. 

In the outer disk (reservoir~C), water is produced mainly through dissociative recombination of H$_3$O$^+$ and radiative association of OH and H. Secondary formation channels are the neutral-neutral reactions involving NO 
and NH$_2$. 
In this reservoir, several processes deplete water vapor: photoionization, charge exchange with He$^+$ and ion-neutral reaction with Si$^+$ (driven by X-rays, cosmic rays and collisional ionization with electrons), 
water photodissociation, and ice formation. The reaction rates here are lower than in reservoir A, of the order of 10$^{-5}$~s$^{-1}$~($\sim$~1~day), but the timescales are still shorter than the disk lifetime. 

Previous studies found differences in terms of nitrogen chemistry for T~Tauri disks. The channels mentioned above are an important formation path for water precursors (as OH) in the outer disk of TW~Hya, and in a tiny region
in the inner rim where water is directly formed from NO and NH$_2$ \citep{kamp}. In the \citet{najita2} chemistry network, atomic nitrogen is a competitor that depletes OH in the disk atmosphere, making it harder to produce 
water. However, this model only includes X-rays and not FUV and also lacks water rotational cooling. Water in the inner disk is formed through the activated channel involving OH and H$_2$, in agreement with our modeling. The 
\citet{glassgold} model does not include FUV and its network is limited to 125 reactions and 25 species, including only H, He, C, and O. This model agrees with our study on the water formation mechanism through neutral-neutral reaction of
OH and H$_2$, and finds that reactions with H$^+$ are the main destruction channels of water in the upper layers. 

In the modeling of Herbig disks, performed with a comparable chemical network, \citet{woitke2} found that dissociative recombination is not important in the inner disk, contrary to the case of T~Tauri disks. The outer disk water
reservoir of Herbigs is then fueled by photodesorption processes from the ices, while the depletion is again due to photodissociation. The main difference from the chemistry in T~Tauri disks is the lack of channels involving 
ions.

\subsection{Water spectroscopy}\vspace{5mm}
\label{WaterSpec}

Water has an extremely rich spectroscopy, making it necessary to select a subsample of lines for this study. We thus focus here on the list of transitions shown in Table~\ref{spectroscopy}, which are based on Spitzer and 
Herschel/HIFI line detections in disks around T~Tauri stars. 
A detailed investigation of each single line regarding the chemistry is only performed for the reference model (parameters are reported in boldface in Table~\ref{global}). 

\begin{table*}
\centering
\caption{Water lines selected for this work}
\begin{tabular}{cccccc}
\hline\hline
Species & Wavelength [$\mu$m] & Instrument & Reservoir & $E_\mathrm{up}$ [K] & Quantum numbers $v_\mathrm{1} v_\mathrm{2} v_\mathrm{3} J K_\mathrm{a}$ $K_\mathrm{c}$\\ \hline
p-H$_{2}$O & 12.407 & Spitzer IRS & A & 4945.0 & 000-000 16-15 3-2 13-14\\ 
o-H$_{2}$O & 12.396 & Spitzer IRS & A & 5781.0 & 000-000 17-16 4-3 13-14\\ 
o-H$_{2}$O & 12.445 & Spitzer IRS & A & 3629.0 &  000-000 11-10 8-5 3-6\\
o-H$_{2}$O & 12.453 & Spitzer IRS & A & 4213.0 & 000-000 13-12 7-4 6-9\\ 
p-H$_{2}$O & 12.832 & Spitzer IRS & A & 3243.0 & 000-000 10-9 8-5 2-5\\ 
p-H$_{2}$O & 12.894 & Spitzer IRS & A & 3310.0 & 000-000 12-11 5-2 7-10\\ 
o-H$_{2}$O & 33.510 & Spitzer IRS & A/B & 2275.0 & 000-000 10-9 4-3 7-6\\ 
p-H$_{2}$O & 269.27 & Herschel HIFI & B/C & 53.43 & 000-000 1-0 1-0 1-0\\ 
o-H$_{2}$O & 538.29 & Herschel HIFI & B/C & 60.96 & 000-000 1-1 1-0 0-1\\ \hline\hline
\end{tabular}
\label{spectroscopy}
\end{table*}

The mid-IR transitions originate from the warmest water reservoir (A, Fig.~\ref{reserv_plot}), consistent with their high upper level energies (Table~\ref{spectroscopy}). The two sub-mm lines are produced in the outer 
cold disk (reservoir C). We chose the mid-IR 12~$\mu$m lines  because they are also observable from the ground with VISIR on the VLT \citep{pontoppidan1} and they are physically unblended. This last aspect is very 
important, because our modeling  is based on individual line radiative transfer. The strongest detections in the Spitzer regime are the blends around 15~$\mu$m and 17~$\mu$m \citep{pontoppidan2,carr1,zhang}, but 
the difference from the 12~$\mu$m lines is within a factor 10. We verified that the behavior of the single components of the blends in the \citet{pontoppidan2} work is the same as that of the individual lines we discussed. 
In the following, we  plot all the line fluxes as a function of the selected parameters. The focus of this study is to understand the discrepancy in detection between Spitzer/IRS and Herschel/HIFI, i.e., between the inner and 
outer disk, and to find a possible connection/explanation with the physical properties of the disks.

\begin{figure*}[htpl]
\centering
\begin{minipage}[l]{0.33\textwidth}
\includegraphics[width=1.03\textwidth]{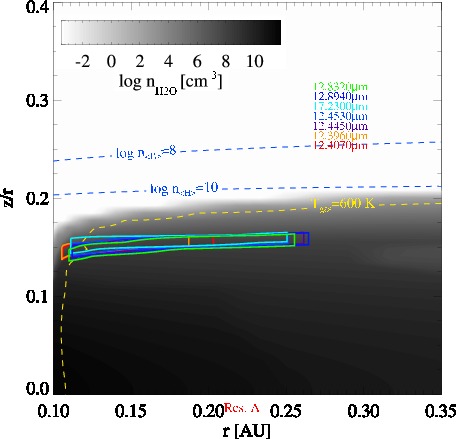}
\end{minipage}
\begin{minipage}[c]{0.33\textwidth}
\includegraphics[width=\textwidth]{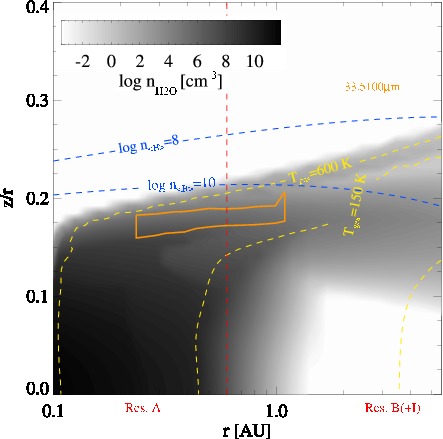}
\end{minipage}
\begin{minipage}[r]{0.33\textwidth}
\includegraphics[width=\textwidth]{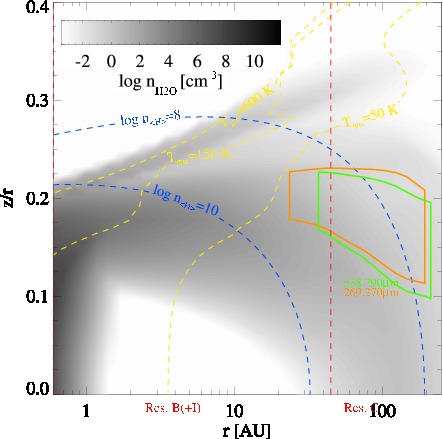}
\end{minipage}
\caption{Water lines from different reservoirs in the standard model; left: reservoir A, center: reservoir A-B, right: reservoir B-C. the blue dashed curved lines are contours for gas densities of 10$^8$ and 10$^{10}$ cm$^{-3}$,
the yellow contours are for gas temperatures of 50~K, 150~K, 600~K. Vertical red dashed lines delimit the reservoirs.}
\label{reserv_plot}
\end{figure*}

Fig.~\ref{reserv_plot} shows the observed and detected Spitzer lines considered here; ($E_\mathrm{up}~\ge~$3000~K). Since reservoir A is very compact compared to the disk size, these transitions largely overlap in their 
emitting volume, and as a consequence, the physical conditions are very similar for all the lines.
As long as the radial and vertical temperature and density gradients in these line emitting regions are small, a simplified slab approach as employed, e.g.,\ in \citet{carr} presents a reasonable approximation.
The transitions below 12.5~$\mu$m are produced in a region that is partially optically thick. A vertical integration of the dust extinction (see Appendix~\ref{app2}) confirms that the $\tau_\mathrm{dust}$~=~1 surface at the 
wavelength of the line is above the mean vertical height of the line emitting region, so that the lines are partially buried in the continuum. 
Lines from reservoir B are produced in an optically thinner region, about 50\% of the emitting region is optically thin.
More than 30\% of the 269.27~$\mu$m and 70\% of the 538.29~$\mu$m line emitting regions (reservoir C) are vertically extended above the $\tau_\mathrm{dust}~=~1$ contour at the respective wavelengths.

The far-IR transitions are very close to LTE because the average density of H$_2$ is close to the critical density (within an order of magnitude), and so we expect them to be sensitive to $T_\mathrm{gas}$ and 
$n_\mathrm{gas}$ variations. Our mid-IR lines, instead, are produced in a very dense region of the disk, and so they form under LTE conditions. Table~\ref{partners} shows that H$_{2}$ is the main collisional partner for all 
considered transitions because the H/H$_{2}$ ratio is always below 0.1. Because of changes in the physical conditions of the different models, the previous statements are strictly valid only for our standard disk model.

\begin{table*}
\caption{Overview and coefficients of the considered transitions}
\centering
\begin{tabular}{|c|c|c|c|c|c|c|c|c|c|c|}
\hline\hline
species & wavelength [$\mu$m] & $A_\mathrm{ul}$ [s$^{-1}$] & $\left<n_\mathrm{H_{2}}\right>C_\mathrm{H_{2},ul}^\mathrm{(a)}$ [s$^{-1}$] & $\left<n_\mathrm{H}\right>C_\mathrm{H,ul}^\mathrm{(a)}$ [s$^{-1}$] & $\left<n_\mathrm{e^-}\right>C_\mathrm{e^-,
ul}^\mathrm{(a)}$ [s$^{-1}$]\\ \hline
p-H$_{2}$O & 12.407 & 4.217 & 4.489$\times10^{12}$ (o \& p) & 2.838$\times10^{9}$ & 8.918$\times10^{-3}$\\ 
o-H$_{2}$O & 12.396 & 7.665 & 4.823$\times10^{12}$ (o \& p) & 3.076$\times10^{9}$ & 1.757$\times10^{-2}$\\ 
o-H$_{2}$O & 12.445 & 2.937$\times10^{-1}$ & 4.982$\times10^{11}$ (o \& p) & 2.990$\times10^{8}$ & 4.089$\times10^{-4}$\\ 
o-H$_{2}$O & 12.453 & 1.160 & 1.660$\times10^{12}$ (o \& p) & 9.963$\times10^{8}$ & 1.431$\times10^{-3}$\\
p-H$_{2}$O & 12.832 & 1.564$\times10^{-1}$ & 1.375$\times10^{12}$ (o \& p) & 8.253$\times10^{8}$ & 2.399$\times10^{-4}$\\ 
p-H$_{2}$O & 12.894 & 2.596$\times10^{-1}$ & 5.856$\times10^{11}$ (o \& p) & 3.154$\times10^{8}$ & 3.956$\times10^{-4}$\\ 
o-H$_{2}$O & 33.510 & 4.691 & 1.877$\times10^{11}$ (o \& p) & 6.689$\times10^{7}$ & 2.862$\times10^{-2}$\\ 
p-H$_{2}$O & 269.27 & 1.852$\times10^{-2}$ & 1.817$\times10^{-3}$ (para) 7.943$\times10^{-3}$ (ortho) & 1.070$\times10^{-6}$ & 4.145$\times10^{-5}$\\ 
o-H$_{2}$O & 538.29 & 3.477$\times10^{-3}$ & 1.348$\times10^{-3}$ (para) 1.274$\times10^{-2}$ (ortho) & 5.985$\times10^{-7}$ & 5.869$\times10^{-5}$\\ \hline
\end{tabular}
\label{partners}
\tablefoot{(a) Collisional rates; the average densities used are computed within the emitting region of the considered line and the $T_\mathrm{gas}$ for the collisional constants are averaged inside the emitting region (see 
Appendix \ref{app2})}%; (b) He collisional constants from \citep{green} for Herschel lines. For Spitzer lines H collisional constant are H$_2$ values scaled by a factor $\sqrt{2}$.}
\end{table*}  

\section{Results}\vspace{5mm}
\label{4}

The previously discussed standard model is a starting point from which we build a series of models, modifying a single parameter per disk model. In the following, we discuss the results from this series.

The line fluxes have been treated as independent transitions, hence, as physically unblended. The results reported here are based on vertical escape probability. This approach agrees within a factor of two 
with a more detailed and computationally more expensive radiative transfer (Fig.~\ref{test}). In particular, far-IR lines agree within 30\%, while mid-IR lines agree within 70\%. 
The lines from vertical escape probability show a systematical stronger value than detailed RT lines. The reason for this mismatch is twofold: the disk inclination (30~$\deg$) causes larger column densities of gas 
and dust in the detailed RT calculation with respect to the vertical escape treatment. In addition, the vertical escape does not take  the line self-absorption into account.
The vertical escape method is a fast and powerful method to compute tens of thousands of lines for different species.
Cases in which the vertical escape probability is unreliable are rare cases of absorption lines.
Those are limited to very few flat models ($\beta~<~0.8$), to models with $a_\mathrm{pow}\!\ge\!4.0,$ or dust-to-gas mass ratio $\ge$~100, and our conclusions are not affected.

\begin{figure}[htpl]
\centering
\includegraphics[width=0.5\textwidth]{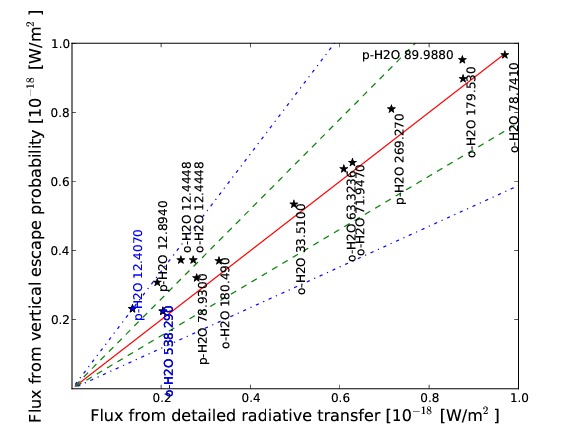}
\caption{Comparison between the vertical escape probability flux and a detailed radiative transfer computation for the standard model. Green dashed and blue dot-dashed lines are confidence intervals  of 30\% and 70\% 
from the red line, respectively.}
\label{test}
\end{figure}
To illustrate the results, the theoretical Spitzer spectrum of each model is computed with a resolution of 600 \citep[the maximum of IRS, of LH and SH modules; ][]{houck}. These spectra give a first qualitative 
result, which can be compared to observations. We also define a ``mid-IR color'' as the ratio of the continuum fluxes at 13.5 and 30~$\mu$m as indicator of the SED slope in the Spitzer wavelength range.

In the following subsections, we discuss the behavior of the selected transitions, considering the representative mid-IR and far-IR lines. We also describe in detail some average physical quantities in the line emitting regions.
The computation of these quantities is described in detail in Appendix~\ref{app2}. We compare the predicted fluxes with the sensitivity limits for Spitzer/IRS \citep[$5\cdot10^{-18}$~W/m$^2$, 1~$\sigma$, deduced from the 
typical upper limits;][]{pontoppidan2}, Herschel/HIFI \citep[$9\cdot10^{-19}$~W/m$^2$, 1~hr, 5~$\sigma$;][]{degraauw}, and finally JWST/MIRI 
\citep[10$^{-20}$~W/m$^2$, 2.7~hrs., 10~$\sigma$;][]{glasse}\footnote{http://www.stsci.edu/jwst/instruments/miri/sensitivity/ \& http://www.stsci.edu/jwst/science/sensitivity}.
Additional plots with the 538.29~$\mu$m o-H$_2$O line flux and line ratios 538.29~$\mu$m/12.407~$\mu$m are included in Appendix~\ref{app3} (Fig.~\ref{HIFI}~\&~\ref{Rat21}).

\subsection{Dust-to-gas mass ratio}\vspace{5mm}
\label{dustgas_par}

A variation of this parameter means changing the dust mass since the gas mass is fixed to 0.01~M$_\mathrm{\odot}$.

The water total column densities in the inner disk are unaffected by the dust content (Fig.~\ref{dg}b). However, the Spitzer line at 12.4~$\mu$m is produced by a volume in which the column density is suppressed up to 14 orders 
of magnitude from the low to high dust-to-gas mass ratio (Fig.~\ref{dg}a). The emitting region of the Spitzer line becomes radially more extended between the low and\ high dust-to-gas case (Fig.~\ref{dg}c). This is because the 
total water reservoir is unaffected, while the emitting region moves upward to the disk surface with increasing dust mass (Fig.~\ref{dg}c); this causes the densities in the emitting region to drop by 8 orders of magnitude. The 
continuum opacity in the line emitting region grows with increasing dust mass (Fig.~\ref{dg}h). 

In summary, mid-IR line fluxes are driven indirectly by the increase in dust opacity; it forces the lines to move upward in the disk into lower density regions (Fig.~\ref{dg}c~\&~h). The continuum optical depth increase 
reduces the column density of emitting water. The deviation beyond a dust-to-gas ratio of 10.0 is due to the extremely optically thick disk, which forces the lines to be produced from more optically thin surface layers beyond 0.5~AU.

In the outer disk, which hosts the ice reservoir, changes in the opacity affect the photoevaporation process, increasing the total water vapor column density (Fig.~\ref{dg}b). The emitting region of the fundamental water 
lines gets progressively close to $A_\mathrm{V}~=~1$. The emitting region's radial displacement with the increase of dust mass is less pronounced than in the mid-IR case (Fig.~\ref{dg}c), but the vertical displacement is 
significant. 
Also, the line flux of the far-IR water lines is driven by an opacity effect, but to a lesser extent compared to the mid-IR. Also, the emitting region moves upward with increasing dust mass, however, the emitting water 
column density is only reduced  by two orders of magnitude between the most extreme models (Fig.~\ref{dg}a). The situation is less extreme, because the outer disk is only marginally 
optically thick (Fig.~\ref{dg}a). 

\begin{figure*}[htpl]
\centering
\begin{minipage}[l]{0.36\textwidth}
(a)\includegraphics[width=1.1\textwidth]{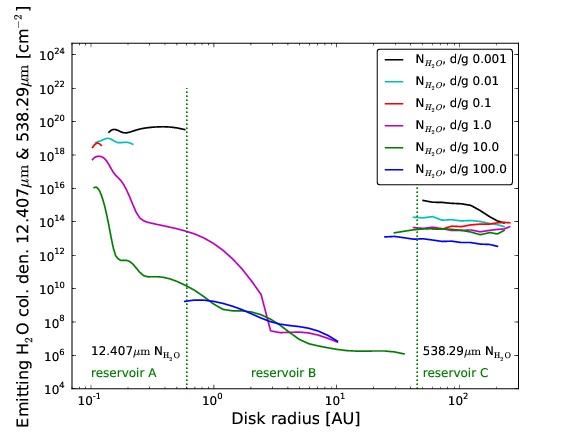}
%(h)\includegraphics[width=\textwidth]{/net/jeans/data/users/antonellini/fits/Theoretical/parametrized/Recon_six/HIFI_dustgas.eps}
\end{minipage}
\begin{minipage}[r]{0.36\textwidth}
(b)\includegraphics[width=1.1\textwidth]{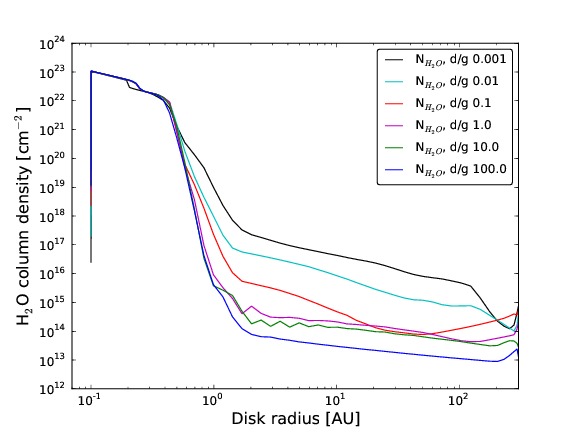}
\end{minipage}
\begin{minipage}[l]{0.36\textwidth}
(c)\includegraphics[width=1.1\textwidth]{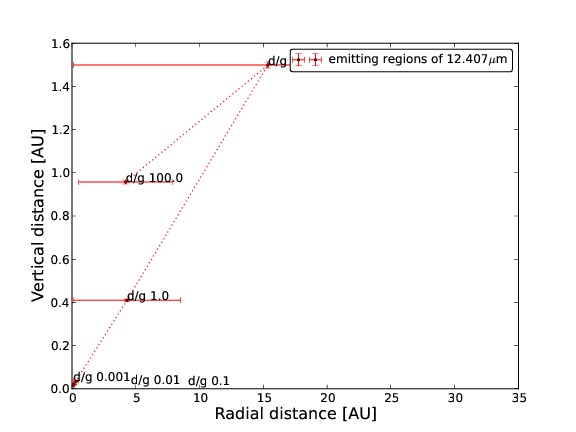}
\end{minipage}
\begin{minipage}[r]{0.36\textwidth}
(d)\includegraphics[width=1.1\textwidth]{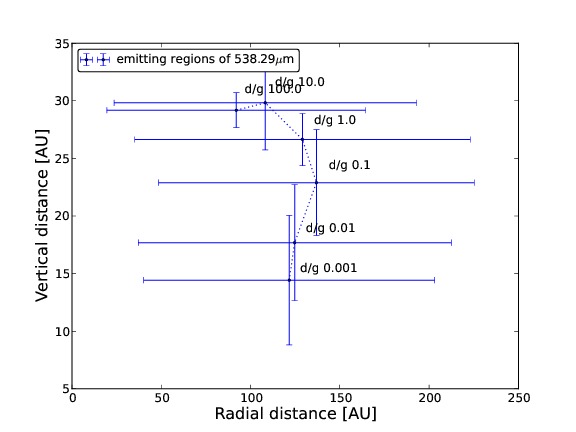}
\end{minipage}
\begin{minipage}[r]{0.36\textwidth}
(e)\includegraphics[width=0.97\textwidth]{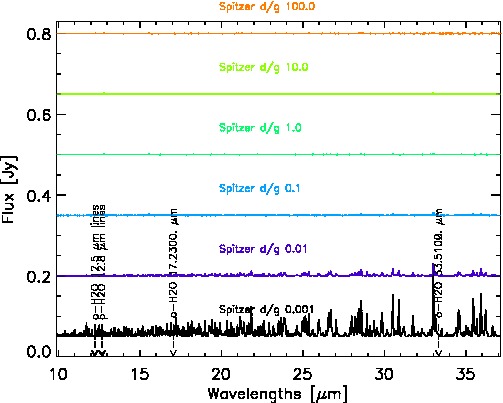}
%(e)\includegraphics[width=\textwidth]{/net/jeans/data/users/antonellini/fits/Theoretical/parametrized/Recon_six/dustgas_Densities.eps}
\end{minipage}
\begin{minipage}[l]{0.36\textwidth}
(f)\includegraphics[width=0.97\textwidth]{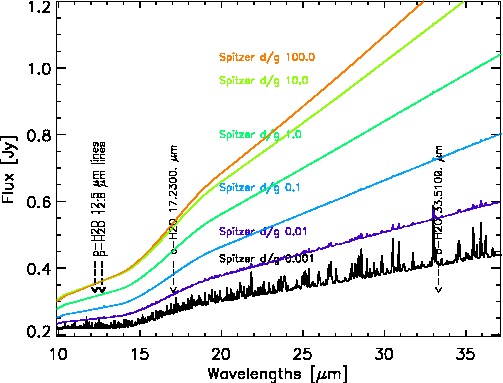}
%(b)\includegraphics[width=\textwidth]{/net/jeans/data/users/antonellini/fits/Theoretical/parametrized/Recon_six/dustgas_Temperatures.eps}
\end{minipage}
\begin{minipage}[l]{0.36\textwidth}
(g)\includegraphics[width=1.1\textwidth]{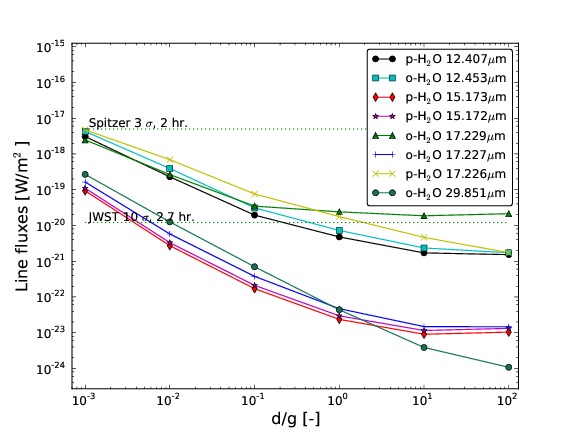}
\end{minipage}
\begin{minipage}[r]{0.36\textwidth}
(h)\includegraphics[width=1.1\textwidth]{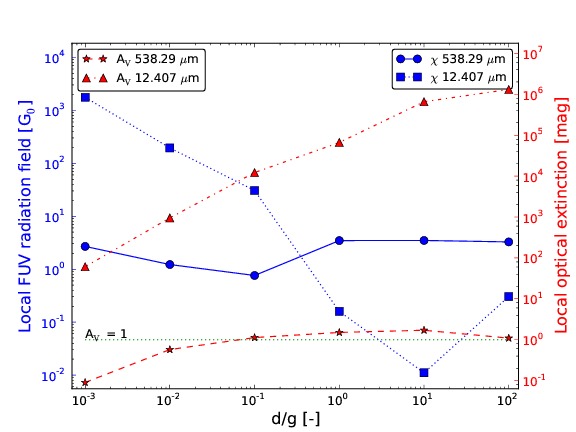}
\end{minipage}
\caption{Plots for dust-to-gas mass ratio models. (a) Emitting region water column density comprises 15-85\% of the radially and vertically integrated flux. Vertical dotted green lines distinguish the reservoirs. 
(b) Total water column densities as a function of radius. (c) Average extension of the line emitting region that comprises 50\% of the line flux radially and vertically for 12.407~$\mu$m. (d) Average extension of the line 
emitting region for 538.29~$\mu$m (50\% of radially and vertically integrated flux). (e) Theoretical Spitzer SH/LH modules spectra ($R$~=~600) for this model series, continuum subtracted and arbitrarily shifted.
(f) Theoretical Spitzer SH/LH modules spectra ($R$~=~600) for this model series. (g) Spitzer line fluxes with sensitivity limits (green dotted  lines) for Spitzer/IRS, JWST/MIRI.
(h) Radiation field and optical extinction in the line emitting regions of 12.407~$\mu$m and 538.29~$\mu$m.}
\label{dg}
\end{figure*}
%(e) Water density and collisional partners densities in the emitting regions of the same lines (critical densities for the standard model conditions are indicated by red dotted lines).
%\caption{Plots for dust-to-gas mass ratio: left d/g = 0.001, Right: d/g = 1.0; from top to bottom: 12.407 $\rm\mu m$ line and continuum opacity, cumulative flux, contour emission (50\%); 538.29 $\rm\mu m$ line and continuum 
%opacity, cumulative flux, contour emission (50\%); Herschel HIFI 538.29 $\rm\mu m$ and 269.27 $\rm\mu m$ line fluxes in all the models; Total water column density in the disk for the different models.}
%; ratio HIFI/Spitzer lines 1 with quadratic fit for show the trend; ratio HIFI/Spitzer lines 2 with quadratic fit for show the trend.}
%(b) Temperatures of dust and gas in the emitting regions of the same%(h) HIFI line fluxes with sensitivity detection for 1 hr. exposure time (dotted green line).

%\begin{figure}[htpl]
%\centering
%\includegraphics[width=0.45\textwidth]{/net/jeans/data/users/antonellini/fits/Theoretical/parametrized/dustgas/unknown/CD.eps}
%\caption{Emitting region water column density for dust-to-gas mass ratio models. This region comprises 15-85\% of the radially and vertically integrated flux. Vertical dotted green lines distinguish the reservoirs.}
%\label{dg1}
%\end{figure}

All Spitzer transitions show a strong dependence on the dust-to-gas mass ratio (Fig.~\ref{dg}e~\&~f). The line-to-continuum ratio changes by a factor 2.5 between the smallest and largest dust-to-gas ratio, and in the 
dust-rich models all line features disappear from the Spitzer/IRS spectra. This is in agreement with previous results of \citet{meijerink}. The continuum fluxes decrease with decreasing dust content, and the 13.5/30~$\mu$m 
continuum flux ratio decreases with increasing dust content in the disk.

%\begin{figure}
%\resizebox{\hsize}{!}{\includegraphics{/net/jeans/data/users/antonellini/fits/Theoretical/parametrized/dustgas/unknown/Spitzer_dustgas.eps}}
%\resizebox{\hsize}{!}{\includegraphics{/net/jeans/data/users/antonellini/fits/Theoretical/parametrized/dustgas/unknown/Sp_lines_dustgas.eps}}
%\caption{Theoretical Spitzer SH/LH ($R$ = 600) for a dust-to-gas model series; top: standard, bottom: continuum subtracted and arbitrarily shifted.} 
%\label{Spitzer_dg} 
%\end{figure}

\subsection{Dust settling}\vspace{5mm}

Settling affects the water spectroscopy much less than the dust-to-gas mass ratio. In fact, the largest systematic difference in water column density can be found in the outer disk, $r\!>$~10~AU (Fig.~\ref{al}b). The Spitzer 
emitting region only shows a change in the water emitting column density  in the most settled model; that is also the only one producing a slightly stronger line flux (Fig.~\ref{al}g).
The most settled model has a higher column density in the water emitting regions of the mid-IR and far-IR lines because of the reduced continuum opacity that boosts the thermally activated water formation channels.
In the inner disk, the density is very high, and therefore the dust is largely unsettled.  In cases of extremely low mechanical mixing only($\alpha_\mathrm{set}\!\le\!10^{-5}$), the disk has such a low opacity that the line emitting region 
moves upward because of increased radial extension in the disk, and the column density of the emitting region is higher than in an unsettled case (Fig.~\ref{al}b). 
In the emitting region of the far-IR lines, the column density is more affected (Fig.~\ref{al}b); for $\alpha_\mathrm{set}\!<\!0.01$, the total water column density grows about an order of magnitude (Fig.~\ref{al}d). Because of  
the migration of the far-IR line emitting region, dust extinction fades quickly from a marginally optically thick regime to an optically thin regime (Fig.~\ref{al}h). The outer disk is affected by settling only for 
$\alpha_\mathrm{set}\!<\!10^{-3}$; this threshold can be found in the evolution of several properties, such as  the opacity (Fig.~\ref{al}h).

In the outer disk the photodesorption of water (Sect.~\ref{wchem}) and photodissociation processes are not strong enough to deplete water significantly. This produces an upward migration of the emitting region for the far-IR 
transition in the most settled models. The Spitzer line emitting region migration (Fig.~\ref{al}c) and larger emitting water column density are negligible given numerical uncertainties. 

\begin{figure*}[htpl]
\centering
\begin{minipage}[l]{0.36\textwidth}
(a)\includegraphics[width=1.1\textwidth]{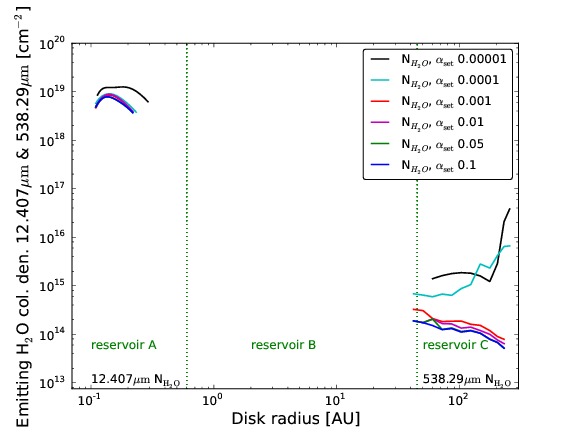}
%(h)\includegraphics[width=\textwidth]{/net/jeans/data/users/antonellini/fits/Theoretical/parametrized/Recon_six/HIFI_Dubrulle.eps}
\end{minipage}
\begin{minipage}[r]{0.36\textwidth}
(b)\includegraphics[width=1.1\textwidth]{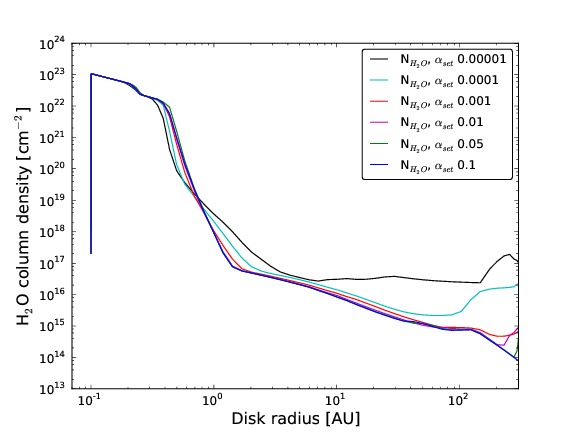}
\end{minipage}
\begin{minipage}[l]{0.36\textwidth}
(c)\includegraphics[width=1.1\textwidth]{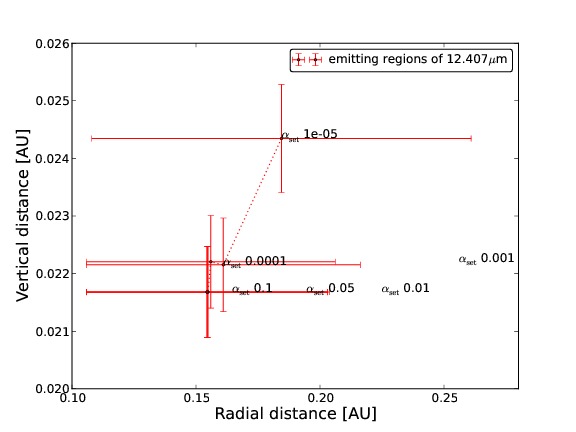}
\end{minipage}
\begin{minipage}[r]{0.36\textwidth}
(d)\includegraphics[width=1.1\textwidth]{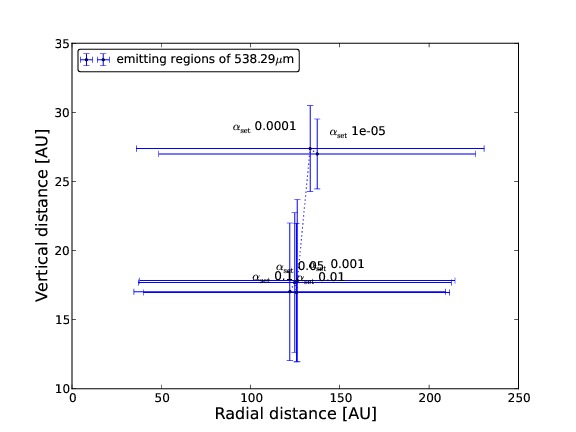}
\end{minipage}
\begin{minipage}[l]{0.36\textwidth}
(e)\includegraphics[width=0.97\textwidth]{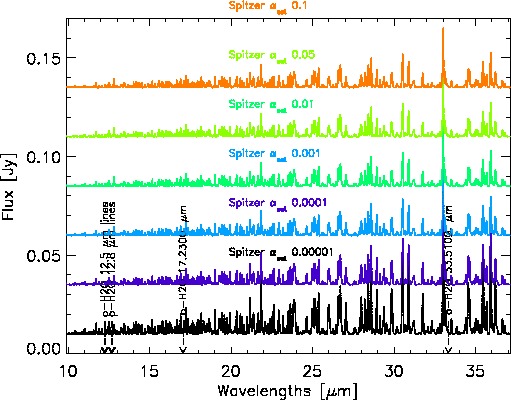}
%(e)\includegraphics[width=\textwidth]{/net/jeans/data/users/antonellini/fits/Theoretical/parametrized/Recon_six/Dubrulle_Densities.eps}
\end{minipage}
\begin{minipage}[r]{0.36\textwidth}
(f)\includegraphics[width=0.97\textwidth]{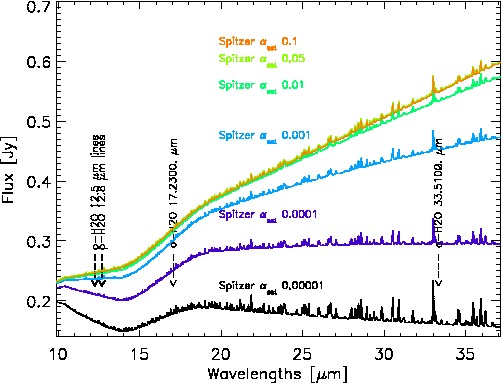} 
%(b)\includegraphics[width=\textwidth]{/net/jeans/data/users/antonellini/fits/Theoretical/parametrized/Recon_six/Dubrulle_Temperatures.eps} 
\end{minipage}
\begin{minipage}[l]{0.36\textwidth}
(g)\includegraphics[width=1.1\textwidth]{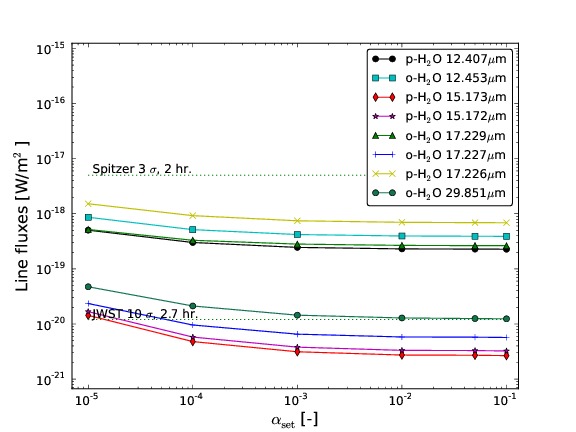}
\end{minipage}
\begin{minipage}[r]{0.36\textwidth}
(h)\includegraphics[width=1.1\textwidth]{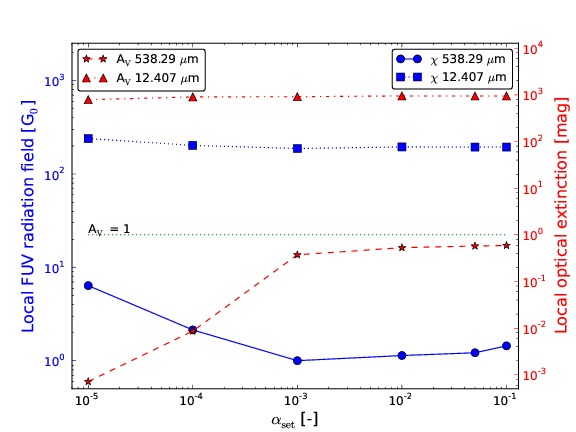}
\end{minipage}
\caption{Plots for mixing coefficient models. For further details see Fig.~\ref{dg}.}
%: (a) Emitting region water column density for the mechanical mixing models. This region comprises 15-85\% of the radially and vertically integrated flux. Vertical dotted green lines distinguish the reservoirs. %(b) Temperatures of dust and gas in the emitting regions of the same 
%(b) Total water column densities as a function of radius.
%lines. (c) Average extension of the line emitting region that comprises 50\% of the line flux radially and vertically for 12.407 $\mu$m. (d) Average extension of the line emitting region that comprises 50\% of the line flux
%radially and vertically for 538.29 $\mu$m. %(e) Water density and collisional partners densities in the emitting regions of the same lines (critical densities for the standard model conditions are indicated by red dotted lines). 
%(e) Theoretical Spitzer SH/LH ($R$ = 600) for $\alpha_\mathrm{set}$ model series, continuum subtracted and arbitrarily shifted.
%(f) Theoretical Spitzer SH/LH ($R$ = 600) for $\alpha_\mathrm{set}$ model series. (g) Spitzer line fluxes with sensitivity limits (dotted green lines) for Spitzer/IRS, JWST/MIRI. %(h) HIFI line fluxes with sensitivity detection for 1 hr. exposure time (dotted green line).
%(h) Radiation field and optical extinction in the line emitting regions of 12.407 $\mu$m and 538.29 $\mu$m.}
\label{al}
\end{figure*}

%\begin{figure}[htpl]
%\centering
%\includegraphics[width=0.45\textwidth]{/net/jeans/data/users/antonellini/fits/Theoretical/parametrized/Dubrulle/unknown/CD.eps}
%\caption{Emitting region water column density for the mechanical mixing models. This region comprises 15-85\% of the radially and vertically integrated flux. Vertical dotted green lines distinguish the 
%reservoirs.}
%\label{al1}
%\end{figure}

The reduction of the mixing parameter produces a moderate enhancement in the mid-IR line-to-continuum ratio (Fig.~\ref{al}e~\&~f), along with a reduction of the continuum flux because of the decrease of dust in the
surface layers of the inner disk. The 13.5/30.0~$\mu$m continuum flux ratio decreases with increasing the mixing coefficient $\alpha_\mathrm{set}$.

%\begin{figure}
%\resizebox{\hsize}{!}{\includegraphics{/net/jeans/data/users/antonellini/fits/Theoretical/parametrized/Dubrulle/unknown/Spitzer_Dubrulle.eps}}
%\resizebox{\hsize}{!}{\includegraphics{/net/jeans/data/users/antonellini/fits/Theoretical/parametrized/Dubrulle/unknown/Sp_lines_Dubrulle.eps}}
%\caption{Theoretical Spitzer SH/LH ($R$ = 600) for $\alpha_\mathrm{set}$ model series; top: standard, bottom: continuum subtracted and arbitrarily shifted.} 
%\label{Spitzer_al} 
%\end{figure}

\subsection{Flaring parameter $\beta$}\vspace{5mm}

In the parametrized modeling, the disk structure can be set through the scale height at a reference radius $H_\mathrm{0}$ and the flaring index $\beta$. Here, we only change  the flaring and fix the 
scale height at the inner disk radius. A value of $\beta\!<\!1.0$ produces a disk collapsed vertically behind the inner radius and completely shadowed by the vertical extent of the inner disk rim.

\begin{figure*}[htpl]
\centering
\begin{minipage}[l]{0.36\textwidth}
(a)\includegraphics[width=1.1\textwidth]{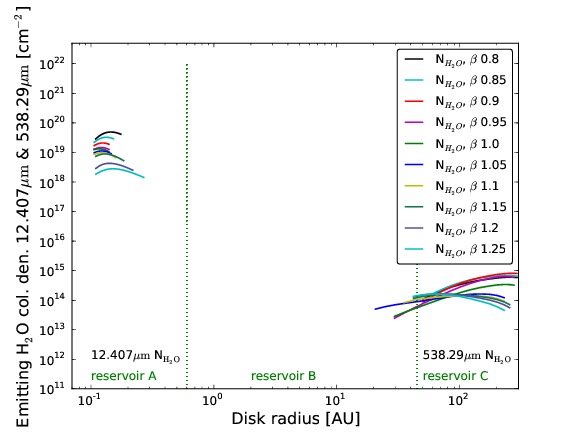}
%(h)\includegraphics[width=\textwidth]{/net/jeans/data/users/antonellini/fits/Theoretical/parametrized/Recon_six/HIFI_beta.eps}
\end{minipage}
\begin{minipage}[r]{0.36\textwidth}
(b)\includegraphics[width=1.1\textwidth]{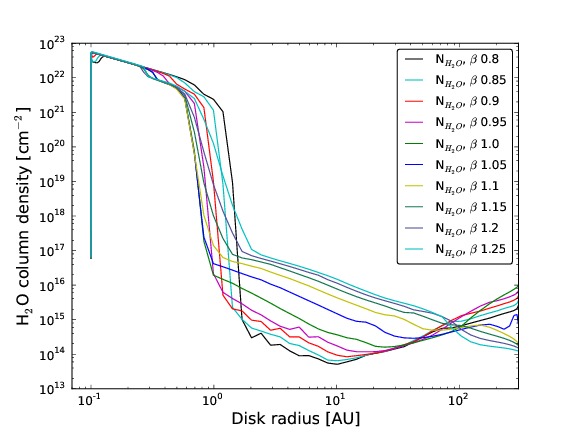}
\end{minipage}
\begin{minipage}[l]{0.36\textwidth}
(c)\includegraphics[width=1.1\textwidth]{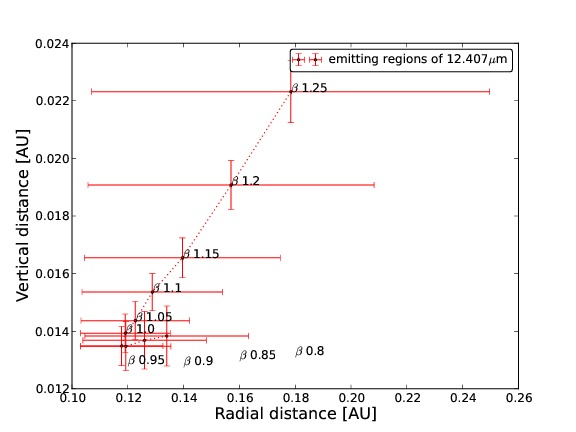}
\end{minipage}
\begin{minipage}[r]{0.36\textwidth}
(d)\includegraphics[width=1.1\textwidth]{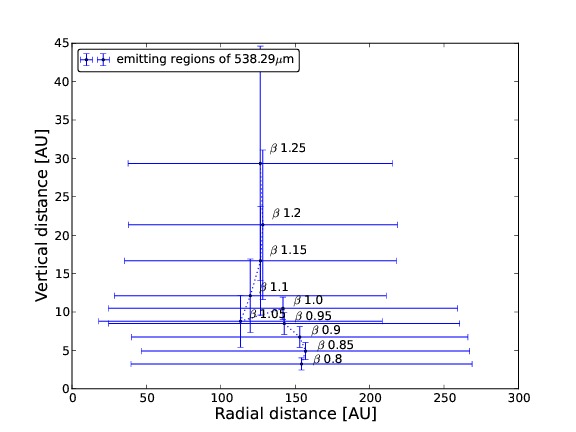}
\end{minipage}
\begin{minipage}[l]{0.36\textwidth}
(e)\includegraphics[width=0.97\textwidth]{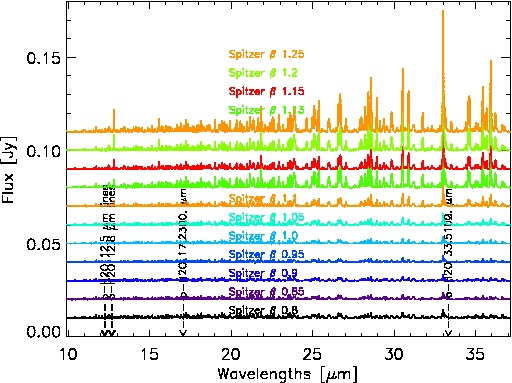}
%(e)\includegraphics[width=\textwidth]{/net/jeans/data/users/antonellini/fits/Theoretical/parametrized/Recon_six/beta_Densities.eps}
\end{minipage}
\begin{minipage}[r]{0.36\textwidth}
(f)\includegraphics[width=0.97\textwidth]{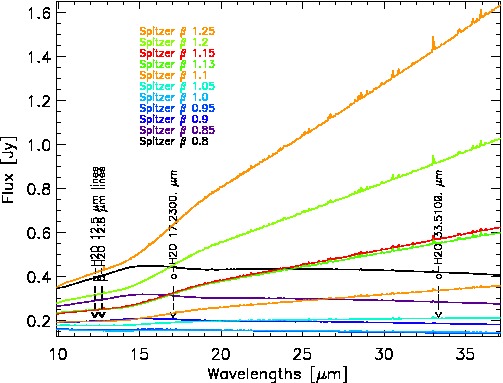}
%(b)\includegraphics[width=\textwidth]{/net/jeans/data/users/antonellini/fits/Theoretical/parametrized/Recon_six/beta_Temperatures.eps}
\end{minipage}
\begin{minipage}[l]{0.36\textwidth}
(g)\includegraphics[width=1.1\textwidth]{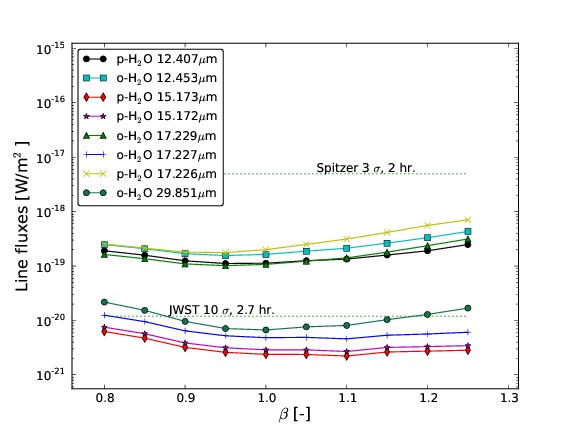}
\end{minipage}
\begin{minipage}[r]{0.36\textwidth}
(h)\includegraphics[width=1.1\textwidth]{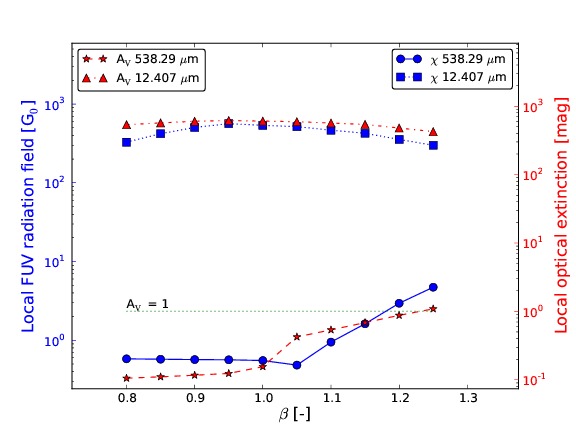} 
\end{minipage}
\caption{Plots for flaring index exponent models. For further details see Fig.~\ref{dg}.}
%: (a) Emitting region water column density for the flaring index models. This region comprises 15-85\% of the radially and vertically integrated flux. Vertical dotted green lines distinguish the reservoirs. %(b) Temperatures of
% dust and gas in the emitting regions of the same 
%(b) Total water column densities as a function of radius.
%(c) Average extension of the line emitting region of 50\% line flux radially and vertically for 12.407 $\mu$m. (d) Average extension of the line emitting region of 50\% line flux radially and vertically for 538.29 $\mu$m.
%%(e) Water density and collisional partners densities in the emitting regions of the same lines (critical densities for the standard model conditions are indicated by red dotted lines). 
%(f) Theoretical Spitzer SH/LH ($R$ = 600) for $\beta$ model series.
%(e) Theoretical Spitzer SH/LH ($R$ = 600) for $\beta$ model series, continuum subtracted and arbitrarily shifted.
%(g) Spitzer line fluxes with sensitivity limits (dotted green lines) for Spitzer/IRS, JWST/MIRI. %(h) HIFI line fluxes with sensitivity limit for 1 hr. exposure time (dotted green line).
%(h) Radiation field and optical extinction in the line emitting regions of 12.407 $\mu$m and 538.29 $\mu$m.}
\label{bt}
\end{figure*}

%\begin{figure}[htpl]
%\centering
%\includegraphics[width=0.45\textwidth]{/net/jeans/data/users/antonellini/fits/Theoretical/parametrized/beta/unknown/CD.eps}
%\caption{Emitting region water column density for the flaring index models. This region comprises 15-85\% of the radially and vertically integrated flux. Vertical dotted green lines distinguish the reservoirs.}
%\label{bt1}
%\end{figure}

An increasing flaring index decreases the water column density of the emitting region by up to two orders of magnitude, and the emitting region becomes up to a factor 1.5 more radially extended (Fig.~\ref{bt}h). As consequence of
the increased radial extension, the emitting region of the Spitzer lines also moves upward with increasing flaring index (Fig.~\ref{bt}c). The line fluxes show a minimum for $\beta~=~1.0$ (Fig.~\ref{bt}g) and increase for 
$\beta~>~1.0$ up to a factor 1.5.
The mid-IR spectral changes can be explained by a pure geometrical effect. Although the variation in the inner disk is small (since the inner disk scale height is fixed), it is enough to increase the heating of the surface 
layers of the inner disk to the central star radiation. 

The outer disk total water column density is more affected than the inner disk, and the total amount of water (beyond 80~AU) is a factor $\sim$~2 larger in the most flared case with respect to the 
flattest case (Fig.~\ref{bt}b). With increasing flaring, the emitting region moves upward and gets vertically more extended (Fig.~\ref{bt}d) because of an increase of the line opacity. The column density of water 
vapor in the emitting region becomes smaller, but the $T_\mathrm{gas}$ becomes larger. This is consistent with the monotonic increase of the Herschel/HIFI line flux. 

With increasing flaring, the disk surface is irradiated more strongly by the central star. This increases the dust temperature through the entire disk (especially in the surface layers) and hence the mid-IR color value 
decreases (Fig.~\ref{bt}f). 

%\begin{figure}
%\resizebox{\hsize}{!}{\includegraphics{/net/jeans/data/users/antonellini/fits/Theoretical/parametrized/beta/unknown/Spitzer_beta.eps}}
%\resizebox{\hsize}{!}{\includegraphics{/net/jeans/data/users/antonellini/fits/Theoretical/parametrized/beta/unknown/Sp_lines_beta.eps}}
%\caption{Theoretical Spitzer SH/LH ($R$ = 600) for $\beta$ model series; top: standard, bottom: continuum subtracted and arbitrarily shifted.} 
%\label{Spitzer_bt} 
%\end{figure}

\subsection{Maximum grain size}\vspace{5mm}

The total column density of water beyond 1~AU is affected by the maximum grain size; the models with bigger grains have a larger water vapor column density (Fig.~\ref{am}b). The lower the opacity, the more depleted is the ice 
reservoir, both vertically and radially, thus enriching the gas phase. The water column density in the emitting region of the Spitzer transition also increases (Fig.~\ref{am}a). That region moves upward and 
becomes more extended with increasing maximum grain size (Fig.~\ref{am}~g). Since the total dust mass remains constant with increasing $a_\mathrm{max}$, the population of small grains is reduced. As a consequence, the 
average continuum opacity (Fig.~\ref{am}h) in the emitting region decreases with the increase of maximum grain size because the radial gas and dust temperature gradient flattens and the emitting region becomes 
radially more extended. This is a consequence of the increased FUV flux through the whole disk. The vertical displacement is a consequence of a radially more extended emitting region because the disk is flared and the emitting 
region moved outward also becomes more extended  vertically (Fig.~\ref{am}c).
The mid-IR line fluxes increase monotonically with increasing maximum dust size because of increased emitting area and lower continuum opacity (Fig.~\ref{am}c~\&~h). 

The emitting region of the Herschel lines (Fig.~\ref{am}d) moves closer to the midplane of the disk for larger $a_\mathrm{max}$, maintaining roughly the same radial extent and position (Fig.~\ref{am}d). This is again due to the 
lower dust optical depth for the models with larger grains. The water column density increase with increasing $a_\mathrm{max}$ translates into larger fluxes since the lines are optically thin. 

\begin{figure*}[htpl]
\centering
\begin{minipage}[l]{0.36\textwidth}
(a)\includegraphics[width=1.1\textwidth]{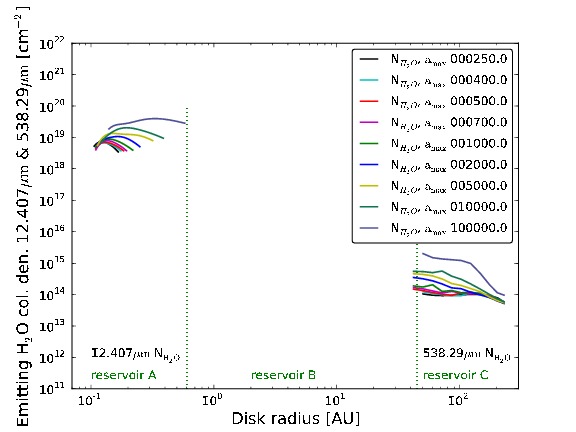}
%(h)\includegraphics[width=\textwidth]{/net/jeans/data/users/antonellini/fits/Theoretical/parametrized/Recon_six/HIFI_amax.eps}
\end{minipage}
\begin{minipage}[r]{0.36\textwidth}
(b)\includegraphics[width=1.1\textwidth]{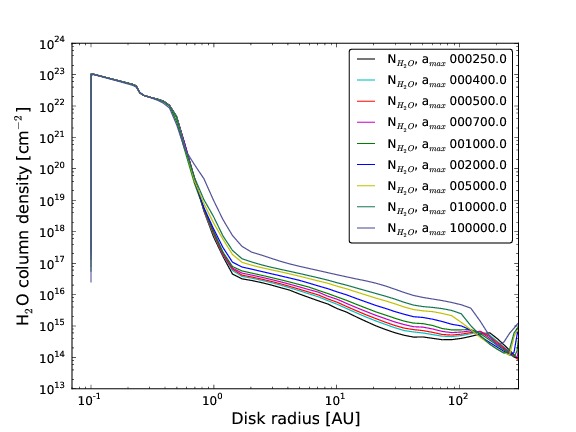}
\end{minipage}
\begin{minipage}[l]{0.36\textwidth}
(c)\includegraphics[width=1.1\textwidth]{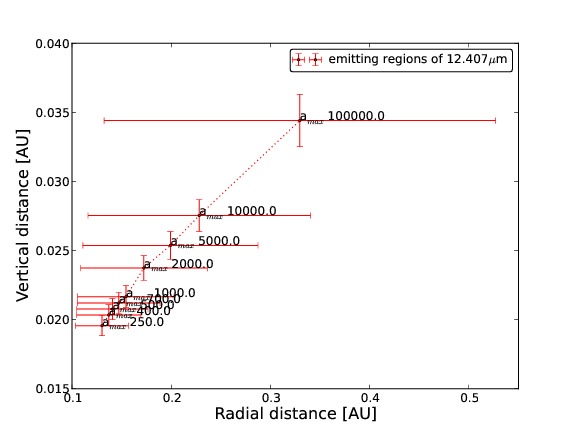}
\end{minipage}
\begin{minipage}[r]{0.36\textwidth}
(d)\includegraphics[width=1.1\textwidth]{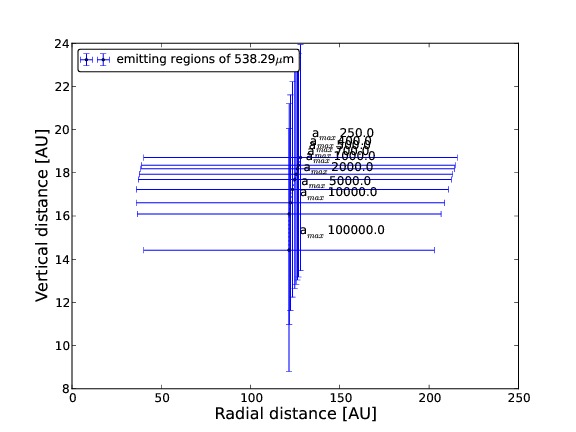}
\end{minipage}
\begin{minipage}[l]{0.36\textwidth}
(e)\includegraphics[width=0.97\textwidth]{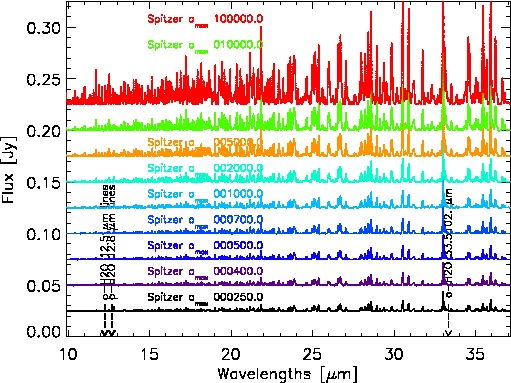}
%(e)\includegraphics[width=\textwidth]{/net/jeans/data/users/antonellini/fits/Theoretical/parametrized/Recon_six/amax_Densities.eps}
\end{minipage}
\begin{minipage}[r]{0.36\textwidth}
(f)\includegraphics[width=0.97\textwidth]{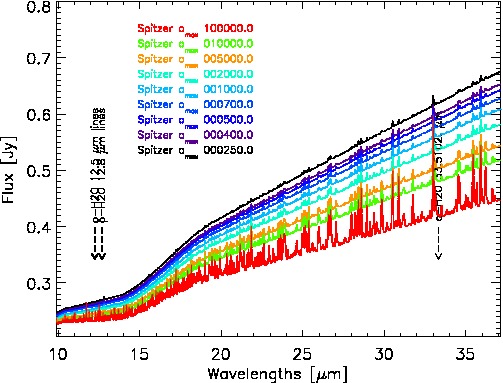}
%(b)\includegraphics[width=\textwidth]{/net/jeans/data/users/antonellini/fits/Theoretical/parametrized/Recon_six/amax_Temperatures.eps}
\end{minipage}
\begin{minipage}[l]{0.36\textwidth}
(g)\includegraphics[width=1.1\textwidth]{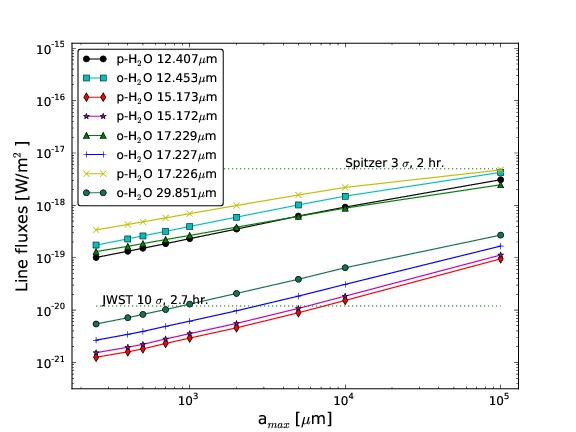}
\end{minipage}
\begin{minipage}[r]{0.36\textwidth}
(h)\includegraphics[width=1.1\textwidth]{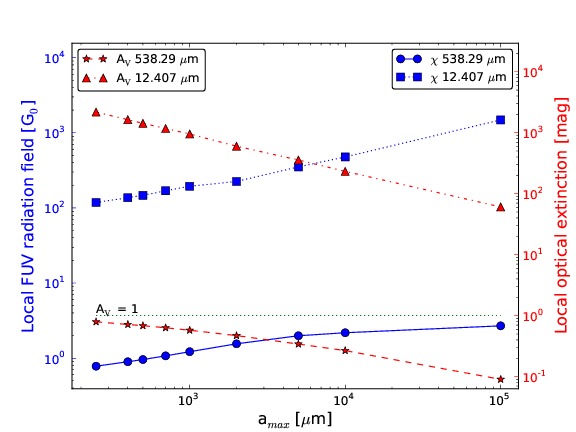}
\end{minipage}
\caption{Plots for maximum dust size models. For further details see Fig.~\ref{dg}.}
%: (a) Emitting region water column density for the maximum dust size models. This region comprises 15-85\% of the radially and vertically integrated flux. Vertical dotted green lines distinguish the reservoirs.%(b) Temperatures of dust and gas in the emitting regions of the same lines.
%(b) Total water  column densities as a function of radius.
%(c) Average extension of the line emitting region that comprises 50\% of the line flux radially and vertically for 12.407 $\mu$m. (d) Average extension of the line emitting region that comprises 50\% of the line flux radially and
%vertically for 538.29 $\mu$m. %(e) Water density and collisional partners densities in the emitting regions of the same lines (critical densities for the standard model conditions are indicated by red dotted lines). 
%(e) Theoretical Spitzer SH/LH ($R$ = 600) for $a_\mathrm{max}$ model series, continuum subtracted and arbitrarily shifted.
%(f) Theoretical Spitzer SH/LH ($R$ = 600) for $a_\mathrm{max}$ model series. (g) Spitzer line fluxes with sensitivity limits (dotted green lines) for Spitzer/IRS, JWST/MIRI. %(h) HIFI line fluxes with sensitivity limit for 1 hr. exposure time (dotted green line).
%(h) Radiation field and optical extinction in the line emitting regions of 12.407 $\mu$m and 538.29 $\mu$m.}
\label{am} 
\end{figure*}

%\begin{figure}[htpl]
%\centering
%\includegraphics[width=0.45\textwidth]{/net/jeans/data/users/antonellini/fits/Theoretical/parametrized/amax/CD.eps}
%\caption{Emitting region water column density for the maximum dust size models. This region comprises 15-85\% of the radially and vertically integrated flux. Vertical dotted green lines distinguish the reservoirs.}
%\label{am1}
%\end{figure}

The modeled Spitzer spectrum (Fig.~\ref{am}e~\&~f) shows a decreasing mid-IR dust continuum for models with larger maximum grain sizes. There is, however, no significant variation of mid-IR color (within a factor 2). 

%\begin{figure}
%\resizebox{\hsize}{!}{\includegraphics{/net/jeans/data/users/antonellini/fits/Theoretical/parametrized/amax/Spitzer_amax.eps}}
%\resizebox{\hsize}{!}{\includegraphics{/net/jeans/data/users/antonellini/fits/Theoretical/parametrized/amax/Sp_lines_amax.eps}}
%\caption{Theoretical Spitzer SH/LH ($R$ = 600) for $a_\mathrm{max}$ model series; top: standard, bottom: continuum subtracted and arbitrarily shifted.} 
%\label{Spitzer_am}
%\end{figure}

\subsection{Dust size power law distribution}\vspace{5mm}

The total water column density inside 1~AU remains constant when varying the power law index of the grain size distribution, while it decreases by several orders of magnitude in the outer disk (Fig.~\ref{ap}b). The water 
column density in the Spitzer line emitting region becomes lower, and the region shrinks radially by more than a factor of 3 (Fig.~\ref{ap}a~\&~c). The migration is, however, not monotonic (Fig.~\ref{ap}c). This is
because in the models with a power law index above 3.0, the continuum optical depth in the Spitzer emitting region increases strongly. Below 3.0, the optical depth stays constant ($\tau_\mathrm{cont}\!\sim\!100$). 
The increase in opacity causes the line emitting region to move inward, where the gas still reaches the line excitation temperatures.
 
In the Herschel/HIFI emitting region, dust and gas are thermally coupled above a threshold of $a_\mathrm{pow}\!=\!3.5$. The emitting region becomes progressively optically thick for larger $a_\mathrm{pow}$, but it never 
reaches $A_\mathrm{V}\!=\!1$.
The line is emitted deeper in the disk in the models with $a_\mathrm{pow}\!=$~2.0 up to 3.0. This is due to enhanced photodesorption, which increases the water column density in the sub-mm line emitting region. The lower 
column densities of water vapour in the outer disk lead to lower fluxes in the sub-mm water lines (Fig.~\ref{ap}a~\&~b). The emitting region of the far-IR lines moves upward for $a_\mathrm{pow}\!>\!3.0$. In 
this case, the opacity is so high that water is not efficiently photodesorbed anymore and the line has to be produced higher up in the disk, where water is still in gaseous form.

\begin{figure*}[htpl]
\centering
\begin{minipage}[l]{0.36\textwidth}
(a)\includegraphics[width=1.1\textwidth]{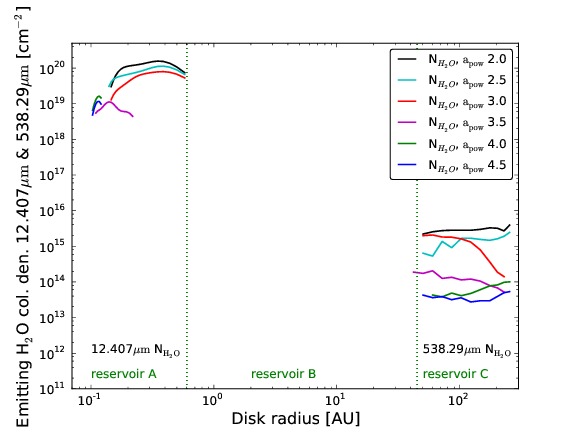}
%(h)\includegraphics[width=\textwidth]{/net/jeans/data/users/antonellini/fits/Theoretical/parametrized/Recon_six/HIFI_apow.eps}
\end{minipage}
\begin{minipage}[r]{0.36\textwidth}
(b)\includegraphics[width=1.1\textwidth]{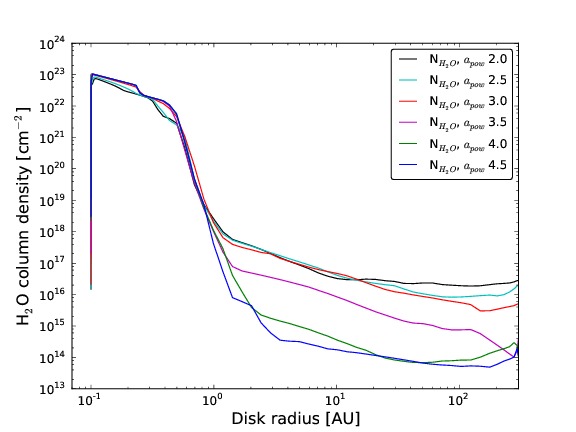}
\end{minipage}
\begin{minipage}[l]{0.36\textwidth}
(c)\includegraphics[width=1.1\textwidth]{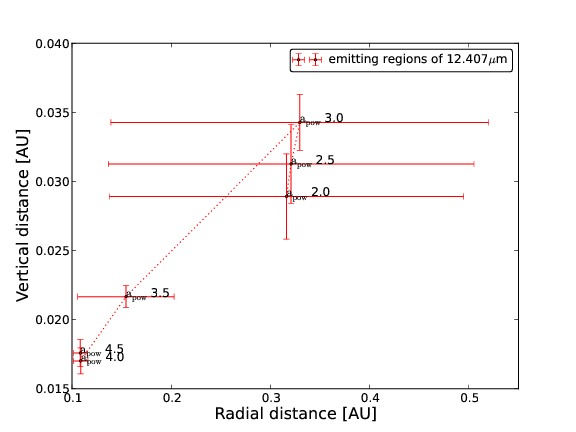}
\end{minipage}
\begin{minipage}[r]{0.36\textwidth}
(d)\includegraphics[width=1.1\textwidth]{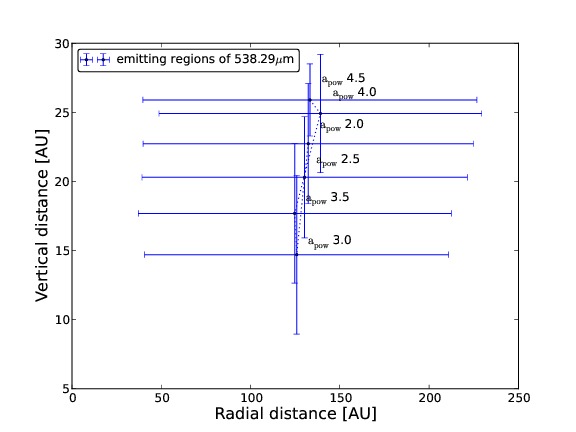}
\end{minipage}
\begin{minipage}[l]{0.36\textwidth}
(e)\includegraphics[width=0.97\textwidth]{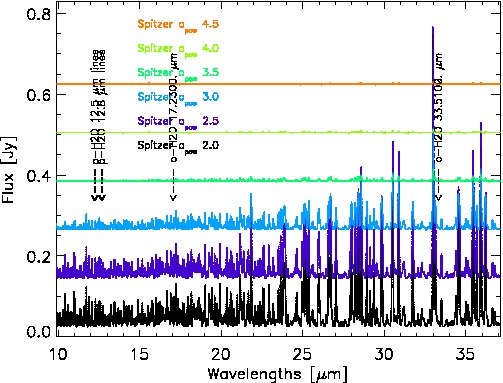}
%(e)\includegraphics[width=\textwidth]{/net/jeans/data/users/antonellini/fits/Theoretical/parametrized/Recon_six/apow_Densities.eps}
\end{minipage}
\begin{minipage}[r]{0.36\textwidth}
(f)\includegraphics[width=0.97\textwidth]{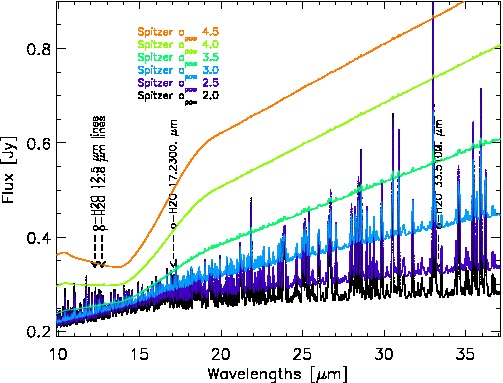}
%(b)\includegraphics[width=\textwidth]{/net/jeans/data/users/antonellini/fits/Theoretical/parametrized/Recon_six/apow_Temperatures.eps}
\end{minipage}
\begin{minipage}[l]{0.36\textwidth}
(g)\includegraphics[width=1.1\textwidth]{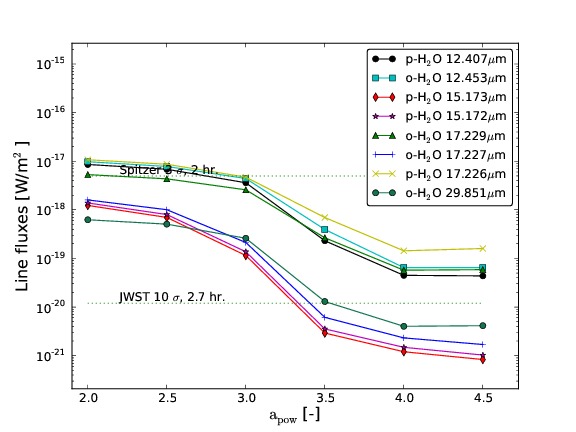}
\end{minipage}
\begin{minipage}[r]{0.36\textwidth}
(h)\includegraphics[width=1.1\textwidth]{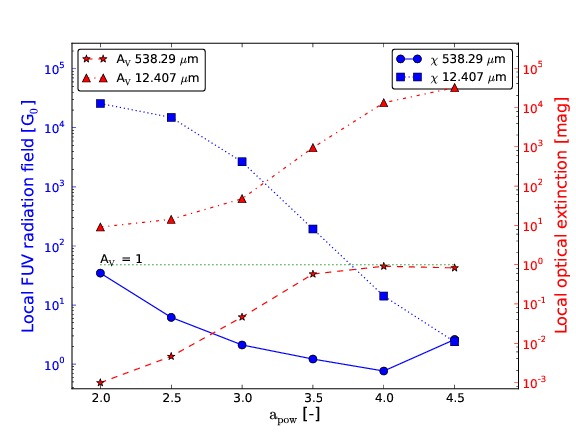}
\end{minipage}
\caption{Plots for dust size distribution power law index models. For further details, see Fig.~\ref{dg}.}
%: (a) Emitting region water column density for the dust size distribution power law index models. This region comprises 15-85\% of the radially and vertically integrated flux. Vertical dotted green lines distinguish the 
%reservoirs. %(b) Temperatures of dust and gas in the emitting regions of the same lines. 
%(b) Total water column densities as a function of radius.
%(c) Average extension of the line emitting region that comprises 50\% of the line flux radially and vertically for 12.407 $\mu$m. (d) Average extension of the line emitting region that comprises 50\% 
%of the line flux radially and vertically for 538.29 $\mu$m. %(e) Water density and collisional partner densities in the emitting regions of the same lines (critical densities for the standard model conditions are indicated by 
%reddotted lines). 
%(e) Theoretical Spitzer SH/LH ($R$ = 600) for $a_\mathrm{pow}$ model series, continuum subtracted and arbitrarily shifted.
%(f) Theoretical Spitzer SH/LH ($R$ = 600) for $a_\mathrm{pow}$ model series. (g) Spitzer line fluxes with sensitivity limits (dotted green lines) for Spitzer/IRS, JWST/MIRI. %(h) HIFI line fluxes with sensitivity detection for 
%1 hr. exposure time (dotted green line).
%(h) Radiation field and optical extinction in the line emitting regions of 12.407 $\mu$m and 538.29 $\mu$m.}
\label{ap}
\end{figure*}

%\begin{figure}[htpl]
%\centering
%\includegraphics[width=0.45\textwidth]{/net/jeans/data/users/antonellini/fits/Theoretical/parametrized/apow/CD.eps}
%\caption{Emitting region water column density for the dust size distribution power law index models. This region comprises 15-85\% of the radially and vertically integrated flux. Vertical dotted green lines distinguish the 
%reservoirs.}
%\label{ap1}
%\end{figure}

The modeled Spitzer spectrum (Fig.~\ref{ap}e~\&~f) shows an increase of the continuum in the near-IR to mid-IR range and a lower line-to-continuum ratio for the models with a larger amount of small dust (high values of 
$a_\mathrm{pow}$). The mid-IR color variation is very high and the 13.5/30~$\mu$m continuum flux ratio decreases by more than a factor 2. The change in continuum flux is a consequence of settling (present in all the models). For 
larger $a_\mathrm{pow}$, the disk has more smaller grains (that are not settled), and so the warm surface layers contain a higher dust mass, which contributes to the continuum. In this case, the dust consists mainly of small 
grains and the opacity is high. 
As discussed in Sect.~\ref{dustgas_par}, the line-to-continuum ratio is then very low. The flat grain size distribution model has a lower opacity (flatter and more than 4 dex weaker in the UV-optical range), and the dust and gas
temperature gradients are radially and vertically steeper than in the case of a steeper grain size distribution. Then, the dust  becomes  colder overall, producing less mid-IR continuum. 

%\begin{figure}
%\resizebox{\hsize}{!}{\includegraphics{/net/jeans/data/users/antonellini/fits/Theoretical/parametrized/apow/Spitzer_apow.eps}}
%\resizebox{\hsize}{!}{\includegraphics{/net/jeans/data/users/antonellini/fits/Theoretical/parametrized/apow/Sp_lines_apow.eps}}
%\caption{Theoretical Spitzer SH/LH ($R$ = 600) for $a_\mathrm{pow}$ model series; top: standard, bottom: continuum subtracted and arbitrarily shifted.} 
%\label{Spitzer_ap}
%\end{figure}

\subsection{ISM radiation field}\vspace{5mm}

The ISM FUV radiation field affects the outer regions of the disk and changes the water column density beyond 30 AU in a complex way (Fig.~\ref{g0}b). In the most irradiated model, the total water column density decreases 
from photodissociation. In the emitting region of the Spitzer lines, the column density is unaffected below $\chi_\mathrm{ISM}\!=\!10^3\!~G_\mathrm{0}$ (Fig.~\ref{g0}a). 
The Spitzer line emitting region for $\chi_\mathrm{ISM}\!>\!10^3\!~G_\mathrm{0}$ extends over the entire disk, overlapping with the sub-mm lines. This happens because of a strong change in the thermal structure of the 
irradiated disk, and the formation of a hot layer of gas extending into the outer disk (beyond 10 AU). The hot gas extends down to the midplane for the $\chi_\mathrm{ISM}\!=\!10^6\!~G_\mathrm{0}$ model. Water is then eroded
through photodissociation in the outer disk and the line is produced at smaller radii and closer to the midplane (Fig.~\ref{g0}d). The increase in $T_\mathrm{gas}$ for the more irradiated models 
($\chi_\mathrm{ISM}\!>\!10^3\!~G_\mathrm{0}$) makes the line emitting regions radially more extended, and the mid-IR line fluxes increase (Fig.~\ref{g0}g). 

In the outer disk, the opacity in the emitting region of the far-IR lines drops of an order of magnitude at $\chi_\mathrm{ISM}\!=\!10^4\!~G_0$, but then it raises again for larger $\chi_\mathrm{ISM}$. This is because of a 
complex variation in the total water column density in the outer disk (Fig.~\ref{g0}b~\&~a), which also  affects  the emitting region's extension and position (Fig.~\ref{g0}c~\&~d).
The far-IR lines first moves outward and then above $\chi_\mathrm{ISM}\!=\!10^3\!~G_\mathrm{0}$ inward and closer to the midplane. The line optical depth drops below 1.0 for $\chi_\mathrm{ISM}\!>\!10^3\!~G_0$ 
(Fig.~\ref{g0}d). 
Increasing the FUV radiation field, the emitting column density becomes larger because of enhanced photodesorption (Fig.~\ref{g0}a). The emitting region, however, is radially less extended because of increasing 
photodissociation, which removes water from the outer disk and even partially depletes the ice reservoir (Fig.~\ref{g0}b).
The gas temperature in the emitting region increases from less than 100 K to 500 K and decouples from the dust temperature for models with $\chi_\mathrm{ISM}\!>\!100\!~G_\mathrm{0}$. The interplay between the increased 
temperature, the larger column density and the decreasing emitting area in the outer disk limits the sub-mm line flux increase to an order of magnitude.

\begin{figure*}[htpl]
\centering
\begin{minipage}[l]{0.36\textwidth}
(a)\includegraphics[width=1.1\textwidth]{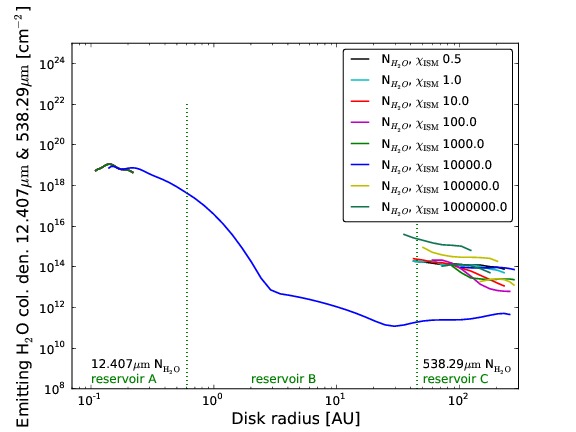}
%(h)\includegraphics[width=\textwidth]{/net/jeans/data/users/antonellini/fits/Theoretical/parametrized/Recon_six/HIFI_G0.eps}
\end{minipage}
\begin{minipage}[r]{0.36\textwidth}
(b)\includegraphics[width=1.1\textwidth]{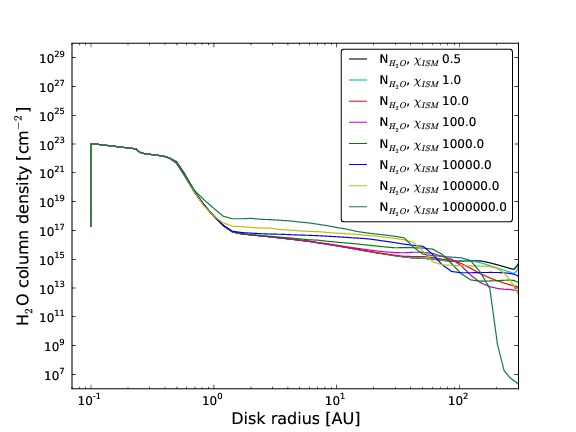}
\end{minipage}
\begin{minipage}[l]{0.36\textwidth}
(c)\includegraphics[width=1.1\textwidth]{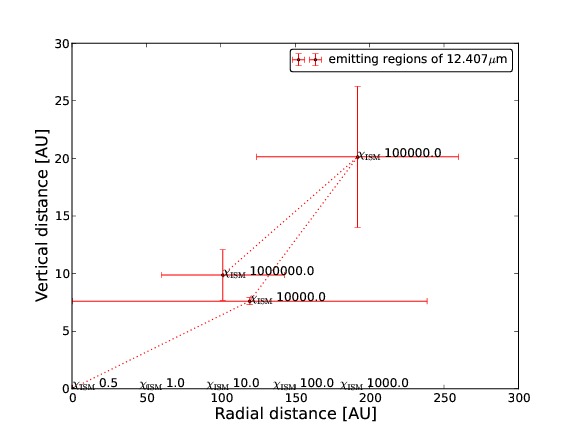}
\end{minipage}
\begin{minipage}[r]{0.36\textwidth}
(d)\includegraphics[width=1.1\textwidth]{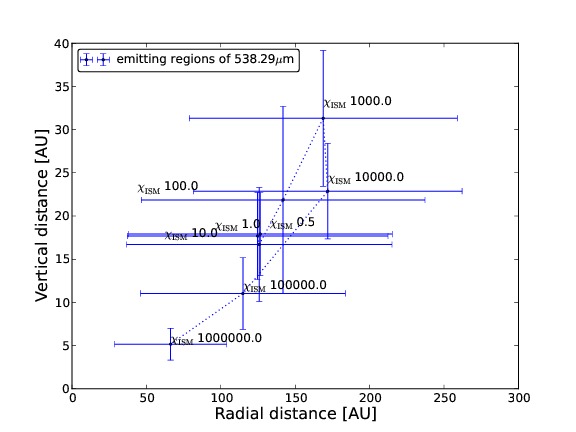}
\end{minipage}
\begin{minipage}[l]{0.36\textwidth}
(e)\includegraphics[width=0.97\textwidth]{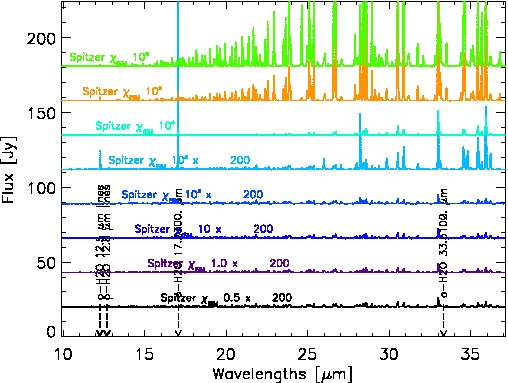}
%(e)\includegraphics[width=\textwidth]{/net/jeans/data/users/antonellini/fits/Theoretical/parametrized/Recon_six/G0_Densities.eps}
\end{minipage}
\begin{minipage}[r]{0.36\textwidth}
(f)\includegraphics[width=0.97\textwidth]{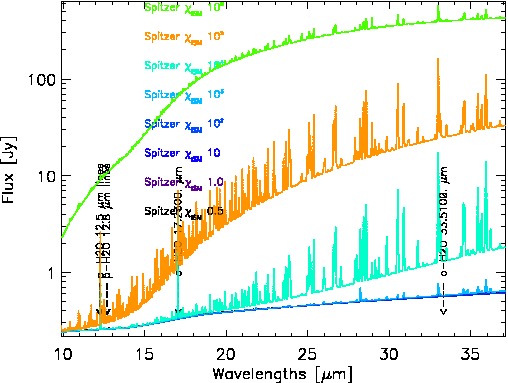}
%(b)\includegraphics[width=\textwidth]{/net/jeans/data/users/antonellini/fits/Theoretical/parametrized/Recon_six/G0_Temperatures.eps}
\end{minipage}
\begin{minipage}[l]{0.36\textwidth}
(g)\includegraphics[width=1.1\textwidth]{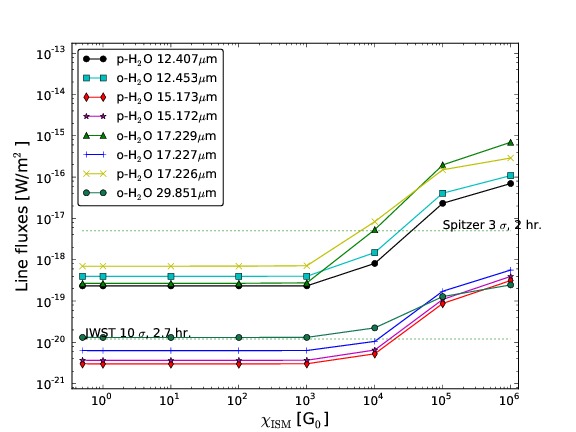}
\end{minipage}
\begin{minipage}[r]{0.36\textwidth}
(h)\includegraphics[width=1.1\textwidth]{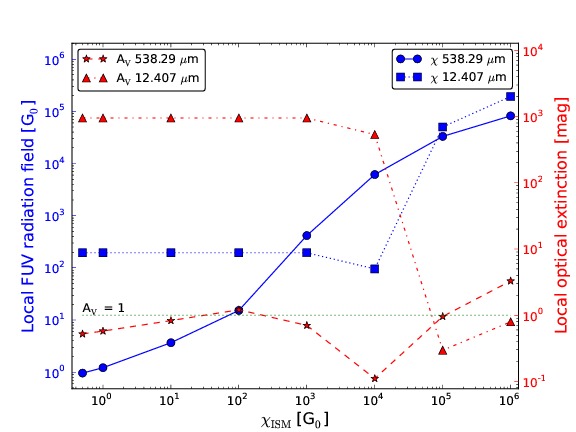}
\end{minipage}
\caption{Plots for ISM radiation field models. For further details see Fig.~\ref{dg}.}
%: (a) Emitting region water column density for the ISM radiation field models. This region comprises 15-85\% of the radially and vertically integrated flux. Vertical dotted green lines distinguish the reservoirs. 
%(b) Temperatures of dust and gas in the emitting regions of the same lines.
%(b) Total water column densities as a function of radius.
%(c) Average extension of the line emitting region that comprises 50\% of the line flux radially and vertically for 12.407 $\mu$m. (d) Average extension of the line emitting region that comprises 50\% of the line flux 
%radially and vertically for 538.29 $\mu$m. %(e) Water density and collisional partners densities in the emitting regions of the same lines (critical densities for the standard model conditions are indicated by red dotted lines).
%(e) Theoretical Spitzer SH/LH ($R$ = 600) for $\chi_\mathrm{ISM}$ model series, continuum subtracted and arbitrarily shifted. Some of the spectra have been multiplied with a factor 20 in order to visualize the line features.
%(f) Theoretical Spitzer SH/LH ($R$ = 600) for $\chi_\mathrm{ISM}$ model series. (g) Spitzer line fluxes with sensitivity limits (dotted green lines) for Spitzer/IRS, JWST/MIRI. 
%(h) Radiation field and optical extinction in the line emitting regions of 12.407 $\mu$m and 538.29 $\mu$m.}
%%(h) HIFI line fluxes with sensitivity limit for 1 hr. exposure time (dotted green line).}
\label{g0}
\end{figure*}

%\begin{figure}[htpl]
%\centering
%\includegraphics[width=0.45\textwidth]{/net/jeans/data/users/antonellini/fits/Theoretical/parametrized/G0/CD.eps}
%\caption{Emitting region water column density for the ISM radiation field models. This region comprises 15-85\% of the radially and vertically integrated flux. Vertical dotted green lines distinguish the reservoirs.}
%\label{g01}
%\end{figure}

The theoretical Spitzer spectrum (Fig.~\ref{g0}e~\&~f) shows an increase of the line-to-continuum ratio for an increasing $\chi_\mathrm{ISM}$ up to 10~$G_\mathrm{0}$. For fluxes larger than 1000~$G_\mathrm{0}$, the continuum 
starts to become stronger, reducing the line-to-continuum ratio. This is due to the fact that $\chi_\mathrm{ISM}\!>\!\chi_\mathrm{star}$ ($\chi_\mathrm{star}\!\sim~\!10^4\!~G_\mathrm{0}$ within 2~AU). Beyond 
10$^3\!~G_\mathrm{0}$,
the mid-IR color becomes much smaller than 1.

%\begin{figure}
%\resizebox{\hsize}{!}{\includegraphics{/net/jeans/data/users/antonellini/fits/Theoretical/parametrized/G0/Spitzer_G0.eps}}
%\resizebox{\hsize}{!}{\includegraphics{/net/jeans/data/users/antonellini/fits/Theoretical/parametrized/G0/Sp_lines_G0.eps}}
%\caption{Theoretical Spitzer SH/LH ($R$ = 600) for $\chi_\mathrm{ISM}$ model series; top: standard, bottom: continuum subtracted and arbitrarily shifted. Some of the spectra have been multiplied with a factor 20 in order to 
%visualize the line features.}
%\label{Spitzer_g0}
%\end{figure}

\subsection{PAHs photoelectric heating}\vspace{5mm}

PAHs affect the total water column density by less than an order of magnitude in the outermost regions of the disk (Fig.~\ref{pa}a). The Spitzer transitions are almost unaffected (Fig.~\ref{pa}b). Their emitting
region moves slightly upward and outward due to the increase in heating caused by PAHs (Fig.~\ref{pa}c). However, this displacement is too small to affect the line fluxes in the mid-IR. The column density in the emitting region
of the far-IR lines increases by less than a factor of two with increasing $f_\mathrm{PAH}$. %(Fig.~\ref{pa}h).
The sub-mm line emitting region moves outward and upward (Fig.~\ref{pa}d), but the gas temperature hardly changes. The line fluxes increase by a factor $\sim$~2. %(Fig.~\ref{pa}h).

\begin{figure*}[htpl]
\centering
\begin{minipage}[l]{0.36\textwidth}
(a)\includegraphics[width=1.1\textwidth]{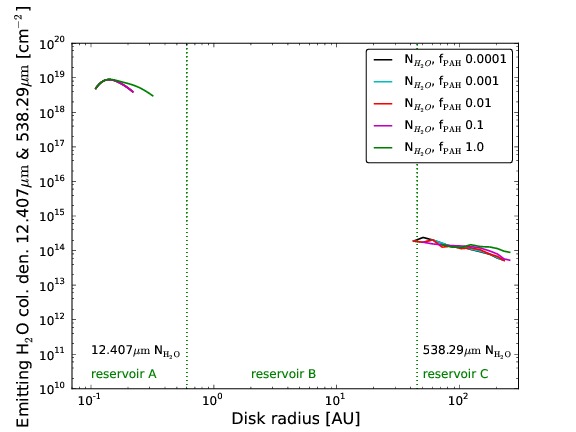}
%(h)\includegraphics[width=\textwidth]{/net/jeans/data/users/antonellini/fits/Theoretical/parametrized/Recon_six/HIFI_fPAH.eps}
\end{minipage}
\begin{minipage}[r]{0.36\textwidth}
(b)\includegraphics[width=0.97\textwidth]{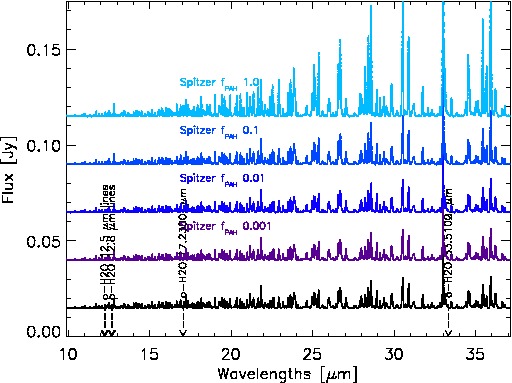}
%(b)\includegraphics[width=\textwidth]{/net/jeans/data/users/antonellini/fits/Theoretical/parametrized/fPAH/ColumnDens.eps}
\end{minipage}
\begin{minipage}[l]{0.36\textwidth}
(c)\includegraphics[width=1.1\textwidth]{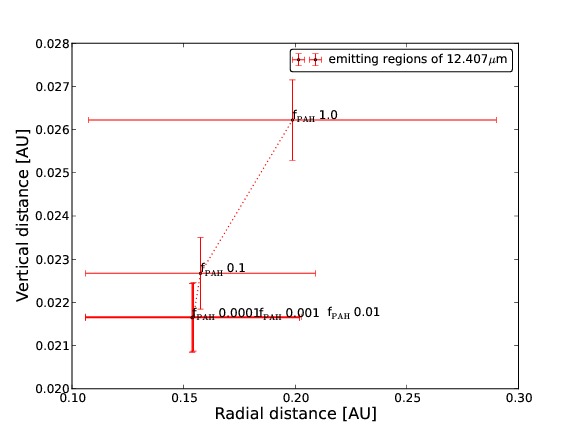}
\end{minipage}
\begin{minipage}[r]{0.36\textwidth}
(d)\includegraphics[width=1.1\textwidth]{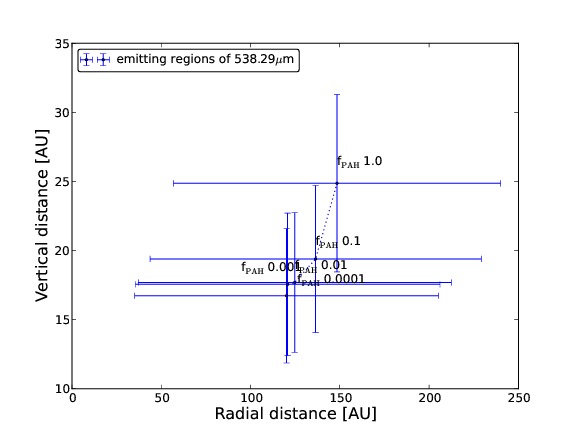}
\end{minipage}
%\begin{minipage}[l]{0.36\textwidth}
%(e)\includegraphics[width=\textwidth]{/net/jeans/data/users/antonellini/fits/Theoretical/parametrized/fPAH/Sp_lines_fPAH.eps}
%(e)\includegraphics[width=\textwidth]{/net/jeans/data/users/antonellini/fits/Theoretical/parametrized/Recon_six/fPAH_Densities.eps}
%\end{minipage}
%\begin{minipage}[r]{0.36\textwidth}
%(f)\includegraphics[width=\textwidth]{/net/jeans/data/users/antonellini/fits/Theoretical/parametrized/fPAH/Spitzer_fPAH.eps}
%(b)\includegraphics[width=\textwidth]{/net/jeans/data/users/antonellini/fits/Theoretical/parametrized/Recon_six/fPAH_Temperatures.eps}
%\end{minipage}
%\begin{minipage}[l]{0.36\textwidth}
%(g)\includegraphics[width=\textwidth]{/net/jeans/data/users/antonellini/fits/Theoretical/parametrized/Recon_six/Spitzer_fPAH.eps}
%\end{minipage}
%\begin{minipage}[r]{0.36\textwidth}
%(h)\includegraphics[width=\textwidth]{/net/jeans/data/users/antonellini/fits/Theoretical/parametrized/Recon_six/fPAH_Radiation.eps}
%\end{minipage}
%\caption{Plots for the fraction of PAHs models. For further details see Fig.~\ref{dg}.}
\caption{Plots for the fraction of PAHs models. (a) Emitting region water column density for dust-to-gas mass ratio models; this region comprises 15-85\% of the radially and vertically integrated flux. Green vertical dotted lines distinguish the reservoirs. %(b) Temperatures of dust and gas in the emitting regions of the same
(b) Theoretical Spitzer SH/LH modules spectra ($R$~=~600) for a dust-to-gas model series, continuum subtracted, and arbitrarily shifted.
(c) Average extension of the line emitting region, which comprises 50\% of the line flux radially and vertically for 12.407~$\mu$m. (d) Average extension of the line emitting region, which comprises 50\% of the line flux radially 
and vertically for 538.29~$\mu$m.}
%(e) Water density and collisional partners densities in the emitting regions of the same lines (critical densities for the standard model conditions are indicated by red dotted lines).
%(f) Theoretical Spitzer SH/LH ($R$ = 600) for a dust-to-gas model series. (g) Spitzer line fluxes with sensitivity limits (dotted green lines) for Spitzer/IRS, JWST/MIRI.} 
%(h) HIFI line fluxes with sensitivity detection for 1 hr. exposure time (dotted green line).
%: (a) Emitting region water column density for the fraction of PAHs models. This region comprises 15-85\% of the radially and vertically integrated flux. Vertical dotted green lines distinguish the reservoirs. %(b) Temperatures of dust and gas in the emitting regions of the same lines.
%(b) Total water column densities as a function of radius.
%(c) Average extension of the line emitting region that comprises 50\% of the line flux radially and vertically for 12.407 $\mu$m. (d) Average extension of the line emitting region that comprises 50\% of the line flux 
%radially and vertically for 538.29 $\mu$m. 
%(e) Theoretical Spitzer SH/LH ($R$ = 600) for $f_\mathrm{PAH}$ model series, continuum subtracted and arbitrarily shifted.
%(f) Theoretical Spitzer SH/LH ($R$ = 600) for $f_\mathrm{PAH}$ model series.
%(g) Spitzer line fluxes with sensitivity limits (dotted green lines) for Spitzer/IRS, JWST/MIRI. %(h) HIFI line fluxes with sensitivity limit for 1 hr. exposure time (dotted green line).
%(h) Radiation field and optical extinction in the line emitting regions of 12.407 $\mu$m and 538.29 $\mu$m.}
\label{pa}
\end{figure*}

%\begin{figure}[htpl]
%\centering
%\includegraphics[width=0.45\textwidth]{/net/jeans/data/users/antonellini/fits/Theoretical/parametrized/fPAH/CD.eps}
%\caption{Emitting region water column density for the fraction of PAHs models. This region comprises 15-85\% of the radially and vertically integrated flux. Vertical dotted green lines distinguish the reservoirs.}
%\label{pa1}
%\end{figure}

Since the dust opacities stay constant, the Spitzer spectra all show the same continuum; the 13.5/30~$\mu$m continuum flux ratio is thus also unaffected. The line-to-continuum ratio grows with increasing PAH abundance because 
of the heating effect on the gas. 

%\begin{figure}
%%\resizebox{\hsize}{!}{\includegraphics{/net/jeans/data/users/antonellini/fits/Theoretical/parametrized/fPAH/Spitzer_fPAH.eps}}
%\resizebox{\hsize}{!}{\includegraphics{/net/jeans/data/users/antonellini/fits/Theoretical/parametrized/fPAH/Sp_lines_fPAH.eps}}
%%\caption{Theoretical Spitzer SH/LH ($R$ = 600) for $f_\mathrm{PAH}$ model series; top: standard, bottom: continuum subtracted and arbitrarily shifted.}
%\caption{Theoretical Spitzer SH/LH ($R$ = 600) for $f_\mathrm{PAH}$ model series, continuum subtracted and arbitrarily shifted.
%The spectra with the continuum does not add any information, since it is completely unaffected between models, and is the same as the standard model.}
%\label{Spitzer_pa}
%\end{figure}

\subsection{Disk gas mass}\vspace{5mm}

The total gas mass in the disk affects almost proportionally the total water column density in the disk (Fig.~\ref{mg}b). All reservoirs are enhanced by the same amount with increasing gas mass. In the Spitzer emitting region, 
the gas density progressively increases for larger $M_\mathrm{gas}$. The water column density in the Spitzer water line emitting region follows a nonlinear trend (Fig.~\ref{mg}a). From $M_\mathrm{gas}$~10$^{-5}$ to 
10$^{-4}$~M$_{\odot}$, the emitting region moves outward and upward (Fig.~\ref{mg}c). Then, for masses larger than $M_\mathrm{gas}\!=\!10^{-4}\!~\mathrm{M_{\odot}}$, the migration of the line emitting region is inverted. 
Below a certain water column density \citep[$N_\mathrm{H_2O}\!=\!1.6-8\cdot10^{17}$~cm$^{-2}$,][]{bethell}, self-shielding becomes inefficient and water is strongly photodissociated. For the Spitzer line fluxes, the 
driving mechanism is first the increase of gas column density in the emitting region and, subsequently, the growth of the emitting region itself. 

The water column density in the outer disk increases with $M_\mathrm{gas}$ (Fig.~\ref{mg}b) and also the optical depth of the lines. This causes the transition to be emitted progressively outward and higher up in the 
outer disk (Fig.~\ref{mg}d). Thus, the increase in the far-IR line flux with $M_\mathrm{gas}$ has the same explanation as the increase in the mid-IR lines.

\begin{figure*}[htpl]
\centering
\begin{minipage}[l]{0.36\textwidth}
(a)\includegraphics[width=1.1\textwidth]{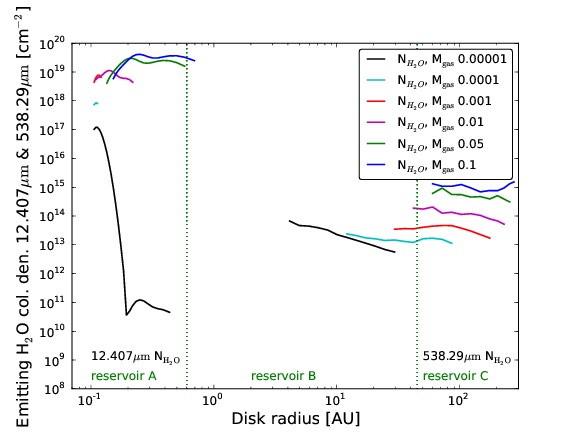}
%(h)\includegraphics[width=\textwidth]{/net/jeans/data/users/antonellini/fits/Theoretical/parametrized/Recon_six/HIFI_Mgas.eps}
\end{minipage}
\begin{minipage}[r]{0.36\textwidth}
(b)\includegraphics[width=1.1\textwidth]{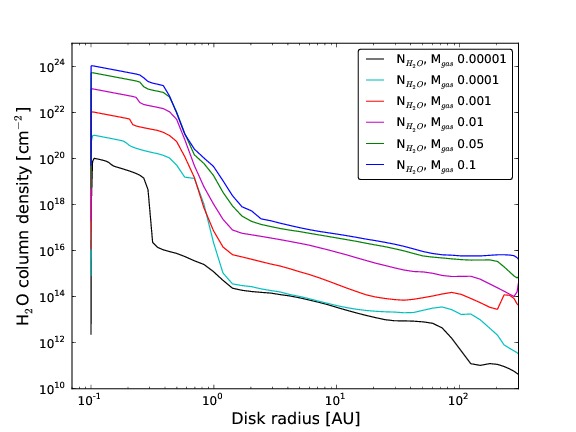}
\end{minipage}
\begin{minipage}[l]{0.36\textwidth}
(c)\includegraphics[width=1.1\textwidth]{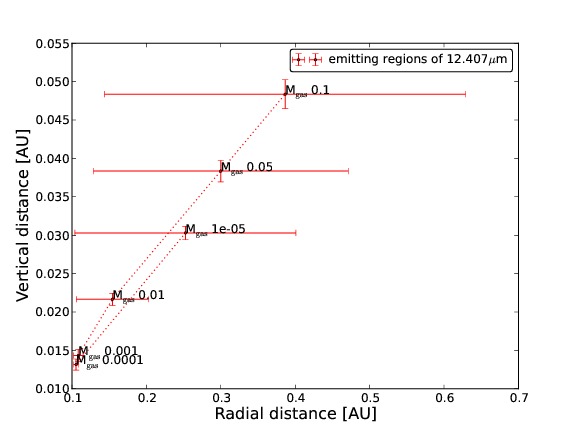}
\end{minipage}
\begin{minipage}[r]{0.36\textwidth}
(d)\includegraphics[width=1.1\textwidth]{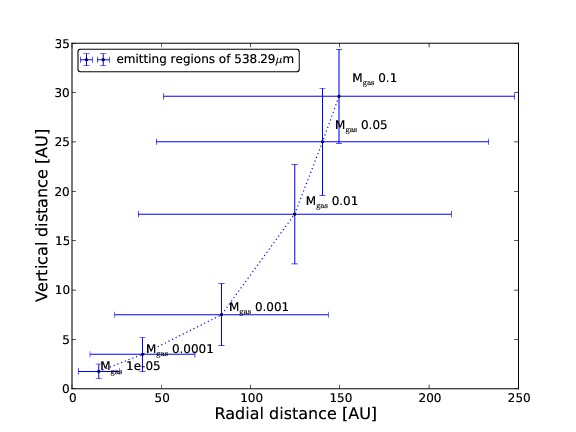}
\end{minipage}
\begin{minipage}[l]{0.36\textwidth}
(e)\includegraphics[width=0.97\textwidth]{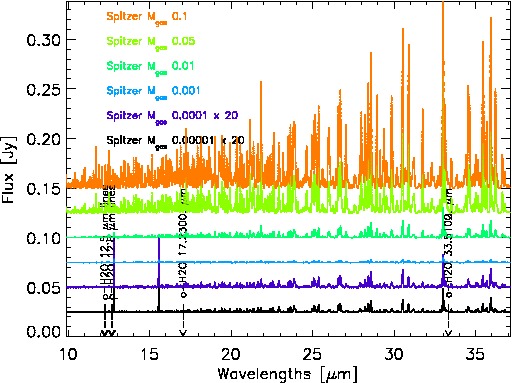}
%(e)\includegraphics[width=\textwidth]{/net/jeans/data/users/antonellini/fits/Theoretical/parametrized/Recon_six/Mgas_Densities.eps}
\end{minipage}
\begin{minipage}[r]{0.36\textwidth}
(f)\includegraphics[width=0.97\textwidth]{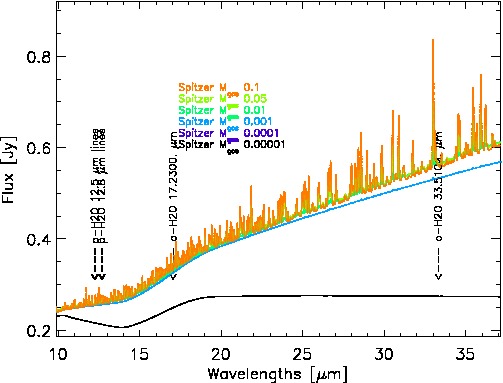}
%(b)\includegraphics[width=\textwidth]{/net/jeans/data/users/antonellini/fits/Theoretical/parametrized/Recon_six/Mgas_Temperatures.eps}
\end{minipage}
\begin{minipage}[l]{0.36\textwidth}
(g)\includegraphics[width=1.1\textwidth]{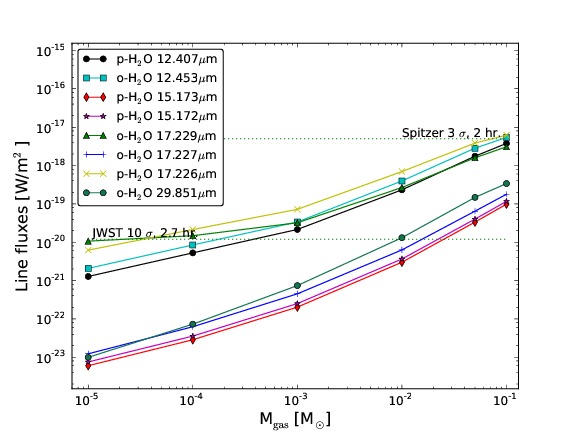}
\end{minipage}
\begin{minipage}[r]{0.36\textwidth}
(h)\includegraphics[width=1.1\textwidth]{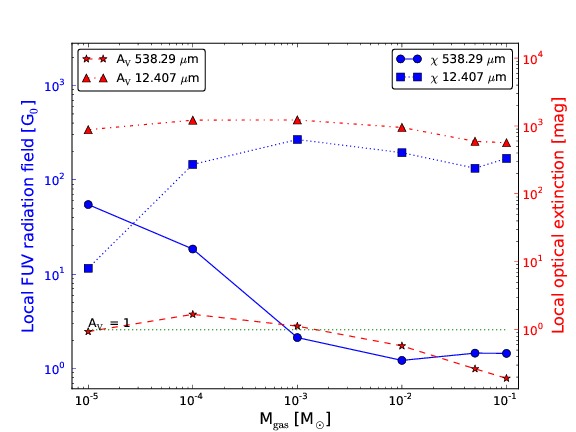}
\end{minipage}
\caption{Plots for the gas mass content models. For further details see Fig.~\ref{dg}.}
%: (a) Emitting region water column density for the gas mass content models. This region comprises 15-85\% of the radially and vertically integrated flux. Vertical dotted green lines distinguish the reservoirs. %(b) Temperatures of dust and gas in the emitting regions of the same lines. 
%(b) Total water column densities as a function of radius.
%(c) Average extension of the line emitting region that comprises 50\% of the line flux radially and vertically for 12.407 $\mu$m. (d) Average extension of the line emitting region that comprises 50\% of the line flux 
%radially and vertically for 538.29 $\mu$m. %(e) Water density and collisional partners densities in the emitting regions of the same lines (critical densities for the standard model conditions are indicated by red dotted lines). 
%(e) Theoretical Spitzer SH/LH ($R$ = 600) for $M_\mathrm{gas}$ model series, continuum subtracted and arbitrarily shifted. Some of the spectra has been magnified of a certain factor, in order to visualize the features.
%(f) Theoretical Spitzer SH/LH ($R$ = 600) for $M_\mathrm{gas}$ model series.
%(g) Spitzer line fluxes with sensitivity limits (dotted green lines) for Spitzer/IRS, JWST/MIRI. 
%(h) Radiation field and optical extinction in the line emitting regions of 12.407 $\mu$m and 538.29 $\mu$m.}
%%(h) HIFI line fluxes with sensitivity limit for 1 hr. exposure time (dotted green line).}
\label{mg}
\end{figure*}

%\begin{figure}[htpl]
%\centering
%\includegraphics[width=0.45\textwidth]{/net/jeans/data/users/antonellini/fits/Theoretical/parametrized/Mgas/CD.eps}
%\caption{Emitting region water column density for the gas mass content models. This region comprises 15-85\% of the radially and vertically integrated flux. Vertical dotted green lines distinguish the reservoirs.}
%\label{mg1}
%\end{figure}

%\begin{figure}
%\resizebox{\hsize}{!}{\includegraphics{/net/jeans/data/users/antonellini/fits/Theoretical/parametrized/Mgas/Spitzer_Mgas.eps}}
%\resizebox{\hsize}{!}{\includegraphics{/net/jeans/data/users/antonellini/fits/Theoretical/parametrized/Mgas/Sp_lines_Mgas.eps}}
%\caption{Theoretical Spitzer SH/LH ($R$ = 600) for $M_\mathrm{gas}$ model series; top: standard, bottom: continuum subtracted and arbitrarily shifted. Some of the spectra has been magnified of a certain factor, in order to 
%visualize the features.}
%\label{Spitzer_mg}
%\end{figure}

The theoretical Spitzer spectrum (Fig.~\ref{mg}e~\&~f) cleary shows  an increase in the line fluxes with the gas content of the disk. The different continuum level is a consequence of the dust settling. In the formalism adopted 
here, the corresponding dust scale height depends on the gas density, which is directly proportional to the mass of the gas in the disk. This means that increasing $M_\mathrm{gas}$  produces less settling, and hence a 
stronger IR continuum with a steeper slope. As a consequence, the mid-IR color becomes smaller.

\section{Discussion}\vspace{5mm}
\label{5}

\subsection{Discrimination between parameters affecting water line fluxes}\vspace{5mm}
\label{before}

The main goal of this work can be summarized in two questions: 
\begin{enumerate}
 \item Why do disks sometimes lack the detection of water?
 \item Why are the water lines from the outer disk absent in disks that have clear detections of inner water reservoirs?
\end{enumerate}
 
\begin{figure}[htpl]
\centering
\includegraphics[width=0.5\textwidth]{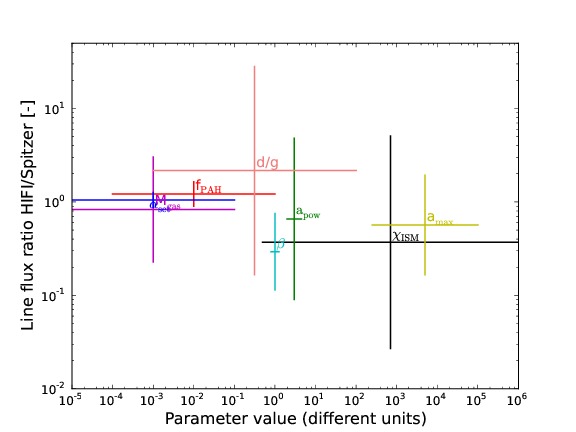}
\caption{Overview of the line ratios Herschel/Spitzer (538.29~$\mu$m/12.407~$\mu$m) for all the parameters as a function of the explored range of values.}
\label{Ratios}
\end{figure}
To answer these questions, we divide the parameters into three categories: those that strongly affect the line ratios (line flux variation $\ge$~1 dex, parameter variation $<$~1 dex); those that produce a relevant 
variation (line flux variation same order of magnitude as parameter variation); and parameters that produce a weak variation (line flux variation less than parameter variation).
An overview of the effects of the parameters on continuum and line fluxes is listed in Table~\ref{summary}.

The Spitzer transitions around 12~$\mu$m are not those typically detected since their flux is weaker compared to other blends. We verified, however, that the same conclusions also hold  for other mid-IR lines around 15, 
17, and 29~$\mu$m. The conclusions about the far-IR lines are limited to the fundamental ortho and para water lines. The line forming regions of other far-IR lines suffer very large radial and vertical dispacements in 
these model series, hence, making the discussion even more complex. They will require a dedicated study.

The Herschel/HIFI lines would be largely undetectable with a standard observational setup and 1 hr of exposure, however, our modeling results suggest that the lines become detectable in many disks at a 5~$\sigma$ 
sensitivity level of 5$\cdot$10$^{-20}$~W/m$^2$. TW~Hya required a total 181~min on-source integration with Herschel/HIFI to detect the 538.29~$\mu$m line \citep{hogerheijde}. %We first discuss the general trends of line 
%fluxes in the mid- and far-IR irrespective of the sensitivity limits of specific instruments.

%\subsection{Categories of parameters}\vspace{5mm}
%\label{before}

%We divided our parameters into three categories, based on their effect on the ratio {\bf{of the mid-IR to the far-IR line}} in relation to the dynamic range explored (Fig.~\ref{Ratios}). We choose this ratio to understand which 
%parameter can explain differences in the detection rates of inner and outer water reservoirs.

In the first category, we find parameters such as the flaring index $\beta$ and the power law dust size distribution exponent $a_\mathrm{pow}$. These parameters are hard to constrain from observations of SEDs alone. The 
Herschel and Spitzer line fluxes are very sensitive to their exact values.
Flaring affects the Herschel/HIFI lines strongly (2 dex) and the Spitzer transitions only marginally (less than a factor 3). The exponent of the dust grain size distribution produces larger changes in the Spitzer regime than in
the Herschel regime. These flux variations are due to geometrical effects (in the outer disk, because of flaring, the emitting column of water and the area of the emitting region increase) and opacity/temperature changes (the 
inner disk in the flared models is directly exposed to the radiation, and so is warmer). In the case of the dust size distribution exponent, the far-IR is always emitted in an optically thin region, and therefore are less affected by any 
opacity change, while in the mid-IR regime the variation is about 4 dex and the line flux is suppressed.
The external part (at larger radii) of reservoir B and all of reservoir C are optically thin in the continuum. This means that the line flux is directly affected by the variation in the dust opacity, and partially by a change 
in the local thermal conditions. For the transitions produced in the inner disk (as the Spitzer ones), the situation is driven by a change in the thermal structure since the lines are optically thick, with the exception o f very peculiar 
models explored in our series of models.

In the second category, we find almost all other parameters: dust-to-gas mass ratio, $d/g$; maximum dust size, $a_\mathrm{max}$; ISM radiation field, $\chi_\mathrm{ISM}$; and gas mass of the disk, $M_\mathrm{gas}$. Excluding the ISM
radiation field, that is an environmental property, all the other properties are good candidates to understand the absence of a correlation between detections in the mid-IR and far-IR wavelength range. In these cases, the 
Spitzer lines are more affected by the parameter variations, and are strongly enhanced (3-4 dex) in disks with $M_\mathrm{dust}\!<\!10^{-4}~\!~M_\mathrm{\odot}$ and/or $M_\mathrm{gas}\!>$~0.01~M$_\mathrm{\odot}$, dust 
grains larger than 400~$\mu$m and few small grains ($a_\mathrm{pow}\!<$~3.5). The continuum is enhanced in the presence of larger dust masses or smaller dust grains that absorb more FUV radiation and become warmer. Herschel 
lines  show the same trend, but the magnitude of the variation is smaller. The effect of these parameters on the spectrum of the outer disk is not as important as in the inner disk.

The last category of parameters, mixing coefficient $\alpha_\mathrm{set}$ and fraction of polycyclic aromatic hydrocarbons $f_\mathrm{PAH}$, do not significantly affect the Spitzer nor the Herschel water emission lines. All water 
transitions respond in the same manner to the mixing coefficient. We stress here that the treatment of the mixing is disconnected from the viscous heating, which is set to zero in 
all our models. The PAHs are more important in the outer disk, which is optically thin, and is thus be warmed up by the increased photoelectric heating from PAHs. PAHs are not treated consistently as an opacity 
source, otherwise the enhanced opacity (expecially in the UV-optical regime) would have produced effects comparable to an increased dust-to-gas mass ratio in the disk.

Since every parameter affecting the line flux also modifies the emitting region position and extension in the disk, the line profiles are consequently affected, and outward displacements of the emitting regions 
produce narrower transitions (Keplerian velocity decreases with radial distances).

Our work finds that the main driver of water spectroscopy in the mid- to far-IR wavelengths range is the continuum opacity. Changes occur because of differences in the disk dust content or the opacity function
($a_\mathrm{max}$, $a_\mathrm{pow}$ variation). These  directly affect the line flux (extinction) or indirectly affect the local thermal conditions (temperature change). 
The second effect is the disk geometry. A flared disk geometry exposes more water to the central stellar radiation field. Hence the total column density is depleted in the outer disk because of photodissociation. An increase in the 
ISM FUV radiation field also produces  a suppression of water in the outermost reservoir C.

As discussed in Section \ref{WaterSpec}, the mid-IR water lines are in LTE, while the outer disk water lines are marginally in non-LTE. Hence the far-IR line fluxes can be affected by uncertanties in the collision cross sections, 
as discussed in \citet{kamp}. We perform a sensitivity check on the standard model making a random variation (increase/decrease) of an order of magnitude in the water collisional constants. The results confirm the robustness of 
the conclusions presented in this paper: the line fluxes of the Spitzer transitions are unaffected, while the Herschel line variations are within 10\%.

The disk water distribution found from the modeling is in qualitative agreement with recent modeling performed by \citet{du}. In particular, we reproduce the change of the column densities of water in the inner and
outer disk due to the variation of dust-to-gas mass ratio and settling. A quantitative comparison is not possible because we use a more physical approach in our models \citep[non-parametric settling;][]{dubrulle}.

\begin{table*}
\caption{General overview of the series of models results. All the variations refer to an increase of the respective parameter value.}
\centering
\begin{tabular}{llllll}%l}
\hline\hline
Parameter$^*$ & IRS lines & mid-IR continuum & HIFI lines & IRS line
profiles & HIFI line profiles \\ \hline%& effects on water lines \\ \hline
% &  &  & &  &  & and continuum\\ \hline
$d/g$ & $--$ & $+$ (bolometrically) & $-$ & narrower & unaffected \\%& mid-IR
%{\bf{undetectable}}, \\
% &  &  &  &  &  &  far-IR {\bf{detectable}}\\ \\
$\beta$ & $+$ & $+$ (bolometrically,  & $++$ & narrower & unaffected \\%&
%far-IR stronger\\
% &  & changes slope  &  &  &  &  \\
% &  & around $17~\mu$m) &  &  &  &  \\ \\
$\alpha_\mathrm{set}$ & $-$ & $+$ (stronger with & $-$ & unaffected &
unaffected \\%& far-IR slightly stronger\\
% &  &  increasing $\lambda$) & &  &  & \\ \\
$a_\mathrm{max}$ & $++$ & $-$ & $+$ & narrower & unaffected \\%& mid-IR
%{\bf{detectable}}, \\
% &  &  &  &  &  & far-IR {\bf{undetectable}}\\ \\
$a_\mathrm{pow}$ & $- -$ & $+$ (variation weaker  & $-$ & wider &
unaffected \\%& mid-IR {\bf{undetectable}}, \\
% &  &  around 15 $\mu$m) & & &  & far-IR lines weaker\\ \\
$\chi_\mathrm{ISM}$ & $++$ $^{(1)}$& $++$ $^{(1)}$ & $+$ & narrower &
unaffected\\ %& far-IR features stronger\\ \\
$f_\mathrm{PAH}$ & x & x & $+$ & unaffected & unaffected \\%& far-IR features
%slightly \\
% &  &  &  &  &  &  enhanced\\ \\
$M_\mathrm{gas}$ & $++$ & $+$ (stronger with & $++$ & narrower & narrower\\ \hline
%& all IR lines stronger\\
% &  &  increasing $\lambda$) &  &  &  & \\ \hline
\end{tabular}
\tablefoot{Symbols meanings: $+$ increase $<$~1~dex, $-$ decrease
$<$~1~dex, $++$ increase $\ge$~1~dex, $- -$ decrease $\ge$~1~dex, x
constant; (1) for $\chi_\mathrm{ISM} > 10^4 G_0$ }
\label{summary}
\end{table*}

\subsection{Comparison with observations}\vspace{5mm}

Observationally, it will be difficult to assess the exact properties of a certain disk. Previous observational studies of mid-IR water spectroscopy used continuum diagnostics in the same spectral region to connect to disk
properties of the targets. We expect to find such correlations, and we want to use them to define the position of each model produced by a parameter space exploration
(that starts from our standard disk model), in a plot of the line flux vs. continuum flux. Here, we decide to use a mid-IR color \citep[13.5/30; similar to what is used in][]{acke2} because it traces slope changes in the Spitzer 
spectrum, and we found that that every parameter produces its own fingerprint. 
For the far-IR, we decided to use a diagnostic connected with the outer disk properties. We choose the SCUBA flux at 850~$\mu$m because it is available for many objects and traces dust mass to first order. 
We search for a direct relation between mid-IR colors, submm photometry, and simulated line fluxes at 12.407~$\mu$m and 538.29~$\mu$m. We distinguished three types of parameters (not to be confused with the categories defined
 in Sect.~\ref{5}): parameters that change the 
dust opacity function ($a_\mathrm{max}$, $a_\mathrm{pow}$), parameters that change only the total opacity ($M_\mathrm{gas}$, $d/g$, $\alpha_\mathrm{set}$), and parameters that do not affect the opacity (ISM radiation field, 
$f_\mathrm{PAH}$, $\beta$).
Every parameter produces its own signature on the SED and line fluxes, and Fig.~\ref{colors} and Table~\ref{rainbow} summarize the main findings. The parameter $M_\mathrm{gas}$ produces a limited variation in the mid-IR 
continuum, and a more important effect on the submm (82\% in sub-mm, 10\% in mid-IR, due to the settling). It produces a strong variation in the line flux both in the mid- and far-IR. Settling only modestly affects  the line 
fluxes in both the regimes, but the mid-IR slope is strongly changed by up to 50\% and the submm continuum flux changes by 87\%. The dust-to-gas mass ratio does not strongly affect the mid-IR continuum (about 25\%); however it 
changes the submm flux by almost a factor 2. Flaring significantly affects  both the mid-IR continuum (90\% relative variation) and the submm (more than 95\% between extreme models), but the Spitzer lines are hardly affected. The
ISM FUV radiation flux produces a stronger change in the Spitzer lines compared to the Herschel/HIFI lines, and a steepening of the IR excess beyond 10~$\mu$m. For the far-IR lines, the flux grows with increasing FUV radiation 
field until a certain limit and then levels off. The fraction of PAHs does not affect the SED, since the PAH opacity is not taken into account in the continuum radiative transfer. 

A comparison of the modeling results with observations toward T~Tauri disks shows that our simplified model predictions define a parameter space that is in good agreement with a number of objects: AA~Tau, DG~Tau, RW~Aur, 
DM~Tau for the mid-IR spectral region (top plot of Fig.~\ref{colors}), DG~Tau, DM~Tau, and TW~Hya for the far-IR spectral region (bottom plot of Fig.~\ref{colors}).
These targets have spectral types and properties compatible with our central star, and are not heavily affected  by the presence of companions, either because they are too faint 
\citep[case of RW~Aur~A+B;][]{white1} or because the members are widely separated \citep[DG~Tau~A+B;][]{mundt}. Appendix~\ref{app4} describes how the 12.4~$\mu$m blend flux was extracted from Spitzer spectra for the above 
targets and how continuum errors were propagated; fluxes are listed in Table~\ref{Obs}.
Fig.~\ref{colors} compares the model results with the observational data: line blends around 12.4~$\mu$m versus our mid-IR color and 538.29~$\mu$m versus the SCUBA 850~$\mu$m flux. 
Our model series does not pretend to describe the situation of each target, which is a complex interplay of properties, but the position of the targets agree nicely with our models and support the predictional power of mid- and 
far-IR observations. The direct comparison with the sensitivities of Spitzer/IRS and JWST/MIRI (or Herschel/HIFI for the far-IR plot) shows the discovery potential of these facilities.

Fig.~\ref{colors} shows that it is not possible to build trends that clearly connect line fluxes with continuum properties because different parameters affect the relation between these two quantities in a 
different manner. Hence, it is logical to expect a large amount of scatter in observations. Previous observational work \citep[e.g.,\ ][]{pontoppidan2} show several plots with scattered observations in the continuum 
properties or line vs continuum fluxes or even continuum versus continuum. Any individual parameter produces a complex displacement in each of these plots, and every disk can be a different combination of parameters.

\subsection{The future with JWST}\vspace{5mm}

The low resolution of the IRS spectrograph ($R$~=~600) does not enable an investigation of the line profiles; the blends usually contain several transitions  \citep{banzatti}. The problem is particularly important when the line profiles 
deviate from Keplerian rotation because of other kinematical components caused by winds/outflows/jets \citep{baldovin-saavedra,baldovin-saavedra1}.
The future facility to study mid-IR water lines is the James Webb Space Telescope \citep{greenhouse}. The larger mirror and higher spectral resolution of the MIRI spectrograph 
\citep[$\diameter~=~6.5~$m, $R~\simeq$~3000;][]{clampin} will overcome the technical limitations of IRS. Fig.~\ref{IRSvsJWST} shows a comparison 
of the mid-IR spectrum of our standard model in the Spitzer/IRS range with the resolution of the IRS and MIRI spectrographs. The higher sensitivity and the spectral coverage (5-28.5~$\mu$m) similar to Spitzer IRS, will allow us 
to detect water lines in almost all the cases modeled in this study, within a 2.7~hr exposure (10~$\sigma$; Fig.~\ref{colors}). Fig.~\ref{IRSvsJWST} also shows that JWST/MIRI can make a decomposition of line 
blends into individual lines feasible, allowing, possibly for the first time, a quantitative analysis of the physical conditions in their emitting region. 

\begin{figure}[htpl]
\centering
\includegraphics[width=0.5\textwidth]{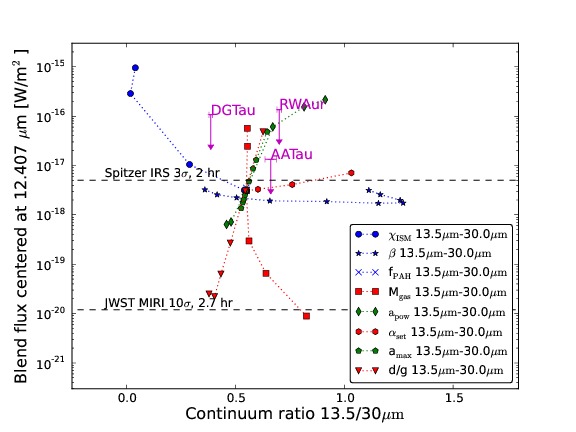}
\includegraphics[width=0.5\textwidth]{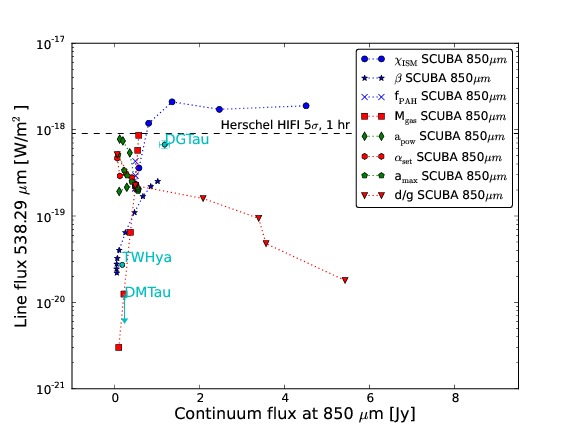}
\caption{Line fluxes vs continuum. Top: 12.407~$\mu$m versus 13.5/30 IRS continuum flux ratios. Bottom: 538.29~$\mu$m versus the 850~$\mu$m SCUBA flux. Color code: blue for parameters not affecting 
the opacity, red for parameters affecting the total continuum opacity, green for parameters affecting the opacity function. Overplotted are the sensitivity limits for Spitzer/IRS, Herschel/HIFI, and JWST/MIRI (references in the 
text). The magenta data points are Spitzer/IRS observations available for stars with spectral type similar to our model central star, and blend fluxes have been extracted from Spitzer spectra following the procedure described in
Appendix~\ref{app4}. Cyan data points are the only Herschel/HIFI observations published for this submm line. All line fluxes have been scaled to 140~pc. Arrows indicate upper limits.}
%The line fluxes from AS 205 and RNO 90 are from \citet{pontoppidan1} %and are spectrally resolved lines from ground based observations.
\label{colors}
\end{figure}

\begin{figure}[htpl]
\centering
\includegraphics[width=0.5\textwidth]{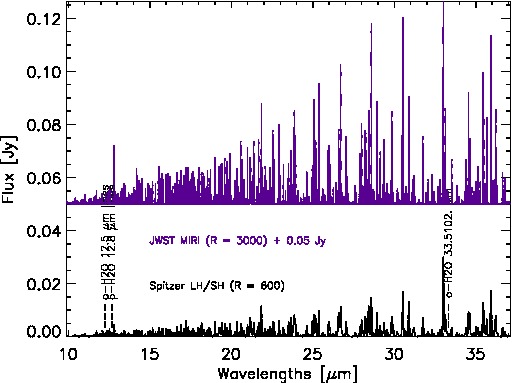}
\includegraphics[width=0.5\textwidth]{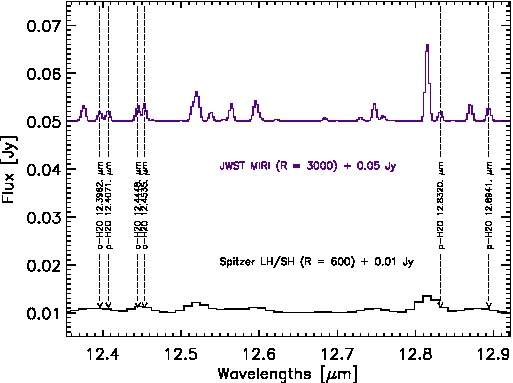}
\caption{Mid-IR spectra of the standard disk model, continuum subtracted, convolved, and properly rebinned with the resolution of Spitzer/IRS LH/SH modules ($R$~=~600) and JWST/MIRI ($R$~=~3000) and arbitrarily shifted by 0.05~Jy.
The bottom panel shows a zoom in the range 12.35-12.39~$\mu$m, JWST spectrum is again shifted by 0.05~Jy. The range of MIRI is artificially extended up to 37~$\mu$m to compare the performances.}
\label{IRSvsJWST}
\end{figure}

\begin{table*}
\caption{Line flux and color variation for the different parameters changes} 
\resizebox{\textwidth}{!}{
\centering
\begin{tabular}{|c|c|c|c|c|c|c|}\hline
\small{Parameter} & type$^{a}$ & \small{Extreme ratios$^{b}$ 538.29$\mu$m} & \small{Flux variation 850 $\mu$m [\%]} & \small{Extreme ratios$^{b}$ 12.407 $\mu$m} & \small{Flux ratio 13.5/30 variation}\\
\hline\hline
$a_\mathrm{pow}$ & 1 & 0.208 & 77.6 & 0.007 & -0.453\\
a$_\mathrm{max}$ & 1 & 3.216 & 85.0 & 24.507 & 0.119\\
$M_\mathrm{gas}$ & 2 & 271.825 & 82.2 & 2600.789 & -0.276\\
$d/g$ & 2 & 0.047 & 98.8 & 5.360 & -0.253\\
$\alpha_\mathrm{set}$ & 2 & 0.456 & 87.5 & 0.454 & -0.484\\
$\beta$ & 3 & 38.236 & 95.2 & 4.070 & -0.911\\
$\chi_\mathrm{ISM}$ & 3 & 12.255 & 89.1 & 274.378 & -0.532\\
$f_\mathrm{PAH}$ & 3 & 2.362 & 0.18 & 1.127 & 0.003\\
\hline
\end{tabular}
}
\tablefoot{$^a$Type 1: parameters affecting global opacity and the opacity function; type 2: parameters affecting global opacity; type 3: parameters not affecting the opacity.
$^b$Defined as ratio between the maximal and minimal value if the flux increases and viceversa if the flux decreases. This gives the maximum variation of line flux because of the parameter range considered.}
\label{rainbow}
\end{table*}

\section{Conclusions}\vspace{5mm}
\label{6}

We explore here the parameter space around a standard T~Tauri disk model  to identify the main drivers of mid- and far-IR water spectroscopy. We primarly analyze two transitions, a mid-IR line at 12.407~$\mu$m and a 
far-IR line at 538.29~$\mu$m, representative of the whole group of transitions considered. In our series of models, we find that the gas mass and dust-to-gas mass ratio simultaneously affect  water lines in the Spitzer and 
Herschel/HIFI wavelength ranges.
Disks with a low gas content or high dust content are undetectable in both regimes. In the first case, the lack of emitting gas explains the nondetections, in the second case, the continuum opacity hides the water reservoirs. 
These two parameters could explain the occurrence of disks without any water features.

Dust properties ($a_\mathrm{max}$, $a_\mathrm{pow}$) only affect  the mid-IR transitions, and can explain the nondetections of Spitzer lines, while the Herschel/HIFI transitions would be detectable in case of several hours of 
integration.
On the other hand, flaring  strongly affects the far-IR regime; so the nondetection of far-IR water lines can also be due to disk geometry.

Models with strong external FUV radiation field ($\chi_\mathrm{ISM}\!>\!10^3\!~G_0$) show strong mid-IR lines with narrow profiles. Herschel/HIFI lines are less affected and increase only by a factor of a few due to the 
external irradiation.

Our choice of implementing the PAHs only as a heating source and not as an opacity source causes the PAHs fraction to only weakly affect both mid- and far-IR lines. Our prescription of the settling is based on a physical
description of the effects of gas vertical mixing on dust grains, but it is not related to the heating associated with the mechanical turbulence in the disk; the two phenomena are disconnected. In that case, the settling affects 
 the inner disk physics very little.

We demonstrate the importance of future telescopes such as JWST in the investigation of water in disks. The higher sensitivity is capable of increasing dramatically the number of disks detected in water. The higher 
spectral resolution will permit us to detect unblended transitions, and to study in detail the physical conditions in the inner water reservoir.

\begin{acknowledgements}
The research leading to these results has received funding from the European Union Seventh Framework Programme FP7-2011 under grant agreement no 284405
\end{acknowledgements}

\clearpage

\bibliographystyle{aa}
\bibliography{Draft}

\clearpage

\appendix

\section{Computation of properties in the line emitting region}\vspace{5mm}
\label{app2}

The different water lines are divided in groups based on their spatial origin. The latter is defined as that part of the disk from which a cumulative radially and vertically integrated flux of 15\% to 85\% of the total flux is 
produced. 
For each of these regions, averaged physical conditions $\left<X\right>$ are derived, such as the density of water, density of the collisional partners (e$^{-}$, H, H$_{2}$), $T_\mathrm{gas}$ and $T_\mathrm{dust}$ by numerically
integrating over the emitting region (Eq.~\ref{(6)}), i.e.,\\

\begingroup
\small
\begin{equation}\label{(6)}
\centering
\left<X\right>=\frac{\int\limits_{r_\mathrm{in}}^{r_\mathrm{out}}\int\limits_{z_\mathrm{low}}^{z_\mathrm{high}}Xn_\mathrm{gas}2\pi rdrdz}{\int\limits_{r_\mathrm{in}}^{r_\mathrm{out}}\int\limits_{z_\mathrm{low}}^{z_\mathrm{high}}n_\mathrm{gas}2\pi rdrdz} 
\end{equation}
\endgroup

In order to determine the continuum extinction, a vertical integration is performed from the disk surface to the midplane. 
$\tau_{\nu}$ is the continuum optical depth at the frequency of the line. It is related to the dust density $\rho_\mathrm{dust}$ and dust mass extinction coefficient $\kappa_{\nu}$. The integration is performed numerically, as follows: 

\begin{equation}\label{(11)}
\centering
\tau_{\nu}(z)=\int_{z_\mathrm{max}(r)}^{z(r,\tau_{\nu}=1)}\rho(r,z)\kappa_{\nu}dz
\end{equation}
Through interpolation, we find the value at which $\tau_{\nu}$ = 1. The optical depth is evaluated at the radial midpoint of the emitting region.\\

\section{Model details}
\label{app3}

The following appendix shows a number of additional plots. These can be useful for the interpretation of the main results and conclusions reported in the paper. 

Here we show the Herschel/HIFI fundamental o-H$_2$O and p-H$_2$O lines fluxes behavior with the different parameters.

\begin{figure*}[htpl!]
\centering
\includegraphics[width=0.4\textwidth]{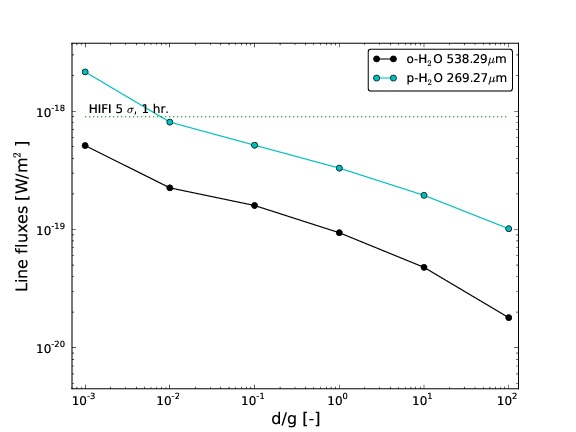}
\includegraphics[width=0.4\textwidth]{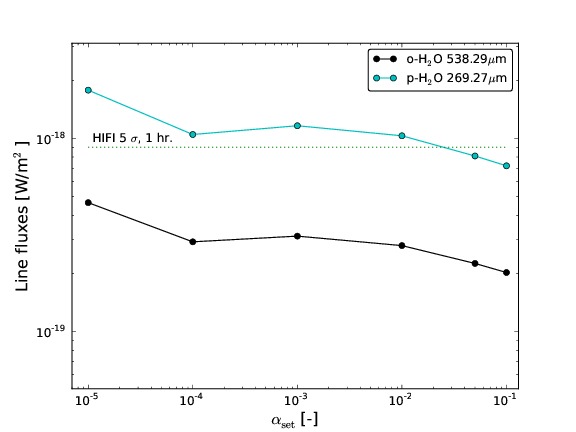}
\includegraphics[width=0.4\textwidth]{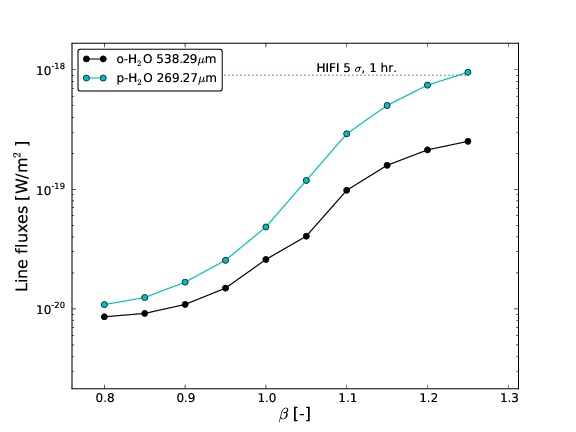}
\includegraphics[width=0.4\textwidth]{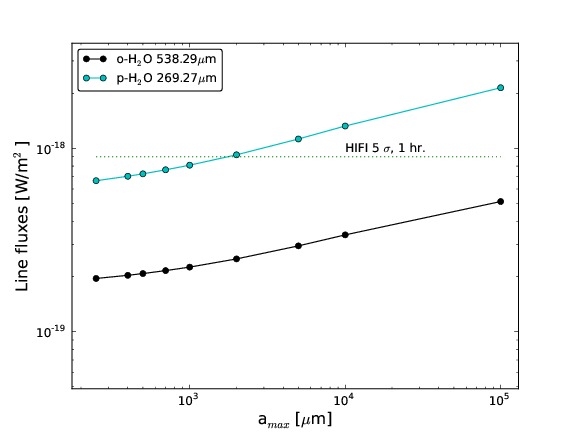}
\includegraphics[width=0.4\textwidth]{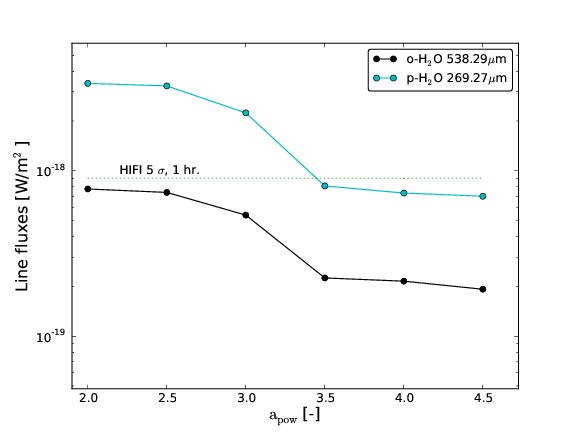}
\includegraphics[width=0.4\textwidth]{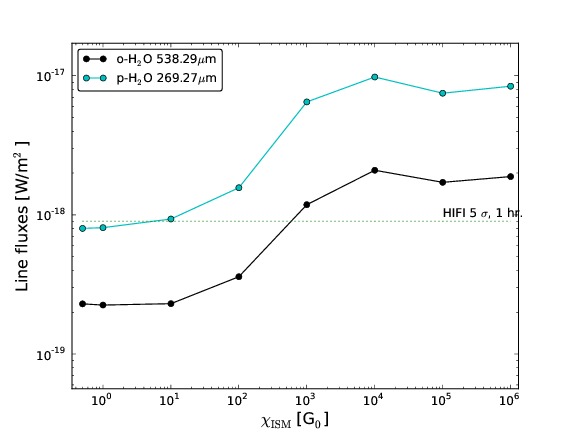}
\includegraphics[width=0.4\textwidth]{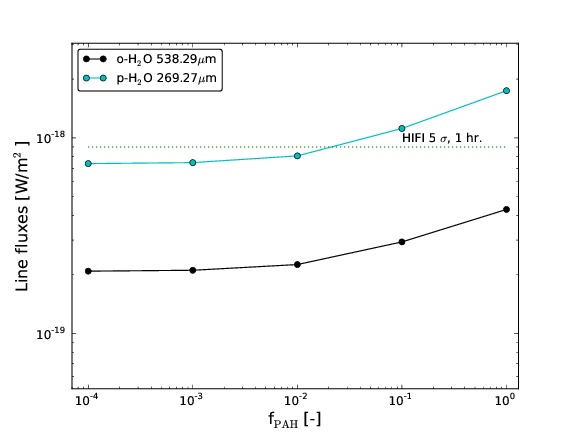}
\includegraphics[width=0.4\textwidth]{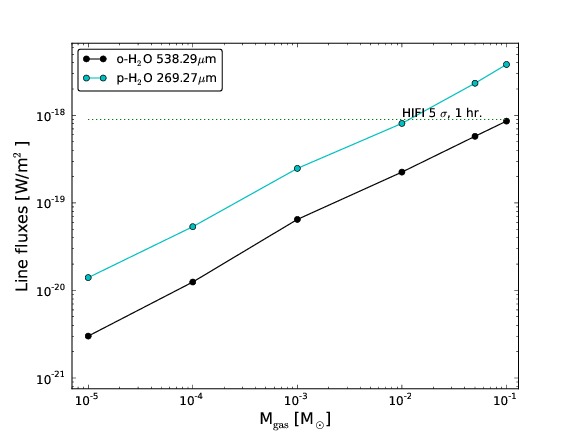}
\caption{Line fluxes behavior for Herschel/HIFI lines 538.29~$\rm\mu m$ and 269.27~$\rm\mu m$ for different model series. First row left: dust-to-gas mass ratio, $d/g$. First row right: mixing parameter, 
$\alpha_\mathrm{set}$. Second row left: disk flaring, $\beta$. Second row right: maximum grain size a$_\mathrm{max}$. Third row left: power law index of grain size distribution, $a_\mathrm{pow}$. Third 
row right: strength of UV radiation field, $\chi_\mathrm{ISM}$. Third row left: PAH abundance, $f_\mathrm{PAH}$. Bottom right: disk gas mass, $M_\mathrm{gas}$.}
\label{HIFI}
\end{figure*}

Here, we also present plots collecting the main properties of the standard model: $T_\mathrm{gas}$, $T_\mathrm{dust}$, the local dust-to-gas mass ratio, local average dust size, and the main heating and 
cooling processes.

\begin{figure*}[htpl!]
\centering
\begin{minipage}[l]{0.36\textwidth}
\includegraphics[width=1.\textwidth]{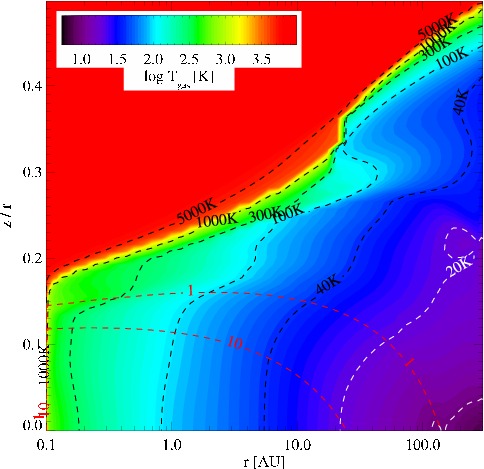}
\end{minipage}
\begin{minipage}[r]{0.36\textwidth}
\includegraphics[width=1.\textwidth]{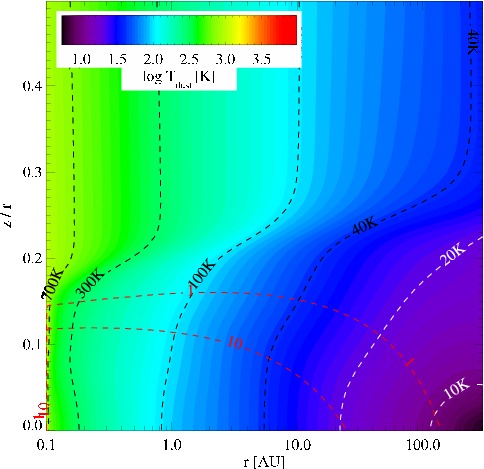}
\end{minipage}
\begin{minipage}[l]{0.36\textwidth}
\includegraphics[width=1.\textwidth]{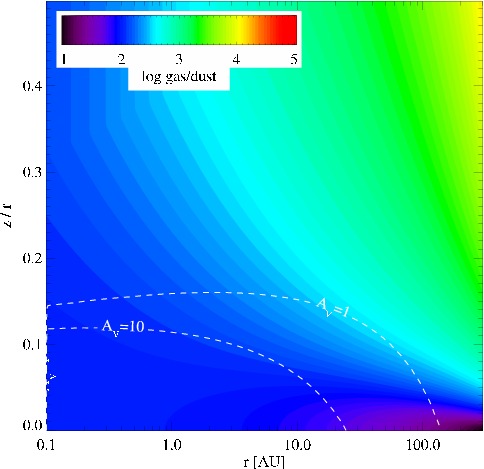}
\end{minipage}
\begin{minipage}[r]{0.36\textwidth}
\includegraphics[width=1.\textwidth]{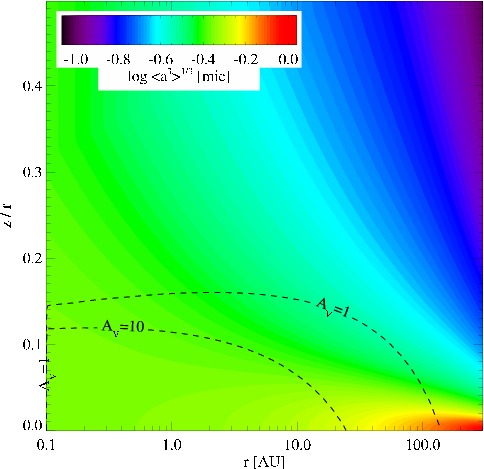}
\end{minipage}
\begin{minipage}[l]{0.36\textwidth}
\includegraphics[width=1.\textwidth]{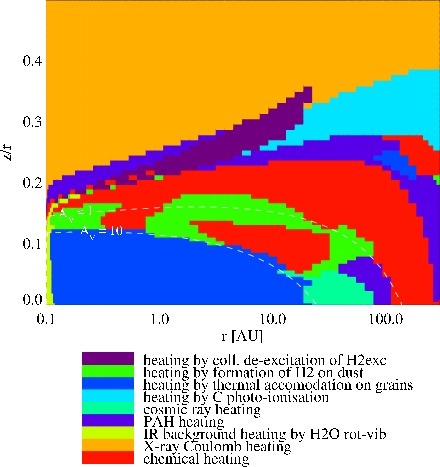}
\end{minipage}
\begin{minipage}[r]{0.36\textwidth}
\includegraphics[width=1.\textwidth]{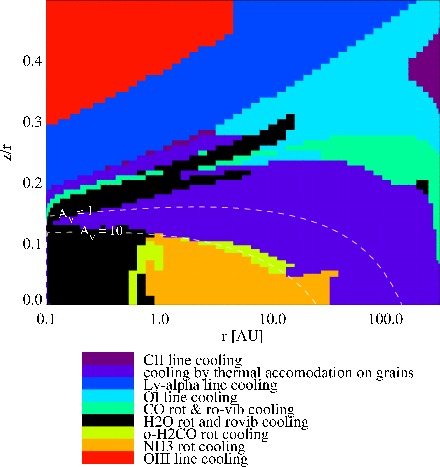}
\end{minipage}
\caption{Standard model, from top to bottom and from left to right: $T_\mathrm{gas}$ with overplotted gas temperature contours (white/black) and $A_\mathrm{V}~=~1,~10$ (red), $T_\mathrm{dust}$ with overplotted temperature 
contours (black/white) and $A_\mathrm{V}~=~1,~10$ (red), dust-to-gas mass ratio, average dust grain size, main cooling processes (CII fine structure lines, dust grain-gas thermal exchange, Ly $\alpha$ line, neutral O line 
cooling, CO rotational and ro-vibrational cooling, water rotational and ro-vibrational cooling, formaldehyde rotational cooling, ammonia rotational cooling, OIII line cooling), main heating processes (collisional excitation of 
vibrationally excited H$_2$, chemical heating due to H$_2$ formation on dust grains, dust grain-gas thermal exchange, heating from the superthermal photo-electrons produced by C, heating by cosmic rays, heating by the 
photo-electrons produced by polycyclic aromatic hydrocarbons, IR heating due to the water transitions, heating by mutual electron collisions, Coulomb heating, produced by X-rays photoionization, reaction free energy heating from
exothermic reactions).}
\label{Standard model}
\end{figure*}

Line ratio plots are used to compare the relative behavior of the mid- and far-IR lines, considering two representative Spitzer water transitions and the Herschel fundamental ortho and para water lines. To show 
 the trend clearly, we add a parabolic regression to the data (magenta lines).

\begin{figure*}[htpl!]
\centering
\includegraphics[width=0.4\textwidth]{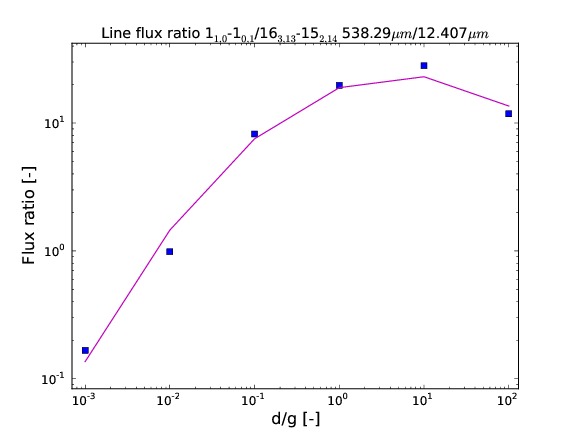}
\includegraphics[width=0.4\textwidth]{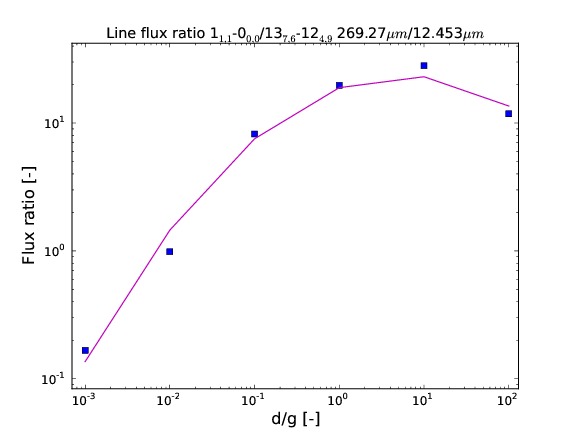}
\includegraphics[width=0.4\textwidth]{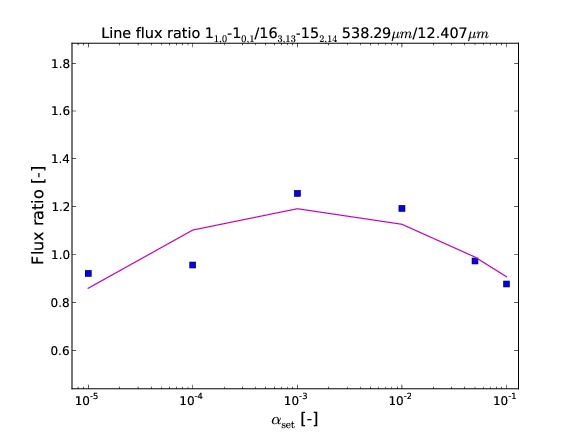}
\includegraphics[width=0.4\textwidth]{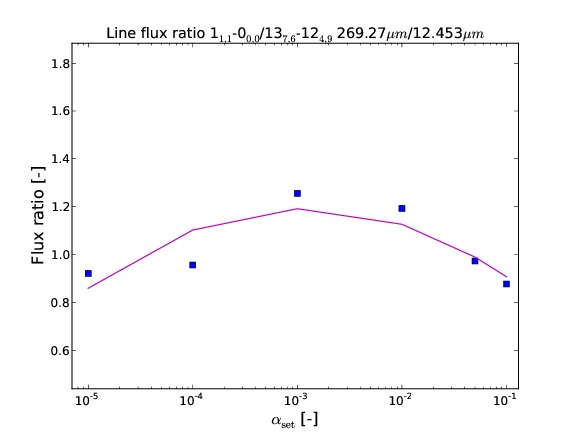}
\includegraphics[width=0.4\textwidth]{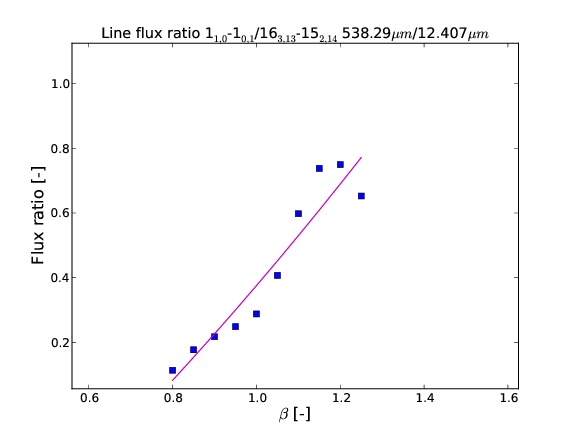}
\includegraphics[width=0.4\textwidth]{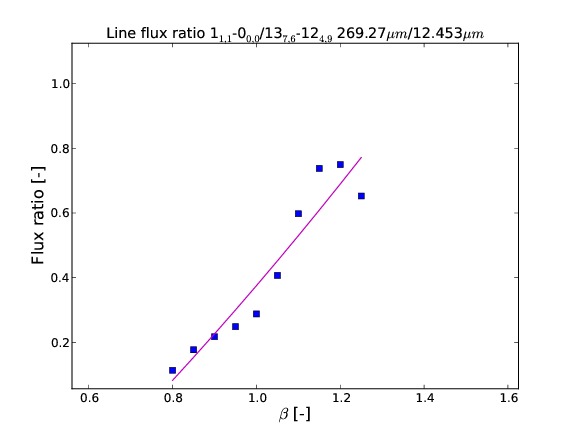}
\includegraphics[width=0.4\textwidth]{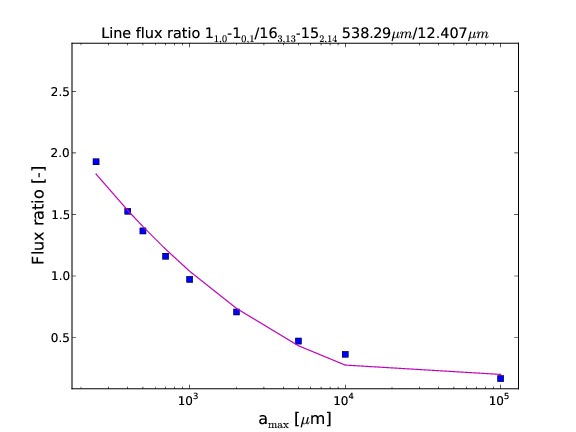}
\includegraphics[width=0.4\textwidth]{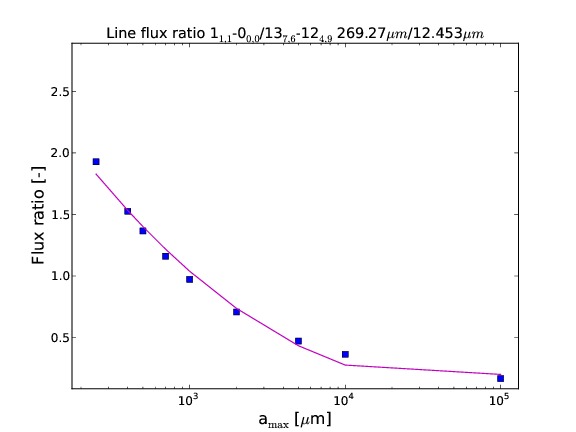}
\caption{Ratios between Herschel/HIFI \& Spitzer/IRS lines 538.29~$\rm\mu m$/12.407~$\rm\mu m$ and 269.27~$\rm\mu m$/12.453~$\rm\mu m$ for different model series. First row: dust-to-gas mass ratio, $d/g$. Second row:
mixing parameter, $\alpha_\mathrm{set}$. Third row: disk flaring, $\beta$. Fourth row: maximum grain size, a$_\mathrm{max}$.}
\label{Rat21}
\end{figure*}

\begin{figure*}[htpl!]
\centering
\includegraphics[width=0.4\textwidth]{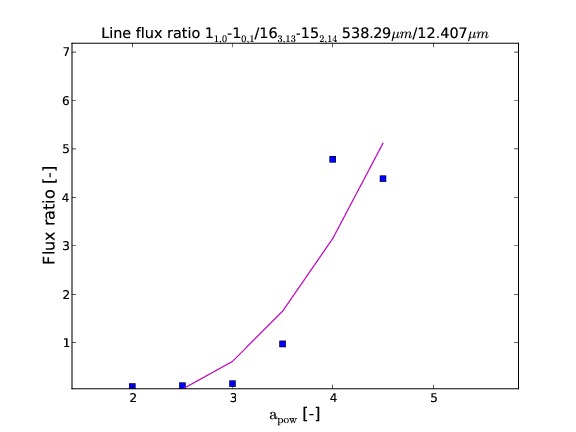}
\includegraphics[width=0.4\textwidth]{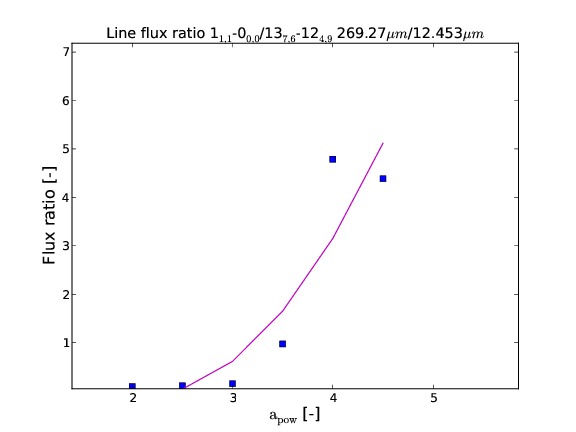}
\includegraphics[width=0.4\textwidth]{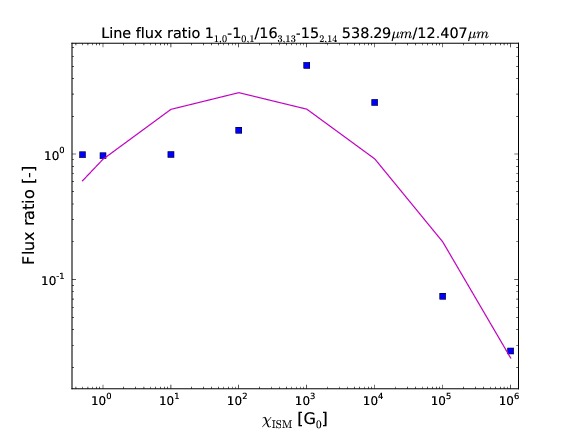}
\includegraphics[width=0.4\textwidth]{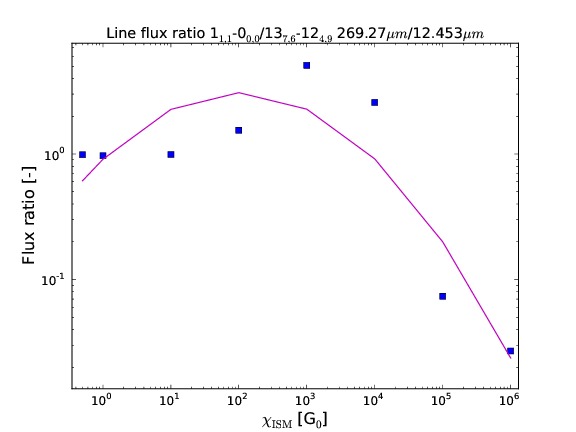}
\includegraphics[width=0.4\textwidth]{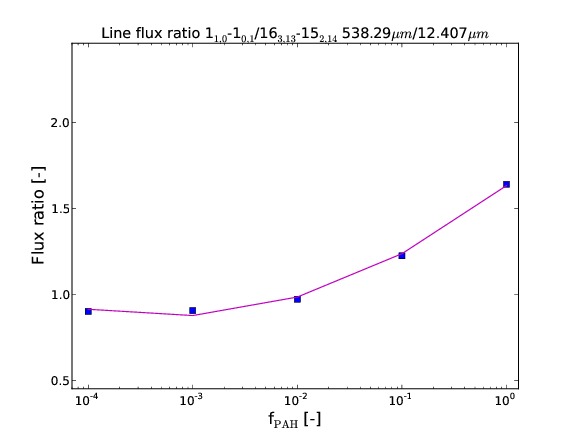}
\includegraphics[width=0.4\textwidth]{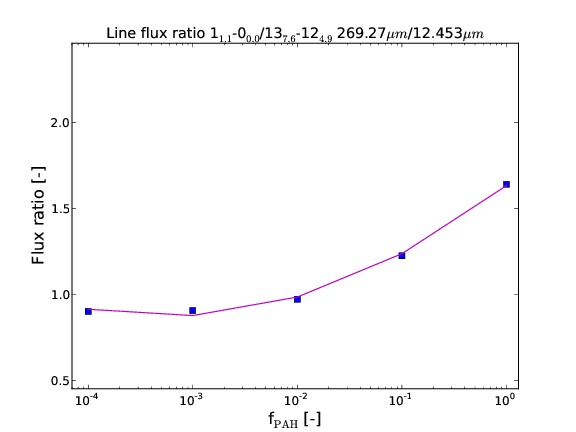}
\includegraphics[width=0.4\textwidth]{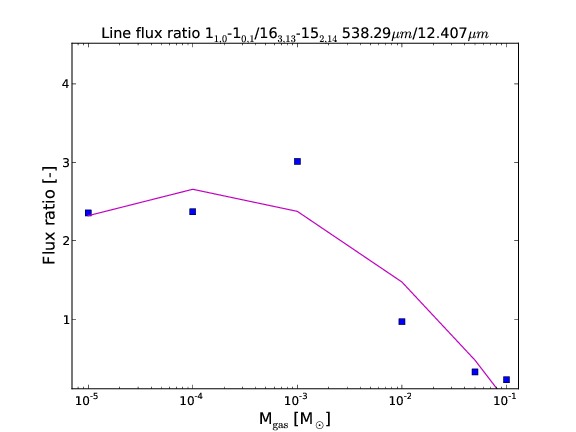}
\includegraphics[width=0.4\textwidth]{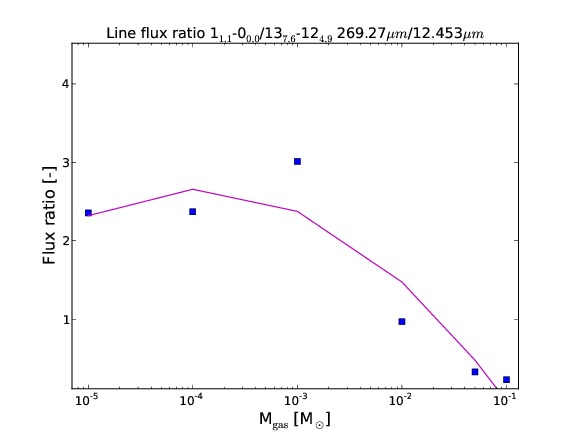}
\caption*{Fig.~\ref{Rat21} Ratios between Herschel/HIFI \& Spitzer/IRS lines 538.29~$\rm\mu m$/12.407~$\rm\mu m$ and 269.27~$\rm\mu m$/12.453~$\rm\mu m$ for different model series. Top: power law index of grain size distribution,
$a_\mathrm{pow}$. Second line: strength of UV radiation field, $\chi_\mathrm{ISM}$. Third line: PAH abundance, $f_\mathrm{PAH}$.\ Fourth: disk gas mass, $M_\mathrm{gas}$.}
\end{figure*}

\section{Mid- and far-IR observations}
\label{app4}

\begin{table*}
%\caption{Observations scaled to a distance of 140 pc}
\caption{Line and continuum observations of the targets discussed in the main paper.}
\centering
\begin{tabular}{llllll}
\hline\hline
Object & $d$ [pc]  & $12.407~\mu$m blend flux$^{**}$ & $13.5/30~\mu$m continuum & $538.29~\mu$m line flux & $850~\mu$m continuum$^{(5)}$ \\
 & &  [10$^{-17}$~W/m$^2$] & flux ratio$^{(1)}$ & [10$^{-20}$~W/m$^2$] &  [Jy]\\ \hline
AA Tau & 140 & $<$1.33 & 0.663$\pm$0.027 & n.a. & n.a. \\
%DM Tau & 140 & $<$0.23 & 0.150$\pm$0.004 & $<$1.15$^{(5)}$ & 0.237$\pm$0.012 \\
DM Tau & 140 & n.a. & n.a. & $<$1.15$^{(4)}$ & 0.237$\pm$0.012 \\
DG Tau & 140 & $<$10.63 & 0.388$\pm$0.008 & 67.00$\pm$7.00$^{(2)}$ & 1.180$\pm$0.118 \\
RW Aur & 144 & $<$13.51$^*$ & 0.702$\pm$0.010$^*$ & n.a. & n.a. \\
%AS 205 & 125 & 1.83$\pm$0.16$^{(1)}$ & 0.552$\pm$0.005 & n.a. & n.a. \\
%RNO 90 & 125 & 0.72$\pm$0.09$^{(1)}$ & 0.567$\pm$0.005 & n.a. & n.a. \\
TW Hya & 56 & n.a. & n.a. & 2.72$\pm$0.12$^{(3)*}$ & 0.182$\pm$0.018$^*$\\ \hline
\end{tabular}
\tablefoot{(1) Ratios have been computed from c2d IRS spectra \citep{evans}; (2) from \citet{podio}; (3) scaled to 140 pc from \citet{hogerheijde}; (4) from \citet{bergin2}; (5) scaled to 140~pc from \citet{difrancesco}; 
(*) flux scaled to 140~pc; (**) 12.407~$\mu$m fluxes extraction procedure is described in the text of this Appendix.}
\label{Obs}
\end{table*}

Far-IR line fluxes have been obtained from the literature. Mid-IR blends have been extracted from IRS spectra fitting a Gaussian function plus a linear continuum, computing the error in the flux as noise of the continuum,

\begin{equation}\label{fit}
F(\lambda) = A \times e^{(\lambda-\lambda_0)^2/(2\sigma)^2} + B \times \lambda + C
\end{equation}
This procedure is formally identical to that used in \citep{pontoppidan2}. Table~\ref{Obs} lists all the observational data shown in Fig.~\ref{colors} with respective references.

The mid-IR color 13.5/30 continuum flux is defined as pure ratio between the Spitzer/IRS fluxes at 13.5~$\mu$m and 30~$\mu$m, similarly to what is done in \citet{acke2} for the 30/13.5 continuum ratio. The observations 
overplotted include the errors propagated accordingly to

\begin{equation}\label{err}
\sigma_{\mathrm{ratio}}=\frac{F_{13.5}}{F_{30}}\cdot\sqrt{\left(\frac{\sigma_{13.5}}{F_{13.5}}\right)^2+\left(\frac{\sigma_{30}}{F_{30}}\right)^2}
\end{equation}

\end{document}